\documentclass[reprint,superscriptaddress,nofootinbib,amsmath,amssymb,aps,prd]{revtex4-2}

\usepackage[utf8]{inputenc}
\usepackage[T1]{fontenc}

\usepackage{microtype}

\usepackage{mathtools, booktabs}

\usepackage{graphicx}

\usepackage[dvipsnames]{xcolor}
\usepackage[colorlinks=true, allcolors=Blue]{hyperref}
\usepackage[capitalise]{cleveref}

\usepackage{mathrsfs, bm, tensor}

\let\t\tensor

\def\dd{\mathrm{d}}
\def\rmd{\mathrm{d}}
\def\ii{\mathrm{i}}
\def\half{\tfrac{1}{2}}

\usepackage{tikz}
\newcommand{\Arrow}[1]{\parbox{#1}{\tikz{\draw[->](0,0)--(#1,0);}}}
\newcommand{\LArrow}[1]{\parbox{#1}{\tikz{\draw[->](#1,0)--(0,0);}}}

\newcommand{\Mvv}{M^{\mathrm{v} \Arrow{.15cm} \, \mathrm{v} } }
\newcommand{\Mdd}{M^{\mathrm{d} \Arrow{.15cm} \, \mathrm{d}} }
\newcommand{\Mdv}{M^{\mathrm{d} \Arrow{.15cm} \, \mathrm{v}} }
\newcommand{\Mvd}{M^{\mathrm{v} \Arrow{.15cm} \, \mathrm{d}} }

\newcommand{\MvvR}{M^{\mathrm{v} \LArrow{.15cm} \, \mathrm{v} } }
\newcommand{\MddR}{M^{\mathrm{d} \LArrow{.15cm} \, \mathrm{d}} }
\newcommand{\MdvR}{M^{\mathrm{d} \LArrow{.15cm} \, \mathrm{v}} }
\newcommand{\MvdR}{M^{\mathrm{v} \LArrow{.15cm} \, \mathrm{d}} }

\newcommand{\MvvH}{\hat{M}^{\mathrm{v} \Arrow{.15cm} \, \mathrm{v} } }
\newcommand{\MddH}{\hat{M}^{\mathrm{d} \Arrow{.15cm} \, \mathrm{d}} }
\newcommand{\MdvH}{\hat{M}^{\mathrm{d} \Arrow{.15cm} \, \mathrm{v}} }
\newcommand{\MvdH}{\hat{M}^{\mathrm{v} \Arrow{.15cm} \, \mathrm{d}} }

\numberwithin{equation}{section}
\renewcommand\theequation{\arabic{section}.\arabic{equation}}

\usepackage{pifont}
\makeatletter\def\@fnsymbol#1{{\ifcase#1\or\ding{168}\or\ding{169}\or\ding{170}\or\ding{171}\fi}}\makeatother

\begin{document}

\title{Gravitational wave memory and its effects on particles and fields}

\author{Abraham I. Harte}
\email{abraham.harte@dcu.ie}
\affiliation{Centre for Astrophysics and Relativity (CfAR), School of Mathematical Sciences, Dublin City University, Glasnevin, Dublin 9, Ireland}

\author{Thomas B. Mieling}
\email{thomas.mieling@univie.ac.at}
\affiliation{University of Vienna, Faculty of Physics and Research Network TURIS, Boltzmanngasse~5, 1090 Vienna, Austria}

\author{Marius A. Oancea}
\email{marius.oancea@univie.ac.at}
\affiliation{University of Vienna, Faculty of Physics, Boltzmanngasse~5, 1090 Vienna, Austria}

\author{Elisabeth Steininger}
\email{elisabeth.steininger@univie.ac.at}
\affiliation{University of Vienna, Faculty of Physics, Vienna Doctoral School in Physics (VDSP) and Research Network TURIS, Boltzmanngasse~5, 1090 Vienna, Austria}

\begin{abstract}
Gravitational wave memory is said to arise when a gravitational wave burst produces changes in a physical system that persist even after that wave has passed. This paper analyzes gravitational wave bursts in plane wave spacetimes, deriving memory effects for timelike and null geodesics, massless scalar fields, and massless spinning particles whose motion is described by the spin Hall equations. We find that all such effects are characterized by four “memory tensors,” three of which are independent. We also show that memory effects for null geodesics can have strong longitudinal components, even in vacuum general relativity. When considering massless particles with spin, we solve the spin Hall equations analytically by showing that there exists a conservation law associated with each conformal Killing vector. For the scattering of fields by gravitational waves, we show that given any solution to the massless scalar field equation in flat spacetime, a weak-field solution in a plane wave spacetime can be generated just by applying an appropriate differential operator---an operator that is constructed from the aforementioned memory tensors. Memory effects for scalar fields are illustrated for both incoming plane waves and higher-order Gaussian beams. We also present a numerical comparison between the spin Hall equations and the full evolution of localized wave packets with angular momentum. Although we work in plane wave spacetimes, which are physically idealized, similar results are also expected to apply for sufficiently small systems affected by distantly generated gravitational waves. Using the Penrose limit, our results may also apply to ultrarelativistic systems at arbitrary locations in arbitrary (even nonradiating) spacetimes.
\end{abstract}

\maketitle

\section{Introduction}

Gravitational waves provide a novel probe of astrophysical phenomena. By encoding information about their sources, they provide access to highly dynamical systems with strong gravitational fields. Gravitational waves can also be lensed or otherwise affected by the spacetime through which they propagate, thus providing information also about mass distributions between sources and observers. Gravitational waves are, however, measured by observing their interactions with different physical systems. As gravity affects almost every conceivable observation, a wide variety of physical systems can, in principle, be used as gravitational wave detectors. While the state of a detector typically varies throughout its interaction with any passing gravitational wave, there are cases in which a detector's state before a wave arrives differs from its state after that wave has left. Such differences are referred to as gravitational wave memory. 

The first prediction of a gravitational memory effect is typically attributed to Zel'dovich and Polnarev in 1974 \cite{Zeldovich1974} (although see also Ref. \cite{newman1966note}), who examined gravitational-wave bursts generated by the flyby of two massive objects in linearized general relativity. They found that the initial and the final separations of pairs of freely falling test particles can differ due to the resulting gravitational waves. Going beyond the weak-field approximation, a nonlinear  memory effect was later obtained by Christodoulou \cite{1991PhRvL..67.1486C} based on mathematical methods used in the proof of the stability of Minkowski spacetime \cite{SEDP_1989-1990____A15_0, ChristodoulouKlainerman1994}. This has as its physical origin the flux of effective stress-energy which is associated with the gravitational waves emitted by a compact system \cite{Wiseman1991, PhysRevD.45.520}. Motivated in part by observational prospects \cite{1987Natur.327..123B,PhysRevLett.117.061102, PhysRevD.101.083026}, gravitational wave memory has since been studied in a wide variety of contexts, primarily using linearized gravity \cite{Tolish2014, Garfinkle2022,Bieri_2024} or the post-Newtonian approximation \cite{PhysRevD.37.1410, PhysRevD.46.4304, Favata_2010, PhysRevD.95.084048}. Memory has additionally been used to better understand the structure of null infinity \cite{spinMem, 2017PhRvD..95d4002F, Satishchandran2019, CMMem, Bieri2021,1992CQGra...9.1639F,Bieri_2024}, and has also been discussed on black hole horizons \cite{Donnay2018, Rahman2020}. 

Much of the literature has focused on studying aspects of the spacetime geometry that may be interpreted as producing permanent changes in the separations of freely falling test particles. However, gravitational waves can produce a variety of other persistent changes in the physical systems through which they pass \cite{HarteStrongLensing, HartePWOptics2, HarteGenMemory, HartePWMemory, Grant2022, GrantMemory}. It is these changes that we investigate here. Since we are interested in how gravitational waves affect spatially compact physical systems---and not on, e.g., the large-scale structure of those waves or on their sources---we focus on the effects of gravitational \textit{plane waves}. More precisely, we work with plane wave spacetimes, which are exact solutions to the Einstein field equations. These are a subclass of the more general family of pp-wave spacetimes, and pp-waves are, in turn, a subclass of the Kundt family of spacetimes \cite{EhlersKundt, Hall, 2023GReGr..55...40R, griffiths2009}. Plane wave spacetimes represent a gravitational-wave analog of the scalar or electromagnetic plane waves which are well known in flat spacetime. 

From a global perspective, plane wave spacetimes do not provide a physically reasonable model of gravitational radiation: They are not asymptotically flat, nor even globally hyperbolic \cite{Penrose1965}, they carry an infinite amount of energy, and they do not have sources. Nevertheless, it is still reasonable to consider these spacetimes as models of gravitational radiation when considering their effects on systems with small spatial extent. At large enough distances, the gravitational waves emitted by compact sources are expected to be well approximated by the plane wave geometry. Separately, plane wave spacetimes also arise as Penrose limits \cite{Penrose1976, BlauWeissPL, BlauNotes}, and might therefore be viewed as describing the ultrarelativistic geometries near arbitrary null geodesics in arbitrary background spacetimes. Regardless, plane waves provide particularly clean models for the effects of gravitational radiation, as it is possible to consider waves whose curvature is nonzero only for a finite time. Any system which is affected by such a wave therefore evolves from one locally flat spacetime to another, allowing its initial and final states to be easily compared.

This paper studies gravitational memory effects on test particles and test fields in plane wave spacetimes. Unlike in much of the previous work on memory in plane wave spacetimes \cite{Zhang2017, Zhang_2018,Zhang_2018(2),Steinbauer_2019,Zhang_2018(3), PhysRevD.98.044037,ZhangSoftGravitons, Elbistan2023, PhysRevD.101.064022,zhang2024displacement,achour2024displacement}, we allow for gravitational wave bursts with arbitrary waveforms (see, however, Refs. \cite{HarteStrongLensing, HartePWOptics2, HartePWMemory, Adamo2017}). One of our main results is that a large class of physical observables can be completely expressed in terms of Jacobi propagators. These propagators are typically introduced to solve the geodesic deviation equation. However, we show that the Jacobi propagators characterize not only the behaviors of \textit{nearby} geodesics, but of \textit{all} geodesics in plane wave spacetimes. They also characterize the behaviors of massless test particles with longitudinal angular momentum and of massless scalar fields. For sandwich wave spacetimes where the curvature is nonzero only for a finite time, we show that the Jacobi propagators can be fully expressed in terms of four constant ``memory tensors,'' three of which are independent. These tensors fully determine all scattering processes and observables we consider.

The first memory effects we discuss are those associated with timelike and null geodesics, which are found to depend on all four of the aforementioned memory tensors. The physical character of these effects can, in general, be considerably more complicated than for the classical memory effect associated with initially comoving pairs of timelike geodesics. Null geodesics, for example, exhibit memory effects that are not only transverse, but also longitudinal, even in vacuum general relativity and even at the linearized level. 

Moving beyond the geodesic approximation, we also consider the scattering of massless particles with longitudinal angular momentum. Such particles move along \textit{nongeodesic} null trajectories determined by the gravitational spin Hall equations \cite{GSHE_rev,HarteOancea}, which are a special case of the Mathisson-Papapetrou equations \cite{Dixon70a}. Physically, the spin Hall equations describe the trajectories of high-frequency localized (electromagnetic \cite{GSHE2020,HarteOancea, Frolov2020, SHE_QM1, PhysRevD.109.064020,Frolov2024}, linearized gravitational \cite{GSHE_GW,SHE_GW, GSHE_lensing, GSHE_lensing2,Frolov2024(2)}, massless Dirac \cite{GSHE_Dirac}, or even some scalar) wave packets  carrying longitudinal angular momentum. Memory effects for \textit{massive} spinning particles have instead been discussed in, e.g., Refs.~\cite{HarteGenMemory, HartePWMemory, Seraj2022, Seraj2023}.

The spin Hall description of wave packets is only approximate, and only describes their ``bulk'' features. However, memory effects can also be considered for fully generic test fields\footnote{Although we focus here on scalar fields, spin-raising procedures can be used to map scalar fields into, e.g., electromagnetic vector potentials or metric perturbations \cite{Mason_Ward, Araneda2022, Kulitskii2023, Tang2023, Audagnotto2024}.}, which we consider from several perspectives. We obtain a general integral formula for the scattering of scalar test fields, and also derive a result which allows fields in flat spacetimes to be easily mapped into solutions in weakly curved sandwich wave spacetimes. Examples are considered as well, including the scattering of ingoing plane waves, and of scalar fields constructed from counterpropagating Hermite–Gauss and Laguerre–Gauss beams. Lastly, we use a numerical implementation of the Kirchhoff integral to compare exact solutions of the wave equation with predictions provided by the spin Hall equations. 

This paper is structured as follows. We start in \cref{Sect:PWs} with a presentation of the theoretical tools used for our study of memory effects. We review the main properties of plane wave spacetimes and introduce the Jacobi propagators. For sandwich wave spacetimes, we show that these propagators can be entirely expressed in terms of four memory tensors that  characterize all scattering processes discussed in this paper. \cref{Sect:Geodesics} discusses geodesic motion. Solutions of the geodesic equations are written in terms of the Jacobi propagators, and in the case of sandwich wave spacetimes, the transverse scattering of geodesics is expressed very simply in terms of the memory tensors. We also discuss longitudinal memory effects, focusing mainly on null geodesics. \cref{Sect:SH} examines the nongeodesic motion of massless spinning particles, solving the spin Hall equations in terms of the Jacobi propagators. Memory effects for massless spinning particles are then expressed in terms of the memory tensors, similar to the geodesic case. In \cref{Sect:FieldProp}, we analyze the dynamics of massless scalar fields in plane wave spacetimes.

There are five appendices. \cref{app:notation} summarizes our notation. \cref{App:Examples} describes example plane waves and their memory tensors. \cref{App:Ward} reviews Ward's \cite{Ward1987} progressing-wave solutions for test scalar fields on plane wave backgrounds, and shows that our Kirchhoff-like formula \eqref{Kirchhoff} is a special case. \cref{Sect:EMmemory} describes memory effects for charged particles in electromagnetic plane waves in flat spacetime, which contrasts with the gravitational memory effects considered elsewhere in this paper. Finally, \cref{app:beam_derivatives} collects certain technical results on Hermite–Gauss and Laguerre–Gauss beams.  

\section{plane wave spacetimes and memory tensors}
\label{Sect:PWs}

This section reviews basic properties of plane wave spacetimes and then introduces a set of four memory tensors which can be used to describe all observables discussed in this paper. \cref{Sect:PWreview} begins by reviewing the basic physical and geometric properties of plane wave spacetimes.  \cref{Sect:Jacobi} then reviews the Jacobi propagators and their properties in a plane wave context. Next, \cref{Sect:SandWave} specializes to sandwich wave spacetimes and introduces the memory tensors. Lastly, \cref{Sect:approxAB} shows that for weak gravitational waves, all four memory tensors are determined by the zeroth, the first, and the second moments of the gravitational waveform. 

\subsection{Review of plane wave spacetimes}
\label{Sect:PWreview}

Exact plane wave spacetimes are commonly specified in one of two classes of coordinate systems: Brinkmann or Rosen. Each of these coordinate systems has its advantages and disadvantages. For example, the Einstein field equations are trivial in Brinkmann coordinates but not in Rosen. By contrast, many geodesics look trivial in Rosen coordinates but not in Brinkmann. Rosen coordinates are also directly related to the transverse-traceless gauge which is commonly used in first-order perturbation theory, so the waveforms that are natural in that context are somewhat more familiar than the waveforms which appear in Brinkmann coordinates. A detailed discussion of these and other differences can be found in Refs.~\cite{BlauNotes, HartePWOptics2}. Further properties of plane wave spacetimes are reviewed in, e.g., Refs.~\cite{EhlersKundt, Beem2017, HarteDrivas, HarteStrongLensing, griffiths2009, 2023GReGr..55...40R}. 

Except in \cref{Sect:PWscatter} and \cref{App:Ward} below, we mainly use Brinkmann coordinates $(u,v,x,y)$ to describe gravitational plane waves. The line element is then given by
\begin{equation}
    \rmd s^2 = - 2 \rmd u\, \rmd v + \rmd x^2 + \rmd y^2 + H_{ij}(u) x^i x^j \rmd u^2 ,
    \label{Brinkmann}
\end{equation}
where the indices $i,j,\ldots$ are associated with the two “transverse” and spacelike coordinates ${x = x^1}$ and ${y = x^2}$. The ``phase'' coordinate $u$ is null, and constant-$u$ hypersurfaces may be viewed as the wavefronts of the gravitational wave. The waveform here is described by $H_{ij} = H_{(ij)}(u)$, and the exact Einstein field equations require only that this be trace-free in vacuum. Therefore, any $2 \times 2$ symmetric, trace-free, matrix function of one variable may be used to describe the waveform of a vacuum gravitational wave. The two free scalar functions that make up that matrix may be viewed as waveforms associated with the two polarization states of the gravitational wave.

Unlike a first-order gravitational waveform $h_{ij}$ which might be familiar from flat-spacetime perturbation theory in transverse-traceless gauge, $H_{ij}$ has units of curvature; these waveforms are related via  \cite{HartePWOptics2}
\begin{equation}
    H_{ij} = \frac{1}{2} \frac{ \rmd^2 h_{ij} }{ \rmd u^2 } + \mathcal{O}(h^2).   
    \label{hToH}
\end{equation}
For an exact plane wave, all nontrivial Riemann components follow from
\begin{equation}
    R_{iuju} = - H_{ij}.
    \label{pwRiemann}
\end{equation}
It follows that if $H_{ij}$ vanishes on a particular $u = \mathrm{constant}$ wavefront, the spacetime there is at least locally flat. If $H_{ij}$ is nonzero in a region but can be diagonalized using a constant (i.e., $u$-independent) rotation, the wave is said to be linearly polarized in that region.

Geometrically, Brinkmann coordinates are a type of Fermi normal coordinate system whose origin is the null geodesic $v=x^i = 0$. As is typical in Fermi coordinates, the presence of curvature results in the metric components growing quadratically as one moves away from the origin. In plane wave spacetimes, this quadratic growth rate is, in fact, exact; it is given by the $H_{ij} x^i x^j \rmd u^2$ term in the Brinkmann line element. In a generic spacetime, quadratic growth in the metric components would instead appear only as a leading-order approximation near the origin. 

This suggests that plane wave metrics can describe the leading-order terms in an expansion of the metric around \emph{generic} null geodesics in \emph{generic} spacetimes. Indeed, it was shown in Ref.~\cite{BlauWeissPL} that if Fermi coordinates are applied to an arbitrary null geodesic in an arbitrary spacetime, there is a sense in which the leading-order metric near that geodesic is always a gravitational plane wave in Brinkmann coordinates. The particular plane wave that results from this process is called the Penrose limit \cite{BlauNotes, Penrose1976} of the reference null geodesic, and the relevant $H_{ij}$ is determined by certain components of the Riemann tensor along the reference geodesic. The Penrose limit thus provides a sense in which plane wave geometries generically encode local aspects of arbitrary spacetimes as seen by ultrarelativistic observers. Of course, plane waves also model local aspects of the far-field gravitational radiation emitted by compact sources, even for observers which are not in any sense relativistic. 

Given a particular plane wave spacetime, the Brinkmann coordinates are not unique. Different frame vectors can be used to construct the relevant Fermi coordinates, and the origins of those coordinates can also be translated. The net effect of these choices is that any Brinkmann coordinate system $(u,v,x,y)$ can be related to any other Brinkmann coordinate system $(\hat{u}, \hat{v}, \hat{x}, \hat{y})$ via the transformations \cite{EhlersKundt}
\begin{subequations}
\label{coordXform}
\begin{align}
    \hat{u} \equiv \frac{1}{\gamma} ( u - u_0 ), \qquad \hat{x}^i \equiv r^{i}{}_{j} [x^j - x_0^j(u)],
    \\
    \hat{v} \equiv \gamma \{ v +\delta_{ij} [\tfrac{1}{2}  x^i_0(u) - x^i] \dot{x}^j_0(u) - v_0\},
\end{align}
\end{subequations}
where $\gamma > 0$, $u_0$, and $v_0$ are arbitrary constants, $r^{i}{}_{j}$ is a constant orthogonal matrix, and $x_0^i(u)$ is any 2-vector which satisfies the differential equation
\begin{equation}
    \ddot{x}^i_0 = H^{i}{}_{j} x_0^j.
    \label{newOrigin}
\end{equation}
Here and below, dots denote derivatives with respect to the phase coordinate $u$. The matrix $r^{i}{}_{j}$ can be used to effect rotations or reflections in the transverse coordinates, while $\gamma$ is a Lorentz factor associated with longitudinal boosts. Varying $u_0$ and $v_0$ instead allows one to rigidly translate the coordinates $u$ and $v$. As elaborated in \cref{Sect:PWgeod} below, the 2-vector $x_0^i(u)$ can be used to recenter the origin around any desired geodesic. A comparison with \cref{Brinkmann} shows that the coordinate transformation \eqref{coordXform} preserves the Brinkmann line element while transforming the waveform into
\begin{equation}
    \hat{H}_{ij} (\hat{u}) = \gamma^2 r^{k}{}_{i} r^{l}{}_{j} H_{kl} ( \gamma \hat{u} + u_0 ).
\end{equation}
Any given Brinkmann waveform is, therefore, unique up to a small number of simple transformations.

All vacuum plane waves are Petrov type N spacetimes wherever they are not flat, and the associated principal null direction is tangent to
\begin{equation}
    \ell^\alpha \equiv - g^{\alpha \beta} \nabla_\beta u.    
    \label{lDef}
\end{equation}
This can be interpreted as describing the direction in which the gravitational wave propagates. A calculation shows that 
\begin{equation}
    \nabla_\alpha \ell^\beta = 0,
\end{equation}
so $\ell^\alpha$ is also a Killing vector field which generates null translations within each wavefront. In fact, plane wave spacetimes admit at least four Killing vector fields in addition to $\ell^\alpha$, and these may be viewed as generators for the transformations which are associated with the $x_0^i(u)$ in \cref{coordXform}. They may be shown to have the form 
\begin{equation}
    [ x^i \dot{\Xi}_i(u) ] \partial_v  + \Xi^i(u) \partial_i,
    \label{Killing}
\end{equation}
where the 2-vector $\Xi^i(u)$ is any solution to the differential equation
\begin{equation}
    \ddot{\Xi}^i = H^{i}{}_{j} \Xi^j.
    \label{JacobiBasic}
\end{equation}
It has been shown in Ref. \cite{Duval_2017} that the symmetry structure of plane wave spacetimes is related to the Carroll group that arises as the $c \to 0$ contraction of the Poincaré group \cite{AIHPA_1965__3_1_1_0}.

The Killing vectors of plane wave spacetimes can be understood more intuitively by considering the globally flat case in which $H_{ij}$ vanishes. In that context, all five Killing vectors reduce to
\begin{equation}
    \partial_v , \qquad \partial_i, \qquad x^i \partial_v + u \partial_i .
    \label{flatKilling}
\end{equation}
The first three of these vector fields clearly generate translations. The final two can be more easily interpreted if inertial coordinates
\begin{equation}
    t \equiv \frac{ 1 }{ \sqrt{2} } (v+u), \qquad  z \equiv \frac{ 1 }{ \sqrt{2} } (v-u)  
    \label{tzDef}
\end{equation}
are introduced to supplement $x$ and $y$, in which case
\begin{equation}
    x^i \partial_v + u \partial_i = \frac{1}{\sqrt{2} } \left[ ( x^i \partial_t + t \partial_i ) - ( z \partial_i - x^i \partial_z ) \right].
    \label{nullRot}
\end{equation}
The two bracketed terms here generate a transverse boost in the $x^i$ direction and a rotation in the $z$-$x^i$ plane. Gravitational plane waves may thus be viewed as being preserved by the composition of these two operations (suitably generalized), but not by either operation on its own. Said differently, applying a transverse boost to a gravitational plane wave is equivalent to rotating its propagation direction. The vector fields in \cref{nullRot} can alternatively be interpreted as “null rotations” whose action preserves the null vector $\ell^\alpha$.

One justification for describing gravitational plane wave spacetimes as “plane waves” is that the five Killing fields \eqref{flatKilling} are exactly those Killing fields that preserve electromagnetic plane waves in flat spacetime \cite{BondiPWs}. More precisely, consider any field strength  
\begin{equation}
    F_{\alpha \beta} = \ell_{[\alpha} \mathcal{A}_{\beta]}(u)
    \label{EMPW}
\end{equation}
in flat spacetime, where $\mathcal{A}_\alpha$ is restricted only to be orthogonal to $\ell^\alpha$. Then $\mathcal{L}_\xi F_{\alpha\beta} = 0$ for all $\xi^\alpha$ in \cref{flatKilling}. Except in \cref{Sect:EMmemory}, we nevertheless focus only on vacuum \emph{gravitational} plane waves in this paper.

Plane wave spacetimes admit many types of symmetries in addition to the Killing fields described above \cite{Maartens1991, Keane2010}. An example which is useful in the following is the homothety
\begin{equation}
    \lambda^\alpha \partial_\alpha = 2 v \partial_v + x^i \partial_i,
    \label{homothety}
\end{equation}
which satisfies $\mathcal{L}_\lambda g_{\alpha \beta} = 2 g_{\alpha \beta}$. This is a special type of conformal Killing vector field which describes a kind of self-similarity in plane wave spacetimes. Indeed, $\lambda^\alpha$ generates the 1-parameter family of finite dilations $(u,v,x^i) \mapsto (u, e^{2\epsilon} v, e^\epsilon x^i)$, which rescale the metric by $e^{2\epsilon}$ while leaving invariant the null geodesic $v = x^i = 0$. An analogous self-similarity is also present in flat-spacetime electromagnetic plane waves, in the sense that $\mathcal{L}_\lambda F_{\alpha \beta} = F_{\alpha \beta}$ when $F_{\alpha\beta}$ is given by \cref{EMPW}. In the gravitational case, $\lambda^\alpha$ is useful because, just like ordinary Killing vector fields, homotheties generate conservation laws for geodesics \cite{Katzin1981}. We find below that homotheties---and more generally, all conformal Killing vector fields---also generate conservation laws for the spin Hall equations.

\subsection{Jacobi propagators}
\label{Sect:Jacobi}

One of the main conclusions of prior work on plane wave spacetimes is that essentially all interesting observables can be written in terms of solutions to the geodesic deviation (or Jacobi) equation \cite{Hollowood_2008, HarteDrivas, HarteStrongLensing,  HartePWOptics2, HartePWMemory, GrantMemory}. Moreover, the many symmetries of these spacetimes allow us to focus only on geodesic deviation in the two-dimensional space which is transverse to the wave's propagation direction. The Jacobi equation then reduces to \cref{newOrigin}, or equivalently to \cref{JacobiBasic}. Solutions to these equations describe not only coordinate freedoms and the symmetries of the spacetime, but also its complete geodesic structure, including gravitational wave memory effects and all standard observables in gravitational lensing. Even effects involving low-frequency wave propagation can be extracted from solutions of the geodesic deviation equation \cite{HarteTails}. We show below that such solutions also describe the motion of massless spinning particles---models for localized massless wave packets which carry angular momentum.

The Jacobi equation is linear, so all of its solutions can be written in terms of two particular solutions to its matrix generalization
\begin{equation}
   \partial_u^2 {E}_{ij} = H_{ik} E^{k}{}_{j}.
    \label{JacobiMatrix}
\end{equation}
It is often convenient to choose these particular solutions to be the two two-point functions $A_{ij}(u,u')$ and $B_{ij}(u,u')$ that satisfy the matrix Jacobi equation together with the initial conditions
\begin{equation}
    [A_{ij}] = [ \partial_u B_{ij} ] = \delta_{ij}, \qquad [\partial_u A_{ij} ] = [ B_{ij} ] =  0.
    \label{JacobiInit}
\end{equation}
Here and below, $[ \ldots ]$ denotes the “coincidence limit” in which $u \to u'$. We refer to $A_{ij}$ and $B_{ij}$ as the “Jacobi propagators” for the given plane wave spacetime. 
In terms of them, any solution to the differential equation \eqref{JacobiBasic} has the form
\begin{equation}
    \Xi_{i} (u) = A_{ij}(u,u') \Xi^{j}(u') + B_{ij}(u,u') \dot{\Xi}^{j}(u'),
    \label{XiDef}
\end{equation}
where $\Xi^{i}(u')$ and $\dot{\Xi}^{i}(u')$ are initial data for that equation. Under the coordinate transformations in \cref{coordXform}, the Jacobi propagators transform via
\begin{align}
    \hat{A}_{ij} = r^{k}{}_{i} r^{l}{}_{j} A_{kl}, \qquad \hat{B}_{ij} =  \gamma^{-1} r^{k}{}_{i} r^{l}{}_{j} B_{kl}.
    \label{ABxForm}
\end{align}
 
In addition to their use in plane wave spacetimes \cite{HarteDrivas, HarteStrongLensing, HartePWMemory, HarteTails}, Jacobi propagators or similar structures have also been applied in more general spacetimes. They have, for example, been essential to understanding the motion of extended bodies \cite{Dixon74, HarteReview}, particularly because they can be used to construct generalized Killing vectors \cite{HarteSyms}. Closely related ideas have also been used to understand gravitational lensing in generic spacetimes \cite{Bilocal1, Bilocal2, Bilocal3, Bilocal4, Uzun, Seitz1994, Korzynski2024}. Although this paper focuses only on plane wave spacetimes, many of the properties of the Jacobi propagators that we discuss remain valid also in more general spacetimes \cite{Bilocal1, Uzun, GrantMemory, Korzynski2024}. 

Given a particular waveform $H_{ij}$, it is at least conceptually straightforward to compute the corresponding Jacobi propagators. Although exact analytic calculations are rarely practical\footnote{One analytic example is discussed in \cref{app:exact_pulse} below, where the Jacobi propagators are computed for a finite-width square wave.}, approximations in powers of the curvature are easily obtained; see \cref{Sect:approxAB} below and also Ref.~\cite{HartePWOptics2}. Numerical calculations which place no restrictions on $H_{ij}$ are straightforward as well. The problem of computing $A_{ij}$ and $B_{ij}$ for a particular $H_{ij}$ can, however, be considered separately from the problem of describing observables in terms of $A_{ij}$ and $B_{ij}$. It is only this latter problem with which most of this paper is concerned. Once the Jacobi propagators associated with a particular plane wave are known, essentially all interesting observables can be straightforwardly determined in terms of them.

Before we turn to this, it is useful to first recall some general identities that are satisfied by all Jacobi propagators. Our starting point is to note that given any two solutions $E_{ij}(u)$ and $\tilde{E}_{ij}(u)$ to the matrix Jacobi equation \eqref{JacobiMatrix}, their Wronskian is necessarily conserved:
\begin{equation}
    E^{k}{}_{i} \partial_u \tilde{E}_{kj} - \tilde{E}_{kj} \partial_u E^{k}{}_{i} = \mathrm{constant}.
    \label{Wronski}
\end{equation}
This can be used to derive a number of useful identities \cite{HarteDrivas, HarteStrongLensing, HartePWMemory, GrantMemory, Korzynski2024}. First, using \cref{Wronski} with $E_{ij} = A_{ij}$ and $\tilde{E}_{ij} = B_{ij}$ results in the constraint
\begin{equation}
    A^{k}{}_{i} \partial_u B_{k j} - B_{k j} \partial_u A^{k}{}_{i} = \delta_{ij},
    \label{ABidentity}
\end{equation}
which implies that there is some overlap in the information which is encoded in $A_{ij}$ and in $B_{ij}$. 

In fact, considerably more can be said about this overlap. Noting that $\partial_{u'} B_{ij}$ satisfies the same matrix Jacobi equation as $A_{ij}$ and $B_{ij}$, the coincidence limits
\begin{equation}
    [\partial_{u'} B_{ij}] = -\delta_{ij}, \qquad [\partial_u \partial_{u'} B_{ij}] = 0
\end{equation}
imply that
\begin{equation}
    A_{ij}(u,u') = - \partial_{u'} B_{ij}(u,u').
    \label{AfromB}
\end{equation}
\emph{Knowledge of $B_{ij}$ alone is therefore sufficient to obtain $A_{ij}$.} Applying a similar argument for $\partial_{u'} A_{ij}$ instead of $\partial_{u'} B_{ij}$ results in
\begin{equation}
    B_{ik}(u,u') H^{k}{}_{j} (u') = - \partial_{u'} A_{ij}(u,u').
    \label{BfromA}
\end{equation}
Although this could be used to derive $B_{ij}$ from $A_{ij}$ when $H_{ij}(u')$ is invertible, this is often not the case. \Cref{AfromB} is therefore more useful than \cref{BfromA}; it implies that we can write everything of interest purely in terms of $B_{ij}$. 

In general, neither $A_{ij}$ nor $B_{ij}$ is symmetric. However, appropriate substitutions into 
\cref{Wronski} can be used to show that the matrices
\begin{equation}
    A_{ki} \partial_u A^{k}{}_{j}, \qquad B_{ki} \partial_u B^{k}{}_{j}, \qquad B_{ik} A_{j}{}^{k}
    \label{SymMatrices}
\end{equation}
\emph{are} necessarily symmetric \cite{HarteStrongLensing}. Lastly, we note that if the arguments of $A_{ij}$ and $B_{ij}$ are swapped, conserved Wronskians can be used to show that
\begin{subequations}
\label{reciprocity}
\begin{align}
    B_{ij} (u,u') &= - B_{ji} (u',u), \label{Btrans} \\
    \partial_u A_{ij} (u,u') &= - \partial_{u'} A_{ji} (u',u).
\end{align}    
\end{subequations}
The first of these identities is a version of Etherington’s reciprocity relation \cite{Etherington, Perlick2000} as applied to plane wave spacetimes (although more general versions of that relation can be obtained using an essentially identical argument).

\subsection{Sandwich waves and memory tensors}
\label{Sect:SandWave}

\begin{figure}
    \centering
    \includegraphics[width=.6\columnwidth]{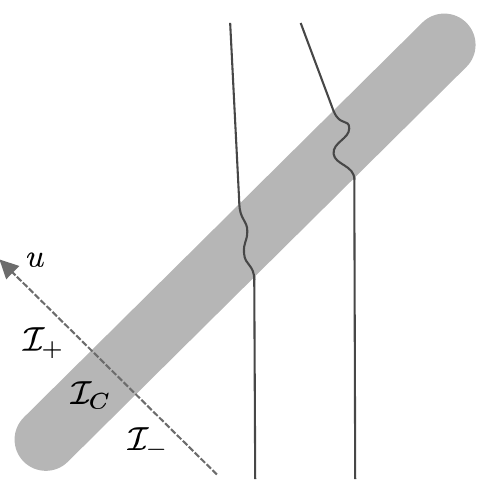}
    \caption{Schematic diagram of a sandwich wave. The geometry is curved only when the phase coordinate $u$ lies in the region $\mathcal{I}_C$. The locally flat regions to the future or to the past of the gravitational wave are characterized by $u$ coordinates in $\mathcal{I}_+$ or in $\mathcal{I}_-$, respectively. A memory effect is also illustrated using the worldlines of two freely falling objects: Interaction with the gravitational wave is seen to change their relative velocity.}
    \label{fig:SandWave}
\end{figure}

A particularly clean model for a gravitational wave burst---or perhaps the Penrose limit of a null geodesic which has been scattered by a compact system---is provided by a plane wave in which $H_{ij}(u)$ vanishes for all $u$ outside some finite  interval $\mathcal{I}_C \subset \mathbb{R}$. All points whose $u$ coordinates lie outside that interval are locally flat, so observers will experience a curved gravitational wave region “sandwiched” between two flat regions, as illustrated in \cref{fig:SandWave}. We refer to the $u$ coordinates associated with the flat region to the future of the gravitational wave by $\mathcal{I}_+ \subset \mathbb{R}$ and the flat region to its past by $\mathcal{I}_- \subset \mathbb{R}$.  Almost all further discussion in this paper is confined to “sandwich waves” such as these (or occasionally to waves whose curvature decays very rapidly outside some finite $u$ interval). Since both the initial and the final geometries are flat in a sandwich wave, it is straightforward to understand different types of scattering processes.

As remarked above, many observables in plane wave spacetimes can be written in terms of the Jacobi propagators $A_{ij}$ and $B_{ij}$. Since it follows from \cref{AfromB} that $A_{ij}$ can be easily computed from $B_{ij}$, our first step is to determine the properties of $B_{ij}$ in sandwich wave spacetimes. To do so, first note that in any region where $H_{ij}$ vanishes, it follows from \cref{JacobiMatrix} that 
\begin{equation}
    B_{ij}(u,u') = C_{ij}(u') + u D_{ij}(u'),
    \label{BflatGen}
\end{equation}
for some $C_{ij}(u')$ and some $D_{ij}(u')$. If the waveform vanishes \emph{throughout} the region between the wavefronts parametrized by $u$ and by $u'$, then the initial data in \cref{JacobiInit} implies that $C_{ij}(u') = - u' \delta_{ij}$ and $D_{ij}(u') = \delta_{ij}$, so 
\begin{equation}
    A_{ij}(u,u') = \delta_{ij}, \qquad B_{ij}(u,u') = (u-u') \delta_{ij}.
    \label{ABflat}
\end{equation}
This is globally valid in Minkowski spacetime. In nontrivial sandwich wave spacetimes, it is valid whenever $u$ and $u'$ are either both in $\mathcal{I}_+$ or both in $\mathcal{I}_-$.

The Jacobi propagators are more interesting when, for example, $u'$ lies in $\mathcal{I}_-$ while $u$ lies in $\mathcal{I}_+$. In that case, $B_{ij}(u,u')$ is given by \cref{BflatGen} and
\begin{equation}
    B_{ij}(u',u) = \tilde{C}_{ij}(u) + u' \tilde{D}_{ij}(u)
\end{equation}
for some $\tilde{C}_{ij}(u)$ and some $\tilde{D}_{ij}(u)$. Using \cref{Btrans} to relate both of these expressions shows that
\begin{equation}
    C_{ij}(u') + u D_{ij}(u') = -\tilde{C}_{ji}(u) - u' \tilde{D}_{ji}(u),
\end{equation}
which can be true only when $C_{ij}(u')$, $\tilde{C}_{ij}(u)$, $D_{ij}(u')$, and $\tilde{D}_{ij}(u)$ are all affine functions of their arguments. It follows that there must exist four “memory tensors” $M^{(\cdots\hspace{-0.025em})}_{ij}$ such that
\begin{align}
    B_{ij}(u,u') = (u-u') \delta_{ij} + \Mvd_{ij} + u \Mvv_{ij}  
    \nonumber
    \\
    ~ - u' \Mdd_{ij}  - u u' \Mdv_{ij} .
    \label{Bexpand}
\end{align}
These tensors are all \emph{constant}. Applying \cref{AfromB}, it also follows that 
\begin{equation}
    A_{ij}(u,u') = \delta_{ij} + \Mdd_{ij} + u \Mdv_{ij},
    \label{Aexpand}
\end{equation}
which does not depend on $u'$. If the roles of $u$ and $u'$ are reversed, so $u'$ lies in $\mathcal{I}_+$ while $u$ lies in $\mathcal{I}_-$, \cref{AfromB,Btrans} can be used to show that \cref{Bexpand} is replaced by
\begin{align}
    B_{ij}(u,u') = (u-u') \delta_{ij} - \Mvd_{ji} + u \Mdd_{ji}
    \nonumber
    \\
    ~ - u' \Mvv_{ji} + u u' \Mdv_{ji},
    \label{Bexpand2}
\end{align}
while \cref{Aexpand} is replaced by
\begin{equation}
    A_{ij}(u,u') = \delta_{ij} + \Mvv_{ji} - u \Mdv_{ji}. 
    \label{Aexpand2}
\end{equation}

Although there is an infinite-dimensional space of possible waveforms, it follows from these results that the effects of those waveforms on the Jacobi propagators are encoded in the finite-dimensional space of possible memory tensors---at least outside $\mathcal{I}_C$. Comparison with \cref{ABflat} shows that all memory tensors vanish when the spacetime is globally flat. More generally, the memory tensors can be computed by integrating the matrix Jacobi equation \eqref{JacobiMatrix} through the curved region $\mathcal{I}_C$. Even when exact analytic solutions are not available, numerical computations are straightforward. Some examples are discussed in \cref{App:Examples}.

While results equivalent to \cref{Bexpand,Aexpand} have been derived before \cite{HarteStrongLensing}, describing the various terms in the Jacobi propagators as “memory tensors” is new. We shall see in \cref{Sect:MemoryIntro} below---in particular in \cref{Smatrix,SDef}---that the superscripts on the memory tensors roughly explain how they affect pairs of geodesics: The “displacement-displacement memory” $\Mdd_{ij}$ determines how initial relative displacements affect final relative displacements, the “velocity-displacement memory” $\Mvd_{ij}$ determines how initial relative velocities affect final relative displacements, the “displacement-velocity memory” $\Mdv_{ij}$ determines how initial relative displacements affect final relative velocities, and the “velocity-velocity memory” $\Mvv_{ij}$ determines how initial relative velocities affect final relative velocities. $\Mdv_{ij}$ has dimension $(\mathrm{length})^{-1}$, $\Mdd_{ij}$ and $\Mvv_{ij}$ are dimensionless, and $\Mvd_{ij}$ has dimension $(\mathrm{length})^1$.

Our displacement-velocity memory $\Mdv_{ij}$ is closely related to what has been referred to elsewhere as the “velocity memory” \cite{GrishchukVelMem, HartePWOptics2, Zhang2017, Zhang_2018, HarteGenMemory, GrantMemory}. However, the final relative velocities of two geodesics depend not only on their initial displacements, but also on their initial velocities. This motivates us to adopt a more precise terminology that distinguishes between contributions associated with initial displacements and contributions associated with initial velocities. What we call velocity-velocity memory $\Mvv_{ij}$ is closely related to what has been referred to before as the “kick memory” \cite{Seraj2021, GrantMemory}. Our displacement-displacement memory is instead related to what has previously been referred to as “the” memory tensor \cite{WhitingMem, Rahman2020, Garfinkle2022}, or sometimes (at least part of) the “displacement memory” \cite{HarteGenMemory}. Lastly, our $\Mvd_{ij}$ is related to what has previously been referred to as the “subleading displacement memory” \cite{HarteGenMemory} or the “drift memory” \cite{GrantMemory},  which is in turn connected to the “spin” \cite{spinMem} and the “center-of-mass” \cite{CMMem} memories. 

Regardless of terminology, the goal in much of this paper is to determine how various observables depend on the four memory tensors. From this perspective, it is important to know which memory tensors are “possible” (for some waveform) and which are not: The space of possible memory tensors is not equivalent to the space of four $2 \times 2$ matrices. Instead, the memory tensors are, for example, constrained by the conserved Wronskian \eqref{ABidentity}. In combination with \cref{Bexpand,Aexpand}, this implies that
\begin{subequations} \label{eq:mem_Constraints}
\begin{align}
    \Mvv_{ij} + \Mdd_{ji} &= \Mdv_{ki} \Mvd_{kj} - \Mdd_{ki} \Mvv_{kj},
    \label{memConstr1}
    \\
    \Mdv_{ij} - \Mdv_{ji} &= \Mdv_{ki} \Mdd_{kj} - \Mdv_{kj} \Mdd_{ki}.
    \label{memConstr2}
\end{align}
\end{subequations}
Additional constraints would result from instead combining the conserved Wronskian with the time-reversed Jacobi propagators given by \cref{Bexpand2,Aexpand2}. Additional constraints can also be generated by antisymmetrizing the symmetric tensors in \cref{SymMatrices}. For example, \cref{Bexpand} and the symmetry of $B_{ki} \partial_u B^{k}{}_{j}$ implies that
\begin{equation}
    \Mvd_{ij} - \Mvd_{ji} = \Mvd_{ki} \Mvv_{kj} - \Mvd_{kj} \Mvv_{ki} .
    \label{memConstr3}
\end{equation}

\Cref{memConstr1} allows the velocity-velocity memory $\Mvv_{ij}$ to be computed from the other three memory tensors. In a weak-field limit where terms quadratic in the curvature are ignored, it implies that 
\begin{equation}
    \Mvv_{ij} = -\Mdd_{ji} + \mathcal{O}(H^2).
\end{equation} %Using \CRef below gave the wrong ordering
Equations \eqref{memConstr2} and \eqref{memConstr3} instead constrain the antisymmetric components of $\Mdv_{ij}$ and $\Mvd_{ij}$, implying that those components vanish in a weak-field limit. In fact, we shall see in \cref{Sect:approxAB} below that all four memory tensors are symmetric (and also trace-free) in such a limit. 

Our definition for the memory tensors in terms of $B_{ij}$ effectively singles out the $u=0$ hyperplane as a “reference” hypersurface. However, there is nothing intrinsically special about this choice, and others can be used instead. Referencing a different $u = \mathrm{constant}$ hypersurface is very similar to performing a constant translation of the $u$ coordinate, which transforms one Brinkmann coordinate system into another. It is therefore interesting to understand how the memory tensors are related to each other in different Brinkmann coordinate systems. Using the \emph{general} Brinkmann-to-Brinkmann coordinate transformation \eqref{coordXform}, comparison of \cref{ABxForm,Bexpand} shows that the memory tensors transform via
\begin{subequations}
\label{memXForm}
\begin{align}
    \MdvH_{ij} &= \gamma r^{k}{}_{i} r^{l}{}_{j} \Mdv_{kl},
    \\
    \MvvH_{ij} &= r^{k}{}_{i} r^{l}{}_{j} (\Mvv_{kl} - u_0 \Mdv_{kl}),
    \\
    \MddH_{ij} &= r^{k}{}_{i} r^{l}{}_{j} (\Mdd_{kl} + u_0 \Mdv_{kl}),
    \\
    \MvdH_{ij} &= \gamma^{-1} r^{k}{}_{i} r^{l}{}_{j} \big[ \Mvd_{kl} -u_0^2 \Mdv_{kl} 
    \nonumber
    \\
    &\qquad \qquad ~ + u_0 ( \Mvv_{kl} - \Mdd_{kl}) \big].
\end{align}
\end{subequations}
Varying the origin of the $u$ coordinate by varying $u_0$ thus mixes the various memory tensors. In some cases, this freedom can be used to eliminate certain memory tensors. However, varying $u_0$ can never affect $\Mdv_{ij}$ or $\Mdd_{ij} + \Mvv_{ij}$, both of which are invariant up to overall scalings and orthogonal transformations. As noted above, $\Mdd_{ij} + \Mvv_{ij}$ always vanishes at least through first order in the curvature. We shall see in \cref{Sect:approxAB} below that in many (though not all) cases, $\Mdv_{ij}$ also vanishes at first order.

Although our primary concern in this paper is with gravitational waves, it can be instructive to compare with the electromagnetic case. \cref{Sect:EMmemory} considers charged-particle scattering in an electromagnetic sandwich wave in flat spacetime, and shows that instead of obtaining four rank-2 memory tensors, there are two memory \emph{vectors} in that context. These are essentially the zeroth and the first moments of the electromagnetic waveform $\mathcal{A}_i$. We now show that at leading order, the gravitational memory tensors are instead given by the zeroth, the first, and the second moments of the gravitational waveform $H_{ij}$.

\subsection{Approximate memory tensors}
\label{Sect:approxAB}

If a gravitational wave is weak, implying that its memory tensors are all “small,” then those tensors can be related to the waveform by perturbatively solving \cref{JacobiMatrix,JacobiInit}. First note that through second order in the curvature $H_{ij}$, a straightforward calculation shows that
\begin{widetext}
\begin{align}
    B_{ij} (u,u') = (u-u') \delta_{ij} + \int_{u'}^u \rmd w\, (u-w)  \left[ (w-u') H_{ij} (w) + \int_{u'}^w \rmd w' \, (w-w') (w' - u') H_{ik}(w) H_{kj} (w') \right] + \mathcal{O}(H^3).
    \label{B1}
\end{align}
Applying \cref{AfromB} then results in
\begin{align}
    A_{ij}(u,u') = \delta_{ij} + \int_{u'}^u \rmd w\, (u-w) \left[ H_{ij}(w) + \int_{u'}^w \dd w'\, \, (w-w') H_{ik}(w) H_{kj}(w') \right] + \mathcal{O}(H^3).
    \label{A1}
\end{align}  
\end{widetext}
Both of these expressions are valid for all $u$, for all $u'$, and for all plane waves in which the relevant integrals exist; they are not restricted only to sandwich waves. One implication is that the first-order contributions to the Jacobi propagators are always symmetric and trace-free in vacuum. However, antisymmetric components and nonzero traces can arise at second order \cite{HartePWOptics2, HartePWMemory}.

If we restrict ourselves to sandwich wave spacetimes, then the leading-order memory tensors can be extracted by comparing \cref{B1} with \cref{Bexpand}. Through first order in the curvature, this results in
\begin{subequations}
\label{memTensLin}
\begin{align}
    \Mdv_{ij} &=  \int_{-\infty}^\infty \!\! \rmd w\, H_{ij}(w ) + \mathcal{O}(H^2),
    \label{Mdv}
    \\
    \Mvv_{ij} &= -\Mdd_{ij} + \mathcal{O}(H^2)
    \nonumber
    \\
    &=  \int_{-\infty}^\infty \!\! \rmd w\, w H_{ij}(w ) + \mathcal{O}(H^2),
    \label{Mvv}
    \\
    \Mvd_{ij} &=  -\int_{-\infty}^\infty \!\! \rmd w\, w^2 H_{ij}(w ) + \mathcal{O}(H^2).
\end{align}
\end{subequations}
Equivalent expressions in terms of a transverse-traceless metric perturbation $h_{ij}$ can  be obtained by substituting \cref{hToH} into these integrals. Regardless of which waveform is employed, it is evident that all four memory tensors are symmetric and trace-free in vacuum and at leading order. They encode the zeroth, the first, and the second moments of the curvature. They are all that are needed for the observables considered in this paper. Higher moments of the curvature can, however, be relevant for certain other “persistent observables” \cite{HarteGenMemory, GrantMemory}. 

If a plane wave spacetime is viewed as a ``local'' idealization of the far-field geometry around a radiating compact system in an asymptotically flat spacetime, then the quadrupole approximation \cite{Wald} suggests that the Brinkmann waveform---which is a curvature, not a transverse-traceless metric perturbation---must be proportional to the fourth derivative of the system's quadrupole moment. If all gravitational waves are sourced, for example, by a violent collision or by an explosion involving multiple masses, then the third derivatives of the quadrupole moment would be expected to vanish at both early and late times \cite{Gibbons1971, HartePWOptics2, ZhangSoftGravitons}. The waveform would thus be given by 
\begin{equation}
    H_{ij}(u) = \partial_u \mathfrak{H}_{ij}(u),
    \label{Hsimp}
\end{equation}
where $\mathfrak{H}_{ij}(u)$ vanishes for all $u \in \mathcal{I}_\pm$.
Substitution into \cref{Mdv} shows that in this case, the displacement-velocity memory vanishes through first order in the curvature. Although Einstein's equations do not constrain $\Mdv_{ij}$ purely in a plane wave context, they are likely to do so for the plane waves which are local approximations of astrophysically relevant gravitational waves.

\begin{figure*}
    \begin{minipage}[t]{0.988\columnwidth}
    \includegraphics[width=\textwidth]{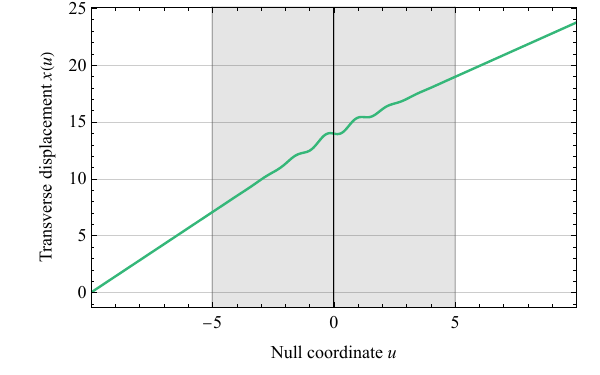}
    %\caption{$x$-component of a null geodesic interacting with a circularly polarized sandwich wave (shaded region).  }
    %\label{fig:null_geo_x}
    \end{minipage}
    \hspace{\columnsep}
    \begin{minipage}[t]{0.988\columnwidth}
    \includegraphics[width=\textwidth]{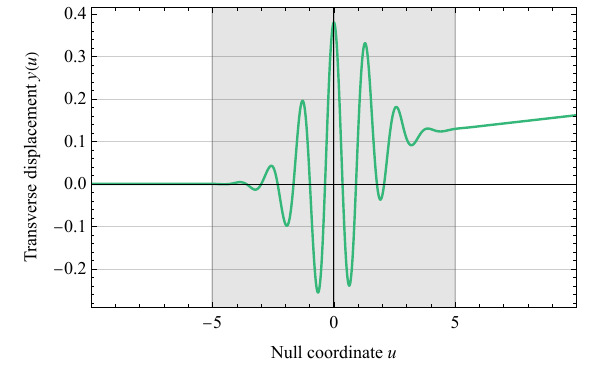}
    %\caption{$y$-component of a null geodesic interacting with a circularly polarized sandwich wave (shaded region).}
    %\label{fig:null_geo_y}
    \end{minipage}
    \raggedright
    \caption{
        Example transverse coordinates of a geodesic in a sandwich wave spacetime. The geodesic has a vanishing initial velocity in the $y$ direction but not in the $x$ direction, and there is a clear (nondiagonal) velocity-velocity memory. However, note the different vertical scales in these plots. The gravitational wave profile used here is given by \cref{gw_profile} in \cref{App:Examples}, with center point $U_0 = 0$, frequency $\nu = 5/U$, curvature length scales $l_{+} = l_{\times} = U/\sqrt 2$, and phases $\phi_{\times} = 0$ and $\phi_{+} = - \pi/2$. This wave might be interpreted as circularly polarized with a Gaussian profile. These plots also employ units in which the characteristic burst width $U$ is equal to unity. The shaded region roughly corresponds to the curved region $\mathcal{I}_C$; at the edge of the shading, the exponential in the waveform is 0.002 times its maximum value.  
        }
    \label{fig:null_geo_x}
\end{figure*}

Nevertheless, vanishing displacement-velocity memory is not typically a feature of the plane waves that arise from Penrose limits. For example, the Penrose limit of a null geodesic with angular momentum $L$ in a Schwarzschild spacetime with mass $M$ is \cite{BlauNotes}
\begin{equation}
    H_{ij}(u) = \frac{ 3M L^2 }{ [r(u)]^5 } 
    \mathrm{diag} [1,-1]_{ij},
    \label{HSchw}
\end{equation}
where $r(u)$ denotes the areal radius of the given geodesic\footnote{This is technically not a sandwich wave. However, if the null geodesic is only scattered---and not captured or bound---its radius grows rapidly as $u \to \pm \infty$. Memory tensors then remain a useful concept. The same could not be said for, e.g., the Penrose limit which is associated with a circular null geodesic in which $r(u)=3M$.}. Since $3ML^2/[r(u)]^5 \geq 0$, substitution into \cref{Mdv} shows that, except in the radial case where $L=0$, the displacement-velocity memory cannot vanish at first order in $H_{ij}$. 

If we again assume a waveform given by \cref{Hsimp}, then the use of \cref{A1} shows that although $\Mdv_{ij}$ vanishes at first order in the curvature, it \emph{cannot} vanish at second order:
\begin{equation}
    \Mdv_{ij} = - \frac{1}{2} \left[ \int_{-\infty}^\infty \!\! \rmd u \, \mathfrak{H}^{kl}(u) \mathfrak{H}_{kl}(u) \right] \delta_{ij} + \mathcal{O}(H^3).
\end{equation}
In these cases, initially comoving geodesics are  focused isotropically in the two transverse directions. That $\Mdv_{ij}$ must be nonzero at second order is closely related to statements in, e.g., Refs.~\cite{Zhang2017, Zhang_2018} that it is impossible for the velocity memory to vanish. 

Unless otherwise noted, we place no constraints below on the leading-order behavior of $\Mdv_{ij}$. We also do not assume that $H_{ij}$ necessarily has the form \eqref{Hsimp}.

\section{Geodesic motion and memory effects}
\label{Sect:Geodesics}

The Jacobi propagators discussed in the previous section govern all properties of geodesics in plane wave spacetimes\footnote{In more general spacetimes, Jacobi propagators only encode the behavior of \emph{nearby} geodesics.}. We now make this explicit by describing how geodesics are scattered in sandwich wave spacetimes. \cref{Sect:PWgeod} begins by reviewing known results on geodesics in plane wave spacetimes. Those results are then specialized to sandwich waves in \cref{Sect:MemoryIntro}, where the memory tensors introduced above are used to describe the transverse properties of scattered geodesics. “Time-reversed” memory tensors, which describe initial states in terms of final states, are also derived there. \cref{Sec:NullMem} completes the treatment of geodesic motion by describing both transverse and longitudinal memory effects. We focus in particular on how null geodesics are affected by gravitational wave memory, which differs considerably from the more typical cases involving slowly moving timelike geodesics. Lastly, \cref{sec:bitensors} explains how memory tensors affect Synge's world function, which provides an alternative way to encode the geodesics of sandwich wave spacetimes. As an application, we describe how memory effects deform light cones.

\subsection{Geodesics in plane wave spacetimes}
\label{Sect:PWgeod}

The geodesic structure of plane wave spacetimes has been extensively described in, e.g., Refs.~\cite{BlauNotes, HarteDrivas}. To briefly review the results needed here, suppose that $x^\alpha(\tau)$ describes an affinely parametrized geodesic. Then, since $\ell^\alpha$ is Killing, $\ell_\alpha \rmd x^\alpha/\rmd \tau = -\rmd u/\rmd \tau$ must be constant. If that constant vanishes, then the geodesic lies on a $u = \mathrm{constant}$ hypersurface and is either spacelike or null. However, the vast majority of geodesics are not of that type, and in all such cases, we are free to identify the affine parameter $\tau$ with $u$. Doing so,
\begin{equation}
    \ell_\alpha \frac{ \rmd x^\alpha}{ \rmd u } = -1.
    \label{uCons}
\end{equation}
The two transverse coordinates of a geodesic are then given by
\begin{equation}
    x^i (u) = A^{i}{}_{j} (u,u') x^j(u') + B^{i}{}_{j} (u,u') \dot{x}^j(u'),
    \label{geodesics}
\end{equation}
where $A_{ij}$ and $B_{ij}$ are the Jacobi propagators introduced in \cref{Sect:Jacobi}. In terms of this transverse motion, the $v$ coordinate of a geodesic may be shown to be
\begin{align}
\begin{split}
    v(u)
        &= v(u') + \kappa(u-u')
        \\
        &\quad+ \tfrac{1}{2} [ x^i(u) \dot{x}_i(u) - x^i(u') \dot{x}_i(u')],
    \label{geodesicv}
\end{split}
\end{align}
where 
\begin{equation}
    \kappa \equiv - \frac{1}{2} g_{\alpha\beta} \frac{ \rmd x^\alpha}{\rmd u} \frac{ \rmd x^\beta}{\rmd u}
    \label{kappaDef}
\end{equation}
is a constant. If a geodesic is timelike, $\kappa > 0$; if it is null, $\kappa = 0$. Nevertheless, there exist families of timelike geodesics in which the limits $\kappa \to 0$ and $\kappa \to \infty$ both approach null geodesics. This is described in more detail below \cref{kappaV}.

\begin{figure}
    \centering
    \includegraphics[width=.99\columnwidth]{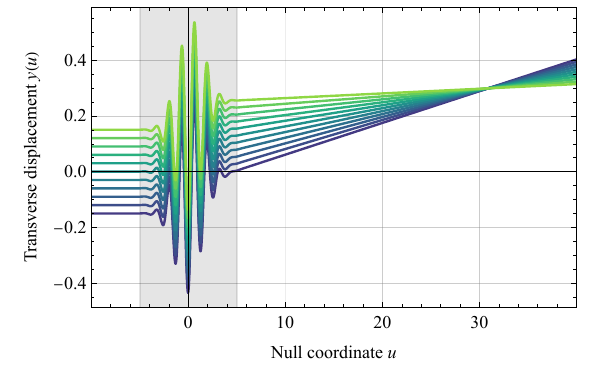}
    \caption{Scattering of a bundle of initially parallel geodesics traveling in the $x$ direction, illustrating a nontrivial displacement-velocity effect. The sandwich gravitational wave (shaded region) is seen to act as a lens for these geodesics. The waveform and other conventions here are the same as in \cref{fig:null_geo_x}.}
    \label{fig:null_geos}
\end{figure}

It follows from \cref{geodesics,geodesicv} that all interesting properties of geodesics in plane wave spacetimes are determined by their two transverse coordinates $x^i$, which are in turn determined by the two Jacobi propagators $A_{ij}$ and $B_{ij}$. Moreover, whether a geodesic is timelike, null, or spacelike affects only its $v$ coordinate. Some examples of geodesics in specific plane wave spacetimes are plotted in \cref{fig:null_geo_x,fig:null_geos}. Additional examples may be found in, e.g., Refs.~\cite{Zhang2017, ZhangSoftGravitons, Zhang_2018, ZhangIonTraps}. 

As noted above, the Brinkmann coordinates used in this paper are essentially Fermi normal coordinates constructed around the null geodesic $x^i(u) = v(u) = 0$. However, there is nothing physically distinct about this origin. If we choose any other null geodesic, appropriate choices for $x_0^i(u)$ and for $v_0$ in the coordinate transformation \eqref{coordXform} may be used to construct Brinkmann coordinates $(\hat{u}, \hat{v}, \hat{x}^i)$ in which that geodesic is given by $\hat{x}^i (\hat{u}) = \hat{v}(\hat{u}) = 0$. Moreover, doing so does not change the waveform. Similar recenterings may also be performed for non-null geodesics, which can always be mapped to $\hat{x}^i(\hat{u}) = 0$ and $\hat{v}(\hat{u}) = \kappa \hat{u} + \mathrm{constant}$. These results provide a sense in which the plane wave spacetimes are indeed planar; the origin is of no intrinsic geometrical significance. The freedom to recenter is also useful below, where it allows us, without loss of generality, to place a geodesic observer at the origin.

One last point to note is that plane wave spacetimes generically admit caustics, or more precisely conjugate points. Given a pair of points in a plane wave spacetime, there typically exists exactly one geodesic that passes between them. However, if two points have phase coordinates $u$ and $u'$ in which 
\begin{equation}
    \det B_{ij}(u,u') = 0,
    \label{detB}
\end{equation}
there are either an infinite number of connecting geodesics or there are none \cite{HarteDrivas}. The pairs of hypersurfaces $u' = \mathrm{constant}$ and $u = \mathrm{constant}$ are then referred to as “conjugate hyperplanes.” All null geodesics emanating from a point on the $u' = \mathrm{constant}$ hypersurface are focused into either a line or (in exceptional cases) a point on the $u=\mathrm{constant}$ hypersurface. Timelike geodesics are instead focused into either two- or one-dimensional subsets of that three-dimensional hypersurface. In the case of a sandwich wave, fixing some $u' \in \mathcal{I}_-$ results in at most two solutions to \cref{detB} in which $u \in \mathcal{I}_+$ \cite{HarteStrongLensing}. An example with one solution is described in \cref{app:exact_pulse} below.

\subsection{Transverse memory effects in sandwich wave spacetimes}
\label{Sect:MemoryIntro}

Before a gravitational wave arrives and after it has left, all geodesics in a sandwich wave spacetime are straight lines in Brinkmann coordinates. However, these lines are not necessarily trivial continuations of one other; geodesics are scattered by the gravitational wave in the curved region $\mathcal{I}_C$. We now show that all possible scatterings are determined by the memory tensors introduced in \cref{Sect:SandWave} above.

If initial data $x'^i = x^i(u')$ and $\dot{x}'^i = \dot{x}^i(u')$ for a particular geodesic is specified at the phase coordinate $u' \in \mathcal{I}_-$, and if $u \in \mathcal{I}_+$, then it follows from \cref{Bexpand,Aexpand,geodesics} that the transverse displacement at late times is
\begin{align}
    x_i (u) = (\delta_{ij} + \Mdd_{ij} + u \Mdv_{ij} )  (x'^j - u' \dot{x}'^j)
    \nonumber
    \\
    ~  + [ \Mvd_{ij}+ u ( \delta_{ij} +  \Mvv_{ij} ) \big] \dot{x}'^j.
    \label{posGeodesic}
\end{align}
Differentiation with respect to $u$ gives the final transverse velocity
\begin{align}
    \dot{x}_i (u) = (\delta_{ij} + \Mvv_{ij}) \dot{x}'^j +  \Mdv_{ij}  ( x'^j -u' \dot{x}'^j) .
    \label{velGeodesic}
\end{align}
These equations are exact, and can be used to predict the results of various memory experiments that compare the initial and the final states of freely falling objects. 

\begin{figure}
    \centering
    \includegraphics[width=0.63\columnwidth]{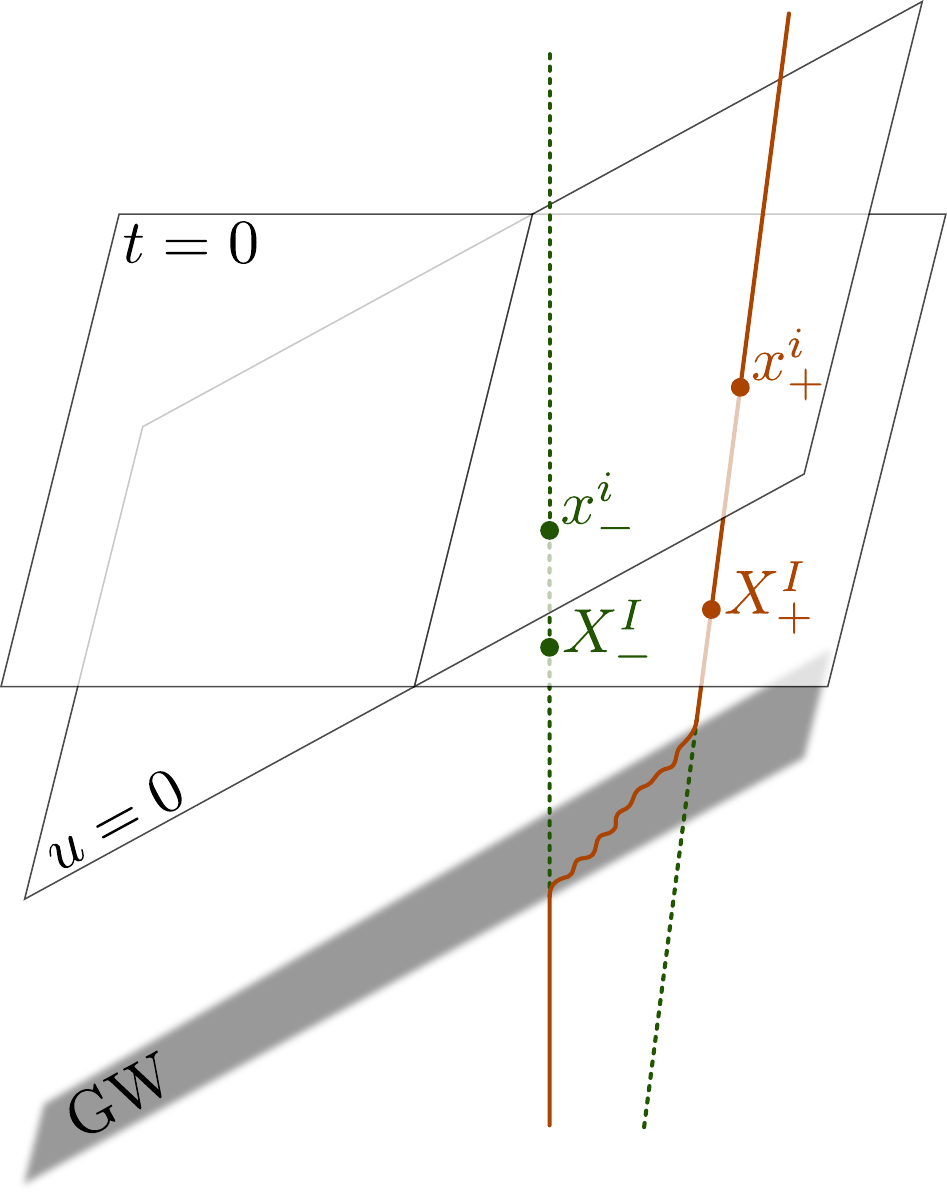}
    \caption{The projected displacements $x^i_\pm$ and $X^I_\pm$. Following \cref{xPM}, the constants $x^i_\pm$ denote the flat projections of the two transverse coordinates of the initial and the final geodesics onto the hypersurface $u=0$. Following \cref{geoInertial}, the constants $X^I_\pm$ denote the flat projections of the three Cartesian coordinates of the initial and the final geodesics onto the hypersurface $t=0$. Dashed lines denote extrapolations of the worldlines which ignore the spacetime curvature.}
    \label{fig:coord}
\end{figure}

One of their consequences is that the memory tensors can be viewed as the components of a linear map which relates the initial and the final transverse states of all geodesics in a sandwich wave spacetime. In particular, the memory tensors determine a “scattering matrix” $\mathcal{S}$ that relates the initial transverse displacements and the initial transverse velocities to the final transverse displacements and the final transverse velocities. In order to construct this matrix, it is first convenient to introduce the initial and the final ``projected displacements''
\begin{equation}
    x^i_+ \equiv x^i(u) - u \dot{x}^i(u), \quad x^i_- \equiv x^i(u') - u' \dot{x}^i(u'),
    \label{xPM}
\end{equation}
which do not depend on precisely how we choose $u$ or $u'$.  Physically, these are the transverse coordinates which the initial and the final geodesics would have if they were extrapolated onto the $u=0$ hypersurface while ignoring any intervening curvature;  see \cref{fig:coord}. If we additionally define\footnote{Despite the notation, the $\dot{x}^i_{\pm}$ are not derivatives of $x^i_{\pm}$.} $\dot{x}_+^i \equiv \dot{x}^i(u)$ and $\dot{x}_-^i \equiv \dot{x}^i(u')$, \cref{posGeodesic,velGeodesic} imply that
\begin{align}
    \begin{pmatrix}
        x^i_+ \\ \dot{x}^i_+ 
    \end{pmatrix} = \mathcal{S}
    \begin{pmatrix}
        x^j_- \\ \dot{x}^j_-  
    \end{pmatrix},
    \label{Smatrix}
\end{align}
where the scattering matrix is
\begin{equation}
    \mathcal{S} = 
    \begin{pmatrix}
        \delta_{ij} + \Mdd_{ij}  &    \Mvd_{ij} \\
        \Mdv_{ij}                    &   \delta_{ij} + \Mvv_{ij}
    \end{pmatrix}.
    \label{SDef}
\end{equation}
This justifies the names for the memory tensors that were introduced in \cref{Sect:SandWave} above: $\Mdd_{ij}$ is the displacement-displacement memory, $\Mdv_{ij}$ is the displacement-velocity memory, and so on. 

The scattering matrix $\mathcal{S}$ determines the final state of a system in terms of its initial state. However, the initial state can instead be determined from the final state by using \cref{Bexpand2,Aexpand2} to show that
\begin{align}
    \begin{pmatrix}
        x^i_- \\ \dot{x}^i_- 
    \end{pmatrix} = \mathcal{S}^{-1}
    \begin{pmatrix}
        x^j_+ \\ \dot{x}^j_+  
    \end{pmatrix},
\end{align}
where
\begin{equation}
    \mathcal{S}^{-1} = 
    \begin{pmatrix}
        \delta_{ij} + \Mvv_{ji}    &   -\Mvd_{ji}   \\
        -\Mdv_{ji}   &  \delta_{ij} + \Mdd_{ji}
    \end{pmatrix}.
    \label{Sinv}
\end{equation}
This inverse can also be derived from \cref{SDef} using the memory tensor identities in \cref{eq:mem_Constraints,memConstr3}. Regardless, it follows that all information contained in the “future-directed” memory tensors $( \Mdd_{ij}, \Mvv_{ij}, \Mvd_{ij}, \Mdv_{ij} )$ can be equivalently encoded in their “past-directed” counterparts $( \MddR_{ij}, \MvvR_{ij}, \MvdR_{ij}, \MdvR_{ij} )$, which are related via
\begin{align}
    \MddR_{ij} = +\Mvv_{ji},& \qquad \MvvR_{ij} = +\Mdd_{ji}, 
    \nonumber
    \\
    \MdvR_{ij} = -\Mvd_{ji},& \qquad \MvdR_{ij} = -\Mdv_{ji}.
\end{align}
The memory tensor $\MdvR_{ij}$, for example, describes how final transverse velocities affect initial transverse displacements. 

No matter which memory tensors are employed, the scattering matrix $\mathcal{S}$ plays a similar role to the \textsc{abcd} ray transfer matrix, which is used in optics to describe the propagation of light rays through optical elements in the paraxial approximation \cite{gerrard1994,torre2005} (see also Ref.~\cite{Uzun} for an implementation of the \textsc{abcd} ray transfer matrix for light propagation in general relativity). Thus, the transverse action of plane gravitational waves on test particles can be compared to that of lenses and other optical elements on light rays. The focusing mentioned above in connection with conjugate hyperplanes is, e.g., reminiscent of the focusing of a nearby point source by a lens. Focusing of initially parallel rays instead occurs when $\det A_{ij}(u,u') = 0$ [rather than $\det B_{ij}(u,u') = 0$], and an example of this is presented in \cref{fig:null_geos}. In both cases, however, focusing by vacuum gravitational waves is typically astigmatic.

The transverse scattering of a geodesic by a gravitational plane wave can be compared with the transverse scattering of a charged particle by an electromagnetic plane wave. It is shown in \cref{Sect:EMmemory} that, unlike their gravitational counterparts, electromagnetic memory effects do not depend on a particle's initial state; they are ``inhomogeneous.'' A somewhat different comparison of motion in electromagnetic and gravitational plane waves may be found in, e.g., \cite{Audagnotto2024}.

\subsection{Transverse and longitudinal memory effects in three-dimensional space}
\label{Sec:NullMem}

The scattering matrix $\mathcal{S}$ provides a simple relation between the gravitational memory tensors and the two transverse coordinates of a scattered geodesic. However, this simplicity can be misleading. The longitudinal dynamics\footnote{Even though it is standard to describe gravitational waves in vacuum general relativity as ``transverse,'' they can exert significant longitudinal effects on some systems---even in a linearized setting. This is particularly evident for geodesics that are moving rapidly with respect to the canonical observer.} can be considerably more complicated, particularly in the inertial coordinates which are most natural in the flat regions to the past and to the future of the gravitational wave. We now provide a full three-dimensional picture of geodesic scattering in sandwich wave spacetimes, including both longitudinal and transverse effects. Unlike in the purely transverse discussion above, there are significant differences between the behaviors of slowly and rapidly moving geodesics (including null geodesics).

Before a gravitational wave arrives, it follows from \cref{geodesicv,xPM} that any geodesic which is not confined to a $u = \mathrm{constant}$ hypersurface may be parametrized by
\begin{subequations}
    \label{geoMinus}
    \begin{gather}
    v(u) = v_- + u ( \kappa + \tfrac{1}{2} \delta_{ij} \dot{x}^i_- \dot{x}^j_- ) ,
    \\
    x^i(u) = x^i_- + u \dot{x}^i_-,
    \end{gather}
\end{subequations}
where $\kappa$, $v_-$, $x^i_-$, and $\dot{x}^i_-$ are all constant. After the wave has left, 
\begin{subequations}
\label{geoPlus}
\begin{gather}
    v(u) = v_+ + u ( \kappa + \tfrac{1}{2}  \delta_{ij} \dot{x}^i_+ \dot{x}^j_+ ) ,
    \\
    x^i(u) = x^i_+ + u \dot{x}^i_+,
\end{gather}
\end{subequations}
where $v_+$, $x^i_+$, and $\dot{x}^i_+$ are also constant. The final transverse state $(x^i_+, \dot{x}^i_+)$ is related to the initial transverse state $(x^i_-, \dot{x}^i_-)$ via \cref{Smatrix}, while $v_+$ and $v_-$ are related via
\begin{equation}
    v_+ = v_- + \tfrac{1}{2} \delta_{ij} ( x^i_+ \dot{x}^j_+ - x^i_- \dot{x}^j_-).
    \label{dv}
\end{equation}
Although these expressions encode all transverse and longitudinal aspects of geodesic scattering in sandwich wave spacetimes, the parameters used here are not particularly intuitive. Real observers would not naturally be inclined to project onto $u = \mathrm{constant}$ hypersurfaces, nor would they be inclined to measure velocities with respect to $u$.

Interpretations can be simplified by adopting the inertial coordinates $t$ and $z$ that are related to $u$ and $v$ through \cref{tzDef}. The timelike worldline $x=y=z=0$ is then a geodesic, even inside the gravitational wave. \emph{We take this worldline to be the trajectory of a canonical observer and measure everything with respect to it.} Doing so does not entail any loss of generality, since the argument at the end of \cref{Sect:PWgeod} implies that given any timelike geodesic, there will exist a Brinkmann coordinate system in which that geodesic lies at the origin. In that context, the coordinates $(t,x,y,z)$ form a local Lorentz frame for the given observer. They are ordinary inertial coordinates both before the gravitational wave arrives and after it has left.

Employing capitalized indices $I, J, \ldots$ to refer to the three spatial coordinates $(x,y,z)$, it is convenient to parameterize the initial and the final geodesics by
\begin{gather}
    x^I(t) = X^I_{\pm} + V^I_\pm t,
    \label{geoInertial}
\end{gather}
where the constants $V^I_\pm$ denote the initial and the final 3-velocities $\rmd x^I/\rmd t$. The constant displacements $X^I_\pm$ represent three-dimensional projections of the initial and the final geodesics onto the $t=0$ hypersurface. This contrasts with the $x_\pm^i$ defined above, which are (i) purely transverse and (ii) projected onto the $u=0$ hypersurface rather than the $t=0$ one; see \cref{fig:coord}.

It is now possible to translate between the “Brinkmann parameters” $(\kappa, v_\pm, x^i_\pm, \dot{x}^i_\pm)$ appearing in \cref{geoMinus,geoPlus}, and the “inertial parameters” $(X^I_\pm, V^I_\pm)$ appearing in \cref{geoInertial}. Defining $V_\pm \in [0,1)$ to be the three-dimensional Euclidean norm of $V^I_\pm$, and $\theta_\pm$ as the angle between the velocity vector and the direction of propagation of the gravitational wave, comparing expressions first shows that
\begin{equation}
    \kappa = \frac{ 1 - V_\pm^2 }{ ( 1 - V_\pm \cos\theta_\pm)^2 } .
    \label{kappaV}
\end{equation}
The fact that $\kappa$ is not affected by the gravitational wave constrains the possible pairs $(V_-, \theta_-)$ and $(V_+, \theta_+)$. For an object that is at rest with respect to the canonical observer (though not only in this case), $\kappa =1$. In an ultrarelativistic limit where $V_\pm \to 1$ at fixed $\theta_\pm \neq 0$, $\kappa \to 0$. However, if $\theta_\pm = 0$, the limit $V_\pm \to 1$ instead results in $\kappa \to \infty$. Both $\kappa \to 0$ and $\kappa \to \infty$ are therefore ultrarelativistic limits, although it is only the former case that is generic. The divergent latter case arises due to the breakdown of the parameterization \eqref{uCons} for null geodesics that propagate in the same direction as the gravitational wave. 

For an arbitrary timelike or null geodesic, further calculations show that the remaining Brinkmann parameters are related to the inertial parameters via 
\begin{subequations}
\label{coordTranslate}
\begin{align}
    v_\pm &= \frac{ \sqrt{2} Z_{\pm} }{ 1- V_\pm \cos\theta_\pm },
    \\
    x_{\pm}^i  &= X_{\pm}^i + \frac{ Z_{\pm} V^i_\pm }{ 1 - V_\pm \cos\theta_\pm },
    \\
    \dot{x}^i_\pm &= \frac{ \sqrt{2} V^i_\pm }{1 - V_\pm \cos\theta_\pm },
\end{align}
\end{subequations}
where $(X^I) = (X^1_\pm, X^2_\pm, Z_\pm)$. The nontrivial nature of these expressions implies that the apparent simplicity of (at least transverse) scattering in terms of the Brinkmann parameters is lost when using inertial parameters. Nevertheless, these latter parameters remain useful due to their familiar interpretations.

\subsubsection{Low-speed memory effects}

The conversions between Brinkmann and inertial parameters simplify considerably at slow speeds relative to the canonical observer. Applying \cref{coordTranslate} for a timelike geodesic in which, say, $V_- =0$,
\begin{equation}
    X_-^i = x_-^i, \qquad Z_- = \frac{1}{\sqrt{2}} v_-.  
\end{equation}
The projected $v$ coordinate thus serves as a proxy for the longitudinal Cartesian coordinate $z$. Allowing for nonzero speeds which are nevertheless small, \cref{kappaV} reduces to
\begin{align}
    \kappa &= 1 + 2 V_\pm \cos \theta_\pm + \mathcal{O}(V_\pm^2).
\end{align}
This is conserved for every scattering process, so gravitational waves cannot affect the longitudinal speeds $V_\pm \cos \theta_\pm$ of slowly moving geodesics. Any effects on their velocities must be transverse to the gravitational wave.

\begin{figure}
    \centering
    \includegraphics[width=.6\linewidth]{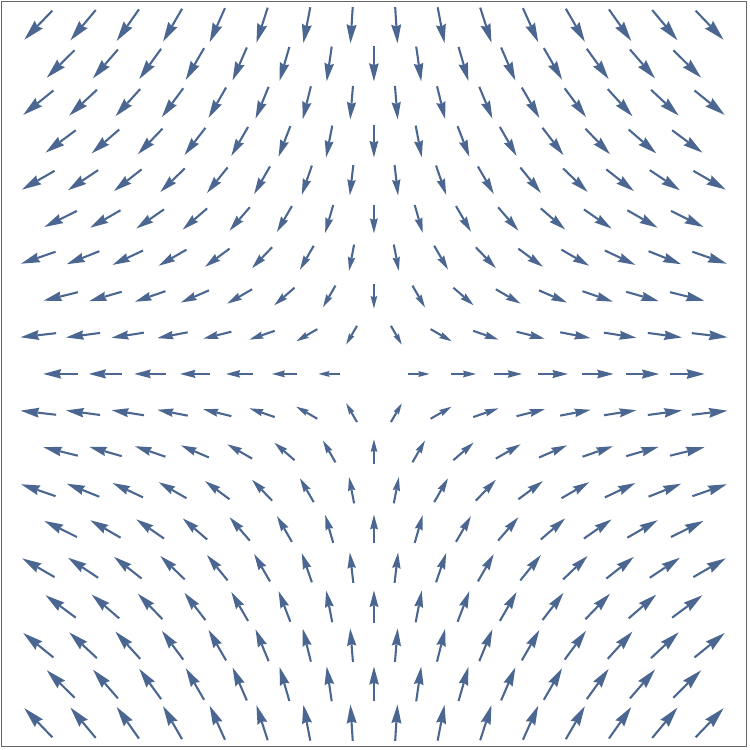}
    \caption[Contribution to the displacement vector field.]{Contribution to the displacement vector field $\Delta X^i$ from $\Mdd_{ij}$ for initially stationary timelike geodesics, through first order in $H_{ij}$. Different points here correspond to different points in the plane which is transverse to the gravitational wave. The vertical and the horizontal axes here are aligned with the eigenvectors of $\Mdd_{ij}$.}
    \label{fig:displMem}
\end{figure}

However, it is not necessarily true that there are no longitudinal effects at all. Again setting $V_- = 0$ for simplicity, an expansion through first order in the curvature $H_{ij}$ yields 
\begin{equation}
    \Delta Z = \frac{1}{2\sqrt{2}} \Mdv_{ij} X_-^i X_-^j + \mathcal{O}(H^2),
    \label{dZSlow}
\end{equation}
where $\Delta Z \equiv Z_+ - Z_-$. Longitudinal displacements can therefore arise when $\Mdv_{ij} \neq 0$. However, this effect is quadratic in the initial displacement and is therefore negligible for sufficiently nearby geodesics. The transverse displacement 
\begin{equation}
       \Delta X_{i} = \left( \Mdd_{ij} - \frac{1}{\sqrt{2}}  Z_- \Mdv_{ij} \right) X^j_-  + \mathcal{O}(H^2)
       \label{dXSlow}
\end{equation}
dominates for slowly moving geodesics that are sufficiently close to the canonical observer. It may also be shown that when the initial velocity vanishes, the transverse components of the final 3-velocity are
\begin{equation}
    \Delta V_{i} = \frac{1}{\sqrt{2}} \Mdv_{ij} X^j_- + \mathcal{O}(H^2).
    \label{dVSlow}
\end{equation}
These expressions are mostly complicated by the displacement-velocity memory $\Mdv_{ij}$. If that memory vanishes through first order in the curvature, as it usually does when a gravitational wave is generated by a compact source, then the only nontrivial effect which remains is the standard displacement memory
\begin{equation}
    \Delta X_{i} = \Mdd_{ij} X^j_- + \mathcal{O}(H^2). 
    \label{displMem}
\end{equation}
This is plotted in \cref{fig:displMem}.

\subsubsection{Memory effects and null geodesics}
\label{Sect:Nullgeodesics}

Memory effects are considerably more complicated at high speeds. There is, e.g., no longer any sense in which longitudinal effects can be ignored. Nevertheless, there is relatively little discussion in the literature on gravitational-wave memory in this regime. What has been discussed---with varying degrees of generality---is the effect of memory on optical observables, including frequency shifts, astrometric deflections, changes in luminosity and angular-diameter distances, and multiple imaging \cite{HarteStrongLensing, HartePWOptics2, vanHaasteren2010, Madison2020}. These observables, of course, involve the effects of a gravitational wave on an observer, a source(s), and the light which passes between them. What does not appear to be available is a direct description of what happens to individual null geodesics as seen by a single canonical observer. We now provide such a description.

\begin{figure*}
    \begin{minipage}[c]{0.988\columnwidth}
    \includegraphics[width=.7\columnwidth]{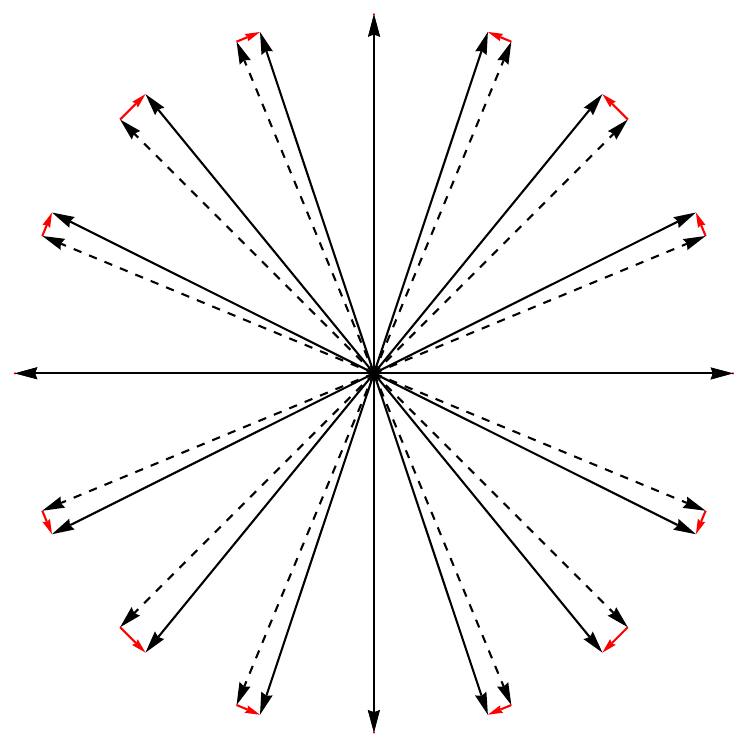}

    \caption[Effect of the displacement-displacement memory on the transverse propagation of null geodesics]{Effect of the displacement-displacement memory on the transverse propagation of null geodesics. Arrows represent propagation directions in the 2-plane which appears, at fixed $t$, to be orthogonal to the propagation of the gravitational wave (according to the canonical observer). The horizontal and the vertical axes are aligned with the eigenvectors of $\Mdd_{ij}$. Dashed arrows denote propagation directions before the gravitational wave arrives. Solid black arrows denote propagation directions after the wave has left. Red arrows denote shifts in the transverse velocities. Displacements of the null geodesics are not shown.}
    
    \label{fig:nullVects2D}
    \end{minipage}
    \hspace{\columnsep}
    \begin{minipage}[c]{0.988\columnwidth}
    \vspace{0.8cm}
    \includegraphics[width=.95\columnwidth]{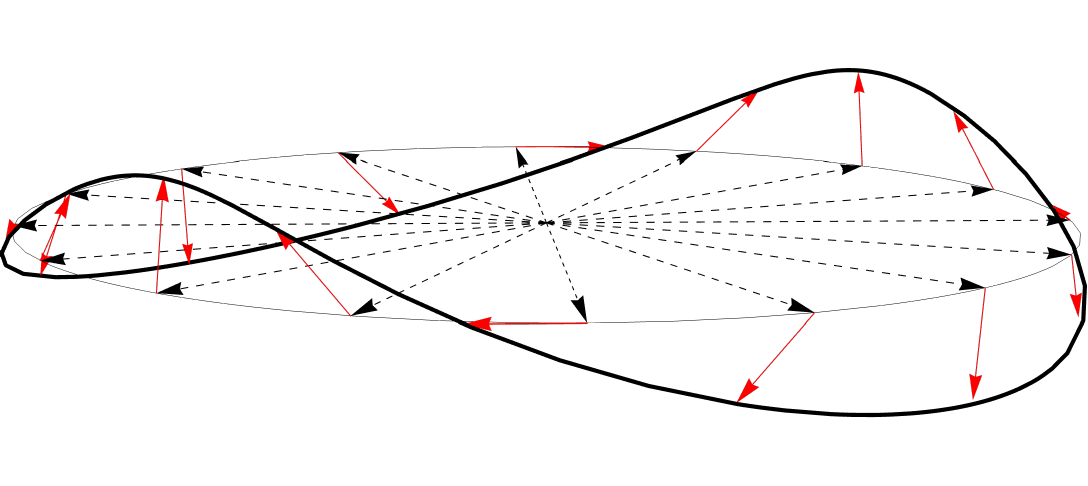}
    \vspace{0.8cm}
    \caption[Effect of the displacement-displacement memory on the propagation of null geodesics in three dimensions]{Effect of the displacement-displacement memory on the propagation of null geodesics in three dimensions. The dashed arrows represent initial propagation directions for the null geodesics, all of which are assumed here to lie in the ${\theta_-= \pi/2}$ plane which is orthogonal to the gravitational wave. The thick black line is formed from the tips of all velocity vectors after the gravitational wave has passed. The red arrows show some particular shifts in the 3-velocities of various geodesics. Displacements are not shown. A top-down projection of this diagram is given in \cref{fig:nullVects2D}.}
    
    \label{fig:nullVects3D}
    \end{minipage}
\end{figure*}

First note that if a null geodesic is initially comoving with a gravitational wave, that wave does not affect it. For all other null geodesics, $\kappa = 0$, $V_\pm = 1$, and \cref{coordTranslate} implies that
\begin{align}
    \delta_{ij} \dot{x}^i_\pm \dot{x}^j_\pm =  2 \cot^2 (\half \theta_\pm) .
    \label{dotXNull}
\end{align}
The magnitudes of the $\dot{x}^i_\pm$ therefore encode the longitudinal angles $\theta_\pm$. This is also true for the magnitudes $\delta_{ij} V^i_\pm V^j_\pm = \sin^2 \theta_\pm$ of the transverse components of the 3-velocities $V^I_\pm$. From this perspective, it is useful to introduce the unit 2-vectors
\begin{equation}
    n^i_\pm \equiv V^i_\pm \csc \theta_\pm,
    \label{nV}
\end{equation}
which describe only the transverse direction of motion.  \Cref{memTensLin,Smatrix,SDef,coordTranslate,dotXNull} then imply that through first order in the curvature, the perturbation to the propagation direction of a null geodesic is described by
\begin{align}
    \Delta \theta = \bigg\{ \Mdd_{ij} n^i_- n^j_- - \frac{1}{\sqrt{2}} \Mdv_{ij} n^i_-  \big[ Z_- n^j_- 
    \nonumber
    \\
    ~ + X^j_- \tan(\half \theta_-) \big] \bigg\} \sin \theta_- + \mathcal{O}(H^2),
    \label{dtheta}
\end{align}
and
\begin{align}
    \Delta n^i =  (n^i_- n^j_- - \delta^{ij} ) \bigg\{ \Mdd_{jk} n^k_- - \frac{1}{\sqrt{2}} \Mdv_{jk}  \nonumber
    \\
    ~ \times  \big[ Z_- n^k_- + X^k_- \tan(\half \theta_-) \big] \bigg\} + \mathcal{O}(H^2).
    \label{dn}
\end{align} 
Note that there is no sense in which $\Delta \theta$ is generically small compared to $\Delta n^i$. Gravitational wave memory therefore affects both the longitudinal and the transverse properties of null geodesics. It is also apparent that null geodesics can exhibit memory effects which are independent of their initial projected displacements $X_-^I$. 

If those displacements are small, or if $\Mdv_{ij}$ is negligible at $\mathcal{O}(H)$, only $\Mdd_{ij}$ can significantly affect the propagation directions. The longitudinal deflection $\Delta \theta$ is then maximized when $n^i_-$ is an eigenvector of $\Mdd_{ij}$. In that case, the transverse deflection $\Delta n^i$ vanishes. If $n^i_-$ instead lies midway between the two eigenvectors of $\Mdd_{ij}$, then it is the transverse deflection that is maximized while the longitudinal deflection vanishes. More generally, $\Delta \theta$ and $\Delta n^i$ can both be nonzero. The transverse velocity changes here are illustrated in \cref{fig:nullVects2D}. A full three-dimensional picture is provided in \cref{fig:nullVects3D}, where it is seen that an initially planar collection of null geodesic rays does not remain planar after it passes through a gravitational wave (even when displacements are ignored). 

\begin{figure}
    \centering
    \includegraphics[width=.85\columnwidth]{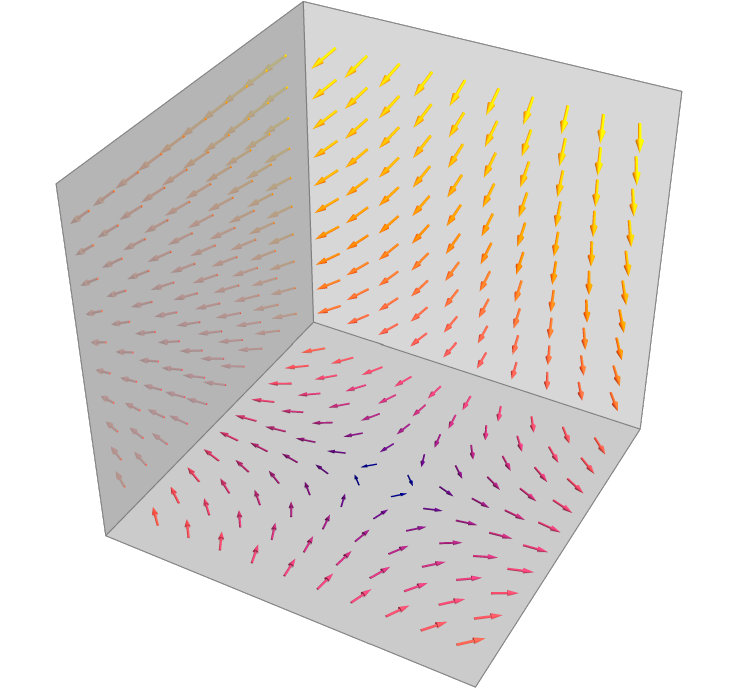}
    \caption[Effect of the displacement-displacement memory on the  displacements of null geodesics]{Effect of the displacement-displacement memory on the displacements of initially comoving null geodesics. Each point here corresponds to a different initial projected position $X^I_-$ for a null geodesic. The vectors at each point represent the displacements $\Delta X^I = X^I_+ -X^I_-$. The bottom plane here is $Z_- =0$, where a standard quadrupolar displacement can be observed; cf. \cref{fig:displMem}. Projections of displacement vectors onto different $Z_- = \mathrm{constant}$ planes result in similar quadrupolar patterns but with different singular points. However, displacements are not purely transverse when $Z_- \neq 0$. Also note that this figure describes only the effects of $\Mdd_{ij}$. The effect of $\Mvd_{ij}$ would be to add a constant displacement at every point, while that of $\Mdv_{ij}$ would be more complicated. Also note that we have chosen a particular $\theta_-$ and $n^i_-$ for all displacements here; somewhat different patterns result for geodesics which propagate in different directions.}
    \label{fig:nullDispl}
\end{figure}

Although 3-velocities of null geodesics are easier to measure than displacements, the latter may be important as well. Similar calculations to the ones which led to \cref{dtheta,dn} [but which also use \cref{dv}] can be used to show that the projected longitudinal displacements of null geodesics are given by
\begin{widetext}
\begin{align}
    \Delta Z = \frac{1 + \cos \theta_-}{\sqrt{2}} \left\{ \left( \Mvd_{ij} + \sqrt{2} Z_- \Mdd_{ij} \right) n^i_- n^j_- + \frac{1}{2} \Mdv_{ij} \left[ X^i_- X^j_- \tan^2 (\half \theta_-)  - Z_-^2 n^i_- n^j_- \right] \right\} + \mathcal{O}(H^2),
    \label{dZNull}
\end{align}
while their transverse counterparts are
\begin{align}
    \Delta X^i = \Mdd_{ij} X^j + \frac{1}{\sqrt{2}} \cot(\tfrac{1}{2}\theta_-) \big[ 2 \delta^{ij} - (1+ \cos \theta_-) n^i_- n^j_- \big] \left( \Mvd_{jk} + \sqrt{2} Z_- \Mdd_{jk} - \frac{1}{2} Z_-^2 \Mdv_{jk} \right) n^k_- 
    \nonumber
    \\
    - \frac{1}{2 \sqrt{2}} \big[ 2 Z_- \delta^{ij} + n^i_- X^j_- \sin \theta_- \big] \Mdv_{jk} X^k_- + \mathcal{O}(H^2).
    \label{dXNull}
\end{align}
\end{widetext}
The first term in $\Delta X^i$ clearly matches the displacement memory \eqref{displMem} that is associated with slowly moving timelike geodesics. However, there is much more in this null setting. The displacement-displacement memory not only maps initial transverse displacements to final transverse displacements; it also maps initial \emph{longitudinal} displacements to final transverse \textit{and} longitudinal displacements. Moreover, these latter effects depend on the direction in which the null geodesic is propagating. The effect of $\Mdd_{ij}$ on the three-dimensional displacement field of a collection of initially comoving null geodesics is shown in \cref{fig:nullDispl}. 

Perhaps more interestingly, our results show that null geodesics can be used to measure the velocity-displacement memory $\Mvd_{ij}$. That affects both $\Delta Z$ and $\Delta X^i$, but not $\Delta \theta$ or $\Delta n^i$, nor any memory effects that are associated with slowly moving geodesics. Unlike the effects due to $\Mdd_{ij}$ or $\Mdv_{ij}$, those of $\Mvd_{ij}$ are independent of the initial displacement. If we consider a collection of initially parallel null geodesics, then their final displacements will be shifted by a constant that depends on $\Mvd_{ij}$, a linear transformation which depends on $\Mdd_{ij}$, and a quadratic displacement which depends on $\Mdv_{ij}$.

\subsection{Memory effects and the world function}
\label{sec:bitensors}

A somewhat different perspective on geodesic memory is provided by examining Synge's world function $\sigma(x,x')$, which returns one half of the squared geodesic distance between the spacetime points\footnote{Here, we use $x$ and $x'$ to denote events in spacetime rather than individual coordinates. When doing so below, the distinction should be clear from context.} $x$ and $x'$. From this biscalar, all properties of (timelike, null, and spacelike) geodesics can be extracted simply by differentiation \cite{Synge1960}. The world function also plays a central role in the discussion of wave propagation in curved spacetimes \cite{Poisson2011}, and we use it for this purpose in \cref{Sect:FieldProp} below.

In terms of the Jacobi propagators $A_{ij}$ and $B_{ij}$, the world function in any plane wave spacetime is known to be given by \cite{Hollowood_2008, HarteDrivas}
\begin{align}
    \sigma = \tfrac{1}{2}(u-u') \big[ -2 (v-v') + \partial_u B_{ik} (B^{-1})^{k}{}_{j} x^i x^j\phantom{\big].}
    \nonumber
    \\
    ~ + (B^{-1})_{ik} A^{k}{}_{j} x'^i x'^j - 2 (B^{-1})_{ij} x'^i x^j\big],
    \label{sigma}
\end{align}
where $(B^{-1})_{ij}$ denotes the matrix inverse of $B_{ij}$. This is exact and holds for all pairs of points which do not contain any intervening conjugate hyperplanes: $u$ and $u'$ must be sufficiently close that $\det B_{ij}(u'',u') \neq 0$ for all $u'' \in (u',u]$. If we now assume that the gravitational wave is weakly curved, \cref{B1,A1} imply that
\begin{align}
    \sigma &= \bar{\sigma} + \frac{ 1 }{ 2( u-u')} \int_{u'}^u \! \rmd w\, H_{ij}(w) \big[ (w-u') x^i  
    \nonumber
    \\
    & ~ + (u-w) x'^i \big] \big[ (w-u') x^j + (u-w) x'^j \big] + \mathcal{O}(H^2),
    \label{sigmaPert}
\end{align}
where
\begin{equation}
    \bar{\sigma} \equiv - (u-u') (v-v') + \frac{1}{2} \delta_{ij} (x^i - x'^i) (x^j - x'^j)
\end{equation}
is the world function in flat spacetime. In a weakly curved sandwich wave where $u'$ lies in $\mathcal{I}_-$ while $u$ lies in $\mathcal{I}_+$, \cref{memTensLin} can now be used to write the perturbed world function in terms of the memory tensors:
\begin{align}
    \sigma = \bar{\sigma} + \frac{1}{2 (u-u')} \big[ \Mdv_{ij} ( u'x^i - u x'^i ) (u' x^j - u x'^j)
    \nonumber
    \\
    ~ + 2 \Mdd_{ij} ( x^i - x'^i) ( u' x^j - u x'^j)  - \Mvd_{ij} ( x^i - x'^i)
    \nonumber
    \\
    ~ \times (x^j - x'^j) \big] + \mathcal{O}(H^2).
    \label{sigmaPert2}
\end{align}

One application of this expression is that it can be used to immediately see how proper times are affected by gravitational-wave memory: If two timelike-separated events are considered, the proper time along the geodesic that connects them is simply $\sqrt{-2 \sigma}$. In general, all memory tensors can contribute to that time. 

The perturbed world function can also be used to easily see how light cones are deformed by gravitational wave memory. A future-pointing light cone whose vertex lies at $(u',v',x'^i)$, before the gravitational wave has arrived, is given by the surface $\sigma = 0$, or equivalently by
\begin{align}
    v = v' + \frac{1}{2(u-u')^2} \Big\{ \big[ (u-u') \delta_{ij} - \Mvd_{ij} \big] (x^i - x'^i)
    \nonumber
    \\
    ~ \times (x^j - x'^j)  + \Mdv_{ij} ( u'x^i - u x'^i ) (u' x^j - u x'^j) 
    \nonumber
    \\
    ~  +  2 \Mdd_{ij} ( x^i - x'^i) ( u' x^j - u x'^j)  \Big\} + \mathcal{O}(H^2)
    \label{vretPert}
\end{align}
after the wave has left. To better interpret this, use the inertial coordinates in \cref{tzDef} and suppose that the light cones emanate from the canonical observer at $x'^i = z' = 0$. The deformed light cone is then given by the surface
\begin{align}
    (t-t')^2 = \bigg[ \delta_{ij} +  \frac{ \sqrt{2} }{ t- t' - z } \Big( 2 t'^2 \Mdv_{ij} + \sqrt{2} t' \Mdd_{ij}
    \nonumber
    \\
    ~  - \Mvd_{ij} \Big) \bigg]  x^i x^j + z^2 + \mathcal{O}(H^2).
        \label{lightCone}
\end{align}
For light emitted long ago, the effect of $\Mdv_{ij}$ dominates over that of $\Mdd_{ij}$ (when the former tensor is nonzero), and the effect of $\Mdd_{ij}$ dominates over that of $\Mvd_{ij}$. 

It also follows that light cones projected onto a transverse plane with constant $t$ and $z$ are necessarily elliptical. More precisely, the eccentricity of each ellipse is $2 \sqrt{\Lambda}$, where $\Lambda$ denotes the positive eigenvalue of
\begin{equation}
     \frac{1}{t-t'-z} (2 t'^2 \Mdv_{ij} + \sqrt{2} t' \Mdd_{ij}  - \Mvd_{ij}). 
\end{equation}
At fixed $t$ and $z$, this eigenvalue generically depends on $t'$, meaning that the cross sections of the observer's forward light cones can have varying eccentricities. These cross sections can also be rotated with respect to one another, as the (generically $t'$-dependent) eigenvector which is associated with $\Lambda$ provides the minor axis of the ellipse. These nontrivial eccentricities and the orientations of the associated cross sections are consequences of gravitational wave memory. \cref{lightCone} can also be used to determine how light cones are projected into, e.g., planes with constant $t$ and $y$. However, the shapes of those projections are more complicated.

\section{Nongeodesic motion, spin, and memory}
\label{Sect:SH}

The geodesic scattering described in the previous section can, in general, only approximate the scattering of an actual extended object. Trajectories can accelerate due to an object's angular momentum, as well as from the quadrupole and higher-order multipole moments of its stress-energy tensor \cite{Dixon74}. We now discuss how angular momentum interacts with gravitational wave memory, focusing on the scattering of massless objects. Physically, this corresponds to considering the trajectories of, e.g., electromagnetic wave packets one order beyond geometric optics.

We begin in \cref{Sect:SHreview} by reviewing the spin Hall equations, which describe the linear-in-spin corrections to the trajectories of massless particles. Next, \cref{Sect:SHconslaws} discusses conservation laws for the spin Hall equations, showing that every conformal Killing vector is associated with a conserved quantity. This result is true not only in plane wave spacetimes, but in any spacetime which admits a conformal Killing vector field. Finally, \cref{Sect:SHscattering} applies these results in sandwich wave spacetimes to describe how gravitational wave memory scatters massless wave packets with angular momentum.

\subsection{Review of the spin Hall equations}
\label{Sect:SHreview}

\begin{figure*}
    \centering
    \begin{minipage}[t]{0.99\columnwidth}
    \includegraphics[width=\textwidth]{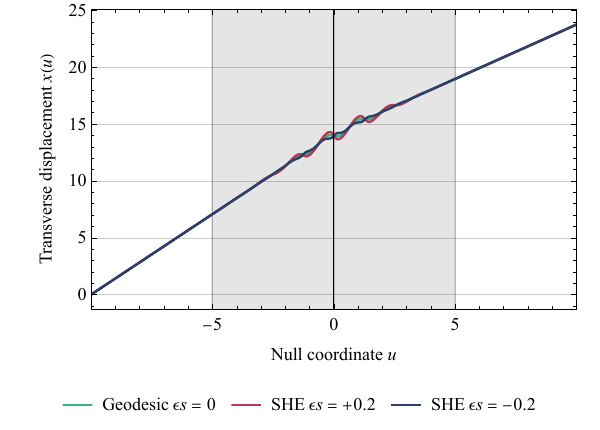}
    %\caption{$x$-component of spin Hall and geodesic trajectories interacting with a circularly polarized sandwich wave (shaded region).}
    %\label{fig:SHE_x}
    \end{minipage}\qquad
    \begin{minipage}[t]{0.99\columnwidth}
    \includegraphics[width=\textwidth]{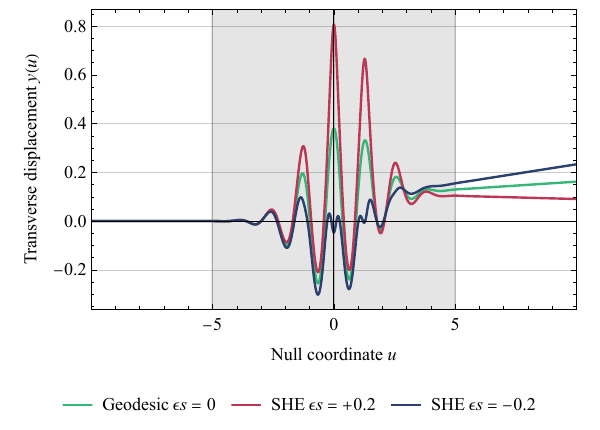}
    %\caption{$y$-component of spin Hall and geodesic trajectories interacting with a circularly polarized sandwich wave (shaded region).}
    %\label{fig:SHE_y}
    \end{minipage}
    \raggedright
    \caption{Transverse motion of massless spinning particles in a gravitational sandwich wave. Three particles are considered with the same initial conditions, one with vanishing angular momentum (which is a geodesic), and two with opposite and nonzero angular momenta. The trajectories of these particles coincide in $\mathcal{I}_-$, but not in $\mathcal{I}_C$ or $\mathcal{I}_+$. More specifically, the final velocities of these particles are seen to depend on $\epsilon s$, illustrating that memory effects can depend on angular momentum. The gravitational waveform and other conventions here are the same as in \cref{fig:null_geo_x}; indeed, the $\epsilon s = 0$ curves here are identical to the curves in that figure. Centroids of the underlying wave packets are defined here by assuming that the $t^\alpha$ in \cref{SSC} is given by $- \nabla_\alpha t$. 
    }
    \label{fig:SHE_x}
\end{figure*}

If angular momentum and other extended-body characteristics are ignored, then there are various senses in which the ``center'' of a high-frequency electromagnetic wave packet moves along a null geodesic. However, wave packets can exhibit significant angular momentum one order beyond geometric optics. To understand the effect of an angular momentum $S^{\alpha \beta} = S^{[\alpha \beta]} (\tau)$, we first define the centroid $x^\alpha(\tau)$ of an extended wave packet by choosing a timelike vector field $t^\alpha$ and then imposing the Corinaldesi–Papapetrou spin supplementary condition \cite{CP_ssc, Costa2015, HarteOancea} 
\begin{equation}
    S_{\alpha\beta} t^\beta = 0.
    \label{SSC}
\end{equation}
If $S^{\alpha\beta}$ is purely longitudinal,\footnote{Here, $S^{\alpha\beta}$ is said to be ``longitudinal''  if it is orthogonal to the linear momentum $p_\beta$. The angular momentum tensor is then dual to a vector which is proportional to $p^\alpha$. In other contexts, $S^{\alpha\beta} p_\beta = 0$ would typically be interpreted as an implicit definition for the centroid of an object, and would then be referred to as the Tulczyjew–Dixon spin supplementary condition \cite{Costa2015}. However, this spin supplementary condition condition does not define a unique worldline when the momentum is null; cf. \cite{HarteOancea} and \cite[p. 70]{PenroseRindler2}. The vanishing of $S^{\alpha\beta} p_\beta$ here is instead interpreted as a physical restriction on the nature of the angular momentum. Wave packets with non-longitudinal angular momentum are possible \cite{HarteOancea}, with examples sometimes described as spatiotemporal vortex beams \cite{PhysRevA.86.033824,PhysRevLett.126.243601}. We nevertheless focus only on the longitudinal case.} then its first nontrivial contribution to the motion of such a centroid may be shown to be described by the spin Hall equations \cite{GSHE2020, HarteOancea, GSHE_rev}
\begin{subequations}
\label{eq:Spin Hall}
\begin{align}
    \frac{ \mathrm{D} p_\alpha }{ \dd\tau }  &= - \frac{1}{2} R_{\alpha \beta \gamma \delta} \frac{ \dd x^\beta }{ \dd\tau } S^{\gamma \delta} ,
    \label{pDot}
    \\
    \frac{ \dd x^\alpha }{ \dd\tau } &= p^\alpha + \frac{ 1 }{ p \cdot t} S^{\alpha \beta} p^\gamma \nabla_\gamma t_\beta,
    \label{momVel}
    \\
    S^{\alpha \beta} &= \left( \frac{\epsilon s }{p \cdot t} \right) \varepsilon^{\alpha \beta \gamma \delta} p_\gamma t_\delta,
    \label{Sform}
\end{align}
\end{subequations}
where $p_\alpha$ denotes the wave packet's linear momentum, which is null\footnote{More precisely, $p_\alpha p^\alpha = \mathcal{O}(\epsilon^2)$. It is not possible for any wave packet with a nonzero angular momentum to have an \emph{exactly} null linear momentum \cite{HarteOancea}.}, $\tau$ is a dimensionless parameter along the worldline, $\epsilon$ and $s$ are constant parameters, and $\varepsilon^{\alpha \beta \gamma \delta}$ denotes the Levi-Civita tensor. The spin Hall equations are intended to hold up to terms of order $\epsilon^2$, where the small parameter $\epsilon$ relates the (large) dominant frequency of a wave packet to its momentum: If $U^\alpha$ denotes the 4-velocity of an observer, then the angular frequency seen by that observer is $\omega = (- p \cdot U)/\epsilon$. The parameter $s$ instead provides a dimensionless magnitude for the angular momentum, in the sense that 
	\begin{equation}
		S^{\alpha\beta} S_{\alpha\beta} = 2 (\epsilon s )^2.
		\label{Spin2}	
	\end{equation}
 For some electromagnetic wave packets, circular polarization results in $s = \pm 1$, depending on the handedness of the polarization state. However, there are other electromagnetic wave packets for which $|s|$ can be much larger than 1. In optics, these are described as wave packets or beams that carry intrinsic orbital angular momentum \cite{AM_Light,BLIOKH20151}.

The spin Hall equations can be used to describe wave packets composed not only of electromagnetic fields \cite{GSHE2020,HarteOancea,Frolov2020,SHE_QM1,PhysRevD.109.064020,Frolov2024}, but also of linearized gravitational fields \cite{GSHE_GW,SHE_GW,GSHE_lensing,GSHE_lensing2,Frolov2024(2)}, massless Dirac fields \cite{GSHE_Dirac}, or even scalar fields (which can carry orbital angular momentum). For these different cases, the spin Hall equations might differ only in the parameter $s$ that determines the magnitude of the angular momentum tensor.  

It may also be noted that the spin Hall equations are a special case of the Mathisson–Papapetrou equations [together with the Corinaldesi–Papapetrou spin supplementary condition \eqref{SSC}], which have long been known to describe the motion of spinning objects in curved spacetimes \cite{HarteOancea}. However, specializing to the case of a nearly massless wave packet with longitudinal angular momentum allows the usual evolution equation for the angular momentum to be solved explicitly, yielding \cref{Sform}. Therefore, all that remains is an evolution equation \eqref{pDot} for the linear momentum and a momentum-velocity relation \eqref{momVel} which relates $p^\alpha$ to $\rmd x^\alpha/\rmd \tau$.

\begin{figure*}
    \centering
    \begin{minipage}[t]{0.99\columnwidth}
    \includegraphics[width=\textwidth]{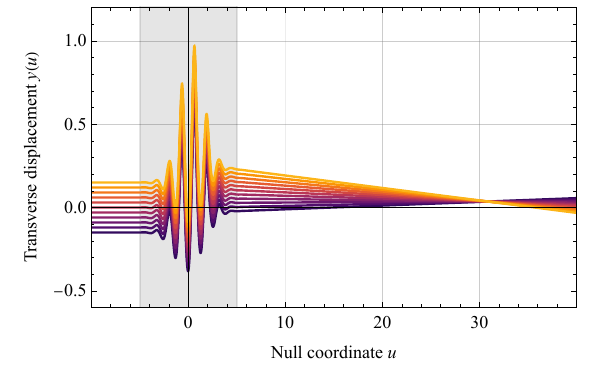}
    \end{minipage}\qquad
    \begin{minipage}[t]{0.99\columnwidth}
    \includegraphics[width=\textwidth]{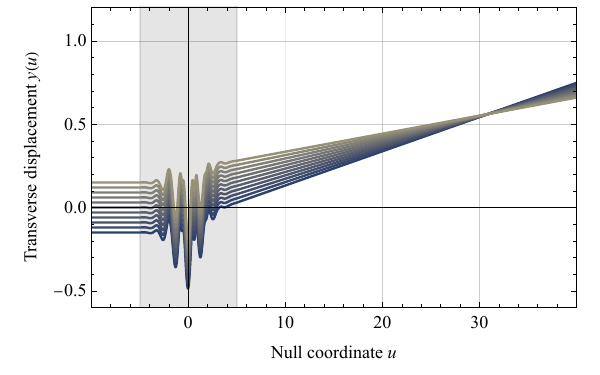}
    \end{minipage}
    \raggedright
    \caption{Bundles of initially parallel massless spinning particles in a gravitational sandwich wave. The left and the right panels here illustrate cases in which $\epsilon s = +0.2$ and $\epsilon s = -0.2$, respectively. Angular momentum is thus seen to affect where a bundle is focused. This figure is analogous to \cref{fig:null_geos} in the geodesic case. The gravitational waveform and other conventions here are the same as in \cref{fig:null_geo_x}.}
    \label{fig:SHE1_bundle}
\end{figure*}

To offer some intuition for these frequency- and angular momentum-dependent effects, some examples of spin Hall rays propagating through a sandwich wave spacetime are presented in \cref{fig:SHE_x,fig:SHE1_bundle}. These are obtained by numerically integrating \cref{eq:Spin Hall}, although the same trajectories can also be obtained using the analytical results derived below in \cref{sym1,sym2}. It is nevertheless clear that memory effects depend, in general, on an object's angular momentum.  

\subsection{Conservation laws and the spin Hall equations}
\label{Sect:SHconslaws}

It is well known in the context of the Mathisson–Papapetrou equations that if $\xi^\alpha$ is a Killing vector field, then the generalized momentum 
\begin{equation}
    \mathcal{P}_\xi \equiv p_\alpha \xi^\alpha + \frac{1}{2} S^{\alpha \beta} \nabla_\alpha \xi_\beta
    \label{consLaw}
\end{equation}
must be conserved \cite{Dixon70a, HarteSyms}. It was already noted in Ref.~\cite{HarteOancea} that since the spin Hall equations are a special case of the Mathisson–Papapetrou equations, these conservation laws continue to hold. However, we now establish a more general result for the spin Hall equations: \cref{consLaw} is conserved not only for ordinary Killing vectors, but for all \emph{conformal} Killing vectors. We require only that
\begin{equation}
    \mathcal{L}_\xi g_{\alpha\beta} = 2\Upsilon \, g_{\alpha\beta}
    \label{confKilling}
\end{equation}
for some scalar field $\Upsilon$. The ordinary Killing case is recovered if $\Upsilon = 0$. 

To motivate why this might be so, note that the generalized momentum is given by
\begin{equation}
    \label{eq:momentum generalized}
    \mathcal{P}_\xi(\tau) = \int_{\Sigma_\tau} \dd S_\alpha\, T^{\alpha}{}_{\beta} \xi^\beta ,
\end{equation}
where $\xi^\alpha$ is now a \emph{generalized} Killing field \cite{HarteSyms} and $T^{\alpha \beta}$ is the stress-energy tensor of the object---here an electromagnetic wave packet. The hypersurfaces $\Sigma_\tau$ are assumed to foliate the worldtube of the object. Using stress-energy conservation, it follows that 
\begin{equation}
    \frac{\dd}{\dd\tau} \mathcal{P}_\xi(\tau) = \frac{1}{2} \int_{\Sigma_\tau} \hspace{-0.5em} \dd S_\gamma\, \tau^\gamma T^{\alpha \beta} \mathcal{L}_\xi g_{\alpha \beta}  ,
\end{equation}
where $\tau^\gamma$ is a time evolution vector field for the foliation. This clearly vanishes in the Killing case. However, ${\t T{^\alpha_\alpha} = 0}$ for any electromagnetic field, and in that case, $\mathcal{P}_\xi$ is constant whenever $\xi^\alpha$ is conformally Killing. 

This motivates our result but does not prove it. First, the spin Hall equations can be applied to, e.g., scalar wave packets, which are not necessarily trace-free. Second, although ordinary Killing fields are also generalized Killing fields, the same cannot be said of proper conformal Killing fields. Third, there are a number of approximations inherent in the spin Hall equations, and we have not made any of these precise. Instead of doing so, we shall establish by direct calculation that given any conformal Killing field, the spin Hall equations imply that $\mathcal{P}_\xi$ is indeed conserved, at least through first order in $\epsilon$.

From \cref{pDot} and from $DS^{\alpha\beta}/\dd\tau = 2 p^{[\alpha}  \rmd x^{\beta]}/\rmd \tau$, which is one of the Mathisson–Papapetrou equations, direct differentiation of \cref{consLaw} results in
\begin{align}
    \frac{ \dd \mathcal{P}_\xi }{ \dd\tau } = \frac{1}{2} \frac{ \rmd x^\beta }{ \rmd \tau } [ p^{\alpha} \mathcal{L}_\xi g_{\alpha\beta} +  S^{\alpha \gamma}   ( \nabla_\beta \nabla_\alpha \xi_\gamma + R_{\alpha \gamma \beta \delta} \xi^\delta) ]. 
\end{align}
This holds for any vector field $\xi^\alpha$. However, if that vector field is conformally Killing \cite[Eq.~(11.4b)]{Hall}, then
\begin{equation}
    \nabla_\gamma \nabla_\alpha \xi_\beta = R_{\beta \alpha \gamma \delta} \xi^\delta + g_{\alpha \beta} \nabla_\gamma \Upsilon - 2 g_{\gamma[ \alpha} \nabla_{\beta]} \Upsilon.
\end{equation}
Using this results in
\begin{equation}
    \frac{ \dd \mathcal{P}_\xi }{\dd \tau} =  \left( \Upsilon 
 p_\alpha - S_{\alpha}{}^{\beta}  \nabla_\beta \Upsilon \right) \frac{\rmd x^\alpha}{\rmd \tau} .
\end{equation}
The null character of $p_\alpha$, together with \cref{momVel,Sform}, show that both terms on the right-hand side of this equation are individually $\mathcal{O}(\epsilon^2)$. It follows that for each conformal Killing field $\xi^\alpha$,
\begin{equation}
    \mathcal{P}_\xi = p_\alpha \left[ \xi^\alpha + \left( \frac{ \epsilon s }{ 2 p \cdot t} \right) \varepsilon^{\alpha \beta \gamma \delta} t_\beta \nabla_\gamma \xi_\delta \right] 
    \label{consLaw2}
\end{equation}
is conserved through first order in $\epsilon$. This is true not only in plane wave spacetimes, but in any spacetime which admits a conformal Killing vector.

However, in the plane wave spacetimes of interest here, there are at least five Killing vector fields.  There is also the homothety \eqref{homothety}, which is a special type of conformal Killing vector in which $\Upsilon = 1$. The spin Hall equations therefore admit at least six conservation laws in plane wave spacetimes.

\subsection{Massless spinning particles in plane wave spacetimes}
\label{Sect:SHscattering}

We may now apply the aforementioned conservation laws to solve the spin Hall equations in plane wave spacetimes. Before doing so, it is first necessary to choose a timelike vector field $t^\alpha$ with which to define the centroid. One simple possibility is to set
\begin{equation}
    t_\alpha = - \nabla_\alpha t = - \frac{1}{\sqrt{2}} \nabla_\alpha (u + v)
    \label{tDef}
\end{equation}
at least in the flat regions $\mathcal{I}_\pm$, where it reduces to the 4-velocity of the canonical observer at $x=y=z=0$. 

The simplest Killing vector in a plane wave spacetime is $\ell_\alpha = - \nabla_\alpha u$, and since this is covariantly constant,
\begin{equation}
    E \equiv -\mathcal{P}_\ell = -p_v = \mathrm{constant}.
\end{equation}
In exceptional cases where $E = 0$, both the momentum and the velocity of the wave packet remain parallel to $\ell^\alpha$. Wave packets that are traveling in the same direction as the background gravitational wave are therefore unaffected either by that wave or by their own angular momentum. The cases in which $E \neq 0$ are more interesting, and we focus on them in the following. 

If $\xi^\alpha$ is used to denote one of the Killing fields with the form given in \cref{Killing}, then a calculation shows that
\begin{equation}
    \nabla_\alpha \xi_\beta = 2\dot{\Xi}_i \ell_{[\beta} \nabla_{\alpha]} x^{i} .
    \label{dxi}
\end{equation}
Substituting this and \cref{tDef} into \cref{consLaw2}, we obtain
\begin{align}
     p_i \left[ \Xi^i - \left( \frac{ \epsilon s  \varepsilon^{ij} }{ \sqrt{2} p \cdot t } \right)   \dot{\Xi}_j \right] - E \dot{\Xi}_i x^i = \mathrm{constant},
\end{align}
where $\varepsilon^{ij} = \varepsilon^{uvij}$ is a two-dimensional permutation symbol. Varying over all possible $\Xi^i$, this encodes four conservation laws. Using \cref{XiDef,AfromB,reciprocity} to evaluate these laws at $u$ and at $u'$ shows that 
\begin{align}
    p_i &= p'^j \partial_u B_{ij} + E \left[ x'^j +  \frac{ \epsilon s}{ \sqrt{2} E  } \left( \frac{ p'_k \varepsilon^{kj} }{ p' \cdot t'} \right)  \right] \partial_{u} A_{ij},
    \label{sym1}
\end{align}
and 
\begin{align}
    x_i = A_{ij} x'^j + \frac{ B_{ij} p'^j}{E} + \frac{ \epsilon s \varepsilon^{jk}}{\sqrt{2} E } \left( \frac{ \delta_{ij} p_k}{ p \cdot t} - \frac{ A_{ij} p'_k }{ p' \cdot t'} \right),
    \label{sym2}
\end{align}
where all instances of the Jacobi propagators are evaluated at $(u,u')$. As in the geodesic case, the final position and the final momentum therefore depend on the initial position and the initial momentum only via the Jacobi propagators $A_{ij}$ and $B_{ij}$. 

In the trivial case where $u$ and $u'$ both lie in one of the flat regions $\mathcal{I}_-$ or $\mathcal{I}_+$, so there is no curvature between the initial and the final states, it follows from \cref{sym1,sym2} that
\begin{equation}
    x^i = x'^i + \frac{1}{ E } (u-u') p'^i , \qquad p_i = p'_i.
\end{equation}
The transverse velocity
\begin{equation}
    \dot{x}^i = \dd x^i / \dd u = p^i /E 
\end{equation}
is therefore constant and the spin is irrelevant; wave packets follow null geodesics in flat regions. However, the initial and the final geodesics may appear to be “different” due to the presence of the intervening gravitational wave. 

This can be understood by assuming that $u$ lies in $\mathcal{I}_+$ while $u'$ lies in $\mathcal{I}_-$. Doing so, we can effectively correct the scattering matrix \eqref{SDef} which was derived above for geodesic motion. Using the projected spatial positions $x^i_\pm$ which were defined by \cref{xPM}, it follows from \cref{Bexpand,Aexpand,sym1} that the momentum after the wave has left is
\begin{align}
    p^+_i = \left[ \delta_{ij} + \Mvv_{ij} + \frac{ \epsilon s }{E} \sin^2 (\tfrac{1}{2} \theta_-) \Mdv_{ik} \varepsilon^{k}{}_{j} \right] p^j_-
    \nonumber
    \\
    ~ + E \Mdv_{ij} x_-^j 
    \label{pPlusSH}
\end{align}
where $\theta_- = 2 \arctan ( [2 E^2/\delta_{ij} p^i_- p^j_-]^{1/2} )$ again denotes the initial angle between the wave packet and the gravitational wave, as seen by the canonical observer at the origin [cf. \cref{dotXNull}]. It follows that $\Mdv_{ij}$ is the only memory tensor that can produce spin-dependent corrections to the transverse momentum. It may also be noted that since $\theta_-$ depends on $p^-_i$, nonzero spin results in a nonlinear mapping from the initial state $(x_-^i, p_-^i)$ to the final state $(x_+^i, p_+^i)$.

Applying a similar calculation to \cref{sym2}, the projected transverse position of a potentially spinning wave packet is given by
\begin{align}
    x_+^i = (\delta_{ij} + \Mdd_{ij} ) x_-^j + \frac{1}{E} \Mvd_{ij} p_-^j + \frac{\epsilon s \varepsilon^{jk} }{ E^2 }
    \nonumber
    \\
     ~  \times \big\{  ( \delta_{ij}  + \Mdd_{ij} ) p_k^- 
 \sin^2 (\tfrac{1}{2} \theta_-) - \delta_{ij}  \big[ ( \delta_{kl} 
 \nonumber
    \\
     ~  + \Mvv_{kl} ) p^l_-    + E \Mdv_{kl} x^l_- \big] \sin^2 (\tfrac{1}{2} \theta_+)\big\} .
     \label{xPlusSH}
\end{align}
Angular momentum therefore affects $x_+^i$ via all memory tensors except $\Mvd_{ij}$ (although $\Mvd_{ij}$ does affect the spin-\emph{independent} contribution to the projected position). Finally, we can write
\begin{align}
    \begin{pmatrix}
        x^i_+ \\ \dot{x}^i_+ 
    \end{pmatrix} = \left(\mathcal{S} + \frac{\epsilon s}{E} \mathcal{S}_1 \right)
    \begin{pmatrix}
        x^j_- \\ \dot{x}^j_-  
    \end{pmatrix},
    \label{Smatrix2}
\end{align}
where $\mathcal{S}$ is given by \cref{SDef} and the spin-dependent corrections to the scattering matrix can be written as
\begin{widetext}
\begin{equation}
    \mathcal{S}_1 = 
    \begin{pmatrix}
        \sin^2 (\tfrac{1}{2} \theta_+) \varepsilon^{i k} \Mdv_{k j}  &    \sin^2 (\tfrac{1}{2} \theta_-) \varepsilon^{k j} ( \delta_{i k} + \Mdd_{i k} ) - \sin^2 (\tfrac{1}{2} \theta_+) \varepsilon^{i k} (\delta_{k j} + \Mvv_{k j}) \\
        0                    &   \sin^2 (\tfrac{1}{2} \theta_-) \varepsilon_{k j} \Mdv_{i k}
    \end{pmatrix}.
    \label{SDef2}
\end{equation}
\end{widetext}
Note, however, that despite the matrix form of \cref{Smatrix2}, the map from initial states to final states is not linear when $\epsilon \neq 0$; the corrected scattering matrix $\mathcal{S}_1$ depends not only on the geometry but also on the particle's state.

Regardless, \cref{pPlusSH,xPlusSH} fully determine the transverse scattering of a spinning wave packet through first order in $\epsilon$. Furthermore, the full 4-momentum $p_\alpha^\pm$ can be determined from its transverse components $p_\alpha^i$ and from the null constraint $p_\alpha^\pm p^\alpha_\pm = 0$. This is not, however, sufficient to determine what happens to the $v$ component of a wave packet's projected position. That follows from the conservation law associated with the homothety $\lambda^\alpha$. Using Eqs.~\eqref{homothety} and \eqref{consLaw2}, as well as $S^{\alpha \beta} \nabla_{\alpha} \lambda_{\beta} = 0$, its associated conservation law is
\begin{align}
    \mathcal{P}_\lambda = x^i p_i - 2 v E = \mathrm{constant}.
\end{align}
It follows that the projected coordinates $v_\pm$ are still related by the geodesic expression \eqref{dv}. \emph{Full} solutions to the spin Hall equations can thus be determined entirely in terms of their transverse components. Moreover, the memory tensors involved in the solution of the spin Hall equations are exactly the same as the ones that are relevant already for geodesics. In this sense, experiments involving the effects of a gravitational wave on light with angular momentum cannot provide more information about the gravitational wave than experiments without angular momentum.

\section{Memory effects and wave propagation in plane wave spacetimes}
\label{Sect:FieldProp}

So far, we have discussed the scattering of geodesics and of massless spinning particles as approximate models for wave packets with angular momentum. Now we remove the “particle” idealization and consider the scattering of test fields by gravitational sandwich waves. For simplicity, we focus on massless scalar fields and compare, where appropriate, with results in the previous sections. Although we shall not do so here, various spin-raising procedures could be applied to our scalar results to understand the behaviors of electromagnetic and other higher-spin fields \cite{Mason_Ward,Adamo2017,Araneda2022,Kulitskii2023,Tang2023,Audagnotto2024}.

Our first result, in \cref{sec:Green_function}, is a general Kirchhoff-type integral formula for scalar fields on plane wave backgrounds. This describes a field in terms of initial data on a given null hypersurface. It is exact and is nontrivially related to a certain representation formula originally due to Ward \cite{Ward1987}. We use our integral formula in \cref{Sect:scatteredField} to show that in a weak field approximation, any solution to the wave equation in flat spacetime can be used to find a solution in a plane wave spacetime. More precisely, scattered fields can be found simply by differentiating ``unscattered'' fields in flat spacetime. 

Sections~\ref{Sect:PWscatter} and \ref{sec:HG_LG_scattering} both investigate the scattering of particular types of scalar waves on plane wave backgrounds. We first consider scalar waves which are initially planar, both exactly and in a weak field approximation. In the latter context, scattered plane waves remain planar whenever $\Mdv_{ij} = \mathcal{O}(H^2)$, and in those cases, most interesting features can be understood from the scattering of null geodesics. We also consider the weak-field scattering of certain localized solutions which are constructed from counterpropagating Hermite–Gauss and Laguerre–Gauss beams. In these cases, we find that gravitational waves excite only a finite number of scalar side modes.

Lastly, \cref{Sect:SHcompare} compares the exact dynamics of high frequency scalar wave packets which carry angular momentum with the predictions of the spin Hall equations. This is done by numerically evaluating an appropriate Kirchhoff integral.

\subsection{A Kirchhoff-like integral for massless scalar fields}
\label{sec:Green_function}

In the absence of sources, it can be convenient to propagate fields forward in time using a Kirchhoff-like integral. For a massless scalar field $\psi$ that satisfies $\nabla^\alpha \nabla_{\alpha} \psi = 0$, as well as a Green function $G(x,x')$ that satisfies
\begin{equation}
    \nabla^\alpha \nabla_{\alpha} G(x,x') = - 4\pi \delta(x,x'),
    \label{Green}
\end{equation}
we first define the current 
\begin{equation}
    J^{\alpha'} (x',x) \equiv \psi(x') \nabla^{\alpha'} G(x',x) - G(x',x) \nabla^{\alpha'} \psi(x').
    \label{JDef}
\end{equation}
Integrating $\nabla_{\alpha'} J^{\alpha'} = - 4\pi \psi(x) \delta(x,x')$ throughout a 4-volume that includes the point $x$ and that has the boundary $\Sigma$, 
\begin{equation}
    \psi(x) = - \frac{1}{4\pi} \oint_\Sigma J^{\alpha'}(x',x) \dd S_{\alpha'}.
    \label{KirchhoffClosedSigma}
\end{equation}
This expresses $\psi$ in terms of its boundary values on $\Sigma$. By appropriately choosing $\Sigma$ and $G$, the integral here can sometimes be reduced to one which involves only a single spacelike hypersurface in the past of the point $x$ \cite{Poisson2011}. Physically, such a result expresses a field in terms of initial data on a spacelike hypersurface, and is referred to as a Kirchhoff integral. 

Unfortunately, such a representation is not globally possible in plane wave spacetimes. This is because the focusing associated with the conjugate hyperplanes mentioned in \cref{Sect:PWgeod}  implies that plane wave spacetimes do not admit Cauchy surfaces; they are not globally hyperbolic \cite{Penrose1965, HarteDrivas}. Separately, it can also be inconvenient to use large spacelike initial-data surfaces, as these cannot be placed entirely before a gravitational wave arrives. 

We can work around these problems by (i) restricting to regions with no conjugate hyperplanes and (ii) allowing for ``incomplete'' initial data which is compatible with more than one field. In particular, we now identify $\Sigma$ with the null hypersurface\footnote{Unlike in \cref{KirchhoffClosedSigma}, this $\Sigma$ is not the boundary of a 4-volume. \cref{Kirchhoff0} nevertheless follows by assuming that a sufficiently large portion of the $u' = \text{constant}$ hypersurface is part of such a boundary, and also that $\psi(x') = \nabla_{\alpha'} \psi(x') = 0$ on remaining parts of that boundary which lie in the support of $G(\cdot , x)$.} $u' = \text{constant}$, and specify initial data only on that hypersurface. Then, if $G(x',x)$ denotes an advanced Green function which vanishes whenever $x'$ is in the future of $x$, then one solution to the massless scalar wave equation may be written as\footnote{The apparent difference in sign between this and \cref{KirchhoffClosedSigma} is due to a change in the time orientation of $\rmd S_\alpha$ between the two expressions.}
\begin{equation}
    \psi(x) = \frac{1}{4\pi} \int_{\Sigma}  J^{\alpha'} (x',x) \dd S_{\alpha'}.
        \label{Kirchhoff0}
\end{equation}
This expresses the field in terms of initial data on the null hypersurface $\Sigma$. Unlike an ordinary Kirchhoff integral, it does not produce the only field which is compatible with the given initial data: One may add to $\psi$, e.g., any function which depends only on $u$ and which vanishes in a neighborhood of $\Sigma$. Physically, these extra solutions correspond to scalar waves which are copropagating with the background gravitational wave. Regardless, \cref{Kirchhoff0} provides \emph{one} solution to the scalar wave equation and we can examine its properties.

Assuming that $x$ is sufficiently close to the $u' = \text{constant}$ hypersurface that there are no intervening conjugate hyperplanes, the advanced Green function may be  shown to have the form \cite{Gunther1965, HarteDrivas}
\begin{equation}
    \label{eq:Green function scalar}
    G(x',x) = \sqrt{ \Delta (x,x') } \delta (\sigma(x,x')) \Theta(x'< x),
\end{equation}
where $\sigma(x,x')$ is the world function described in \cref{sec:bitensors}, 
\begin{align}
    \Delta(u,u') = \frac{ (u-u')^2 }{ \det B_{i j} (u,u') }
    \label{eq:van Vleck determinant}
\end{align}
denotes the van~Vleck determinant \cite{Poisson2011}, and $\Theta(x' < x)$ is a Heaviside-type distribution which is equal to one if $x'$ is in the past of $x$ and which vanishes otherwise. The van~Vleck determinant reduces to unity in the coincidence limit $u' \to u$. It would instead diverge if $u$ and $u'$ are conjugate in the sense of \cref{detB}, although such cases are not relevant here. In a weak-field vacuum limit, $\Delta = 1 + \mathcal{O}(H^2)$.

Now consider a sandwich wave and suppose that $u'$ lies in $\mathcal{I}_-$ so all initial data is specified in the flat region before the gravitational wave arrives. Then, since $\Delta$ depends only on $u$ and $u'$, Eqs. \eqref{sigma}, \eqref{Kirchhoff0}, and \eqref{eq:Green function scalar} imply that when $u > u'$,
\begin{align}
    \psi(u,v,x^i) = \frac{ \Delta^{1/2}(u,u') }{ 2 \pi (u-u') } \int \rmd^2 x' \partial_{v'} \psi (u', v_\mathrm{ret}, x'^j),
    \label{Kirchhoff}
\end{align}
where $v_\mathrm{ret} = v_\mathrm{ret}( v,u,u',x^i,x'^j)$ denotes the value of $v'$ which corresponds to the retarded event on $\Sigma$ with given $x'^j$. It is found by solving
\begin{equation}
    \sigma \big( u,v,x^i; u', v_\mathrm{ret}, x'^j \big)  = 0,
\end{equation}
which results in
\begin{align}
    v_\mathrm{ret} = v - \frac{ 1 }{ 2 } \big[ (\partial_u B B^{-1} )_{ij} x^i x^j + (B^{-1} A)_{ij}x'^i x'^j
    \nonumber
    \\
    ~   - 2 (B^{-1})_{ij} x'^i x^j \big].
    \label{vRet}
\end{align}
Substituting this into \cref{Kirchhoff} gives an exact expression for a scalar field in terms of initial data on $\Sigma$. A direct calculation can be used to verify that \cref{Kirchhoff} is indeed a solution. We show in \cref{App:Ward} that although it is not obvious, this integral representation is related to one which was previously obtained by Ward \cite{Ward1987}. However, it differs from other specializations of Ward's result which have been used in, e.g., \cite{Adamo2017, Adamo2017_2, Adamo2020}. 

One physical consequence of \cref{Kirchhoff} is that scalar waves which are scattered in plane wave spacetimes depend on the spacetime geometry only via the Jacobi propagators $A_{ij}(u,u')$ and $B_{ij}(u,u')$. In a sandwich wave context where $u' \in \mathcal{I}_-$ and $u \in \mathcal{I}_+$, all nontrivial effects can thus be described in terms of the same memory tensors which arose in our study of scattered geodesics. No new information about a gravitational wave can be learned by measuring memory effects in wave optics rather than geometric optics.

\subsection{Fields in plane wave spacetimes from fields in flat spacetime}
\label{Sect:scatteredField}

Assuming that a gravitational wave is weak, we may now apply the integral representation \cref{Kirchhoff} perturbatively in the gravitational waveform $H_{ij}$. Doing so allows us to generate (scattered) scalar fields in plane wave spacetimes simply by applying an appropriate differential operator to (unscattered) scalar fields in flat spacetime. The form of this operator will illustrate how each memory tensor contributes to scattering.

When the spacetime curvature is weak, its only nontrivial effect on the $\psi$ in \cref{Kirchhoff} is via the $v_\mathrm{ret}$ given by \cref{vRet}. Furthermore, use of \cref{sigmaPert} shows that if $v_\mathrm{ret} = \bar{v}_\mathrm{ret} + v^{(1)}_\mathrm{ret} + \mathcal{O}(H^2)$, the flat-spacetime limit of $v_\mathrm{ret}$ is
\begin{equation}
    \bar{v}_\mathrm{ret} = v - \frac{\delta_{ij} ( x^i - x'^i) (x^j-x'^j) }{ 2(u-u')} ,
    \label{vBar}
\end{equation}
while its first-order correction is
\begin{align}
    v_\mathrm{ret}^{(1)} = - \frac{ 1 }{ 2( u-u')^2} \int_{u'}^u \! \rmd w\, H_{ij}(w) \big[ (w-u') x^i  
    \nonumber
    \\
    ~ + (u-w) x'^i \big] \big[ (w-u') x^j + (u-w) x'^j \big].
    \label{vRet1}
\end{align}
Assuming that $v^{(1)}_\mathrm{ret}$ remains sufficiently small, it follows from \cref{Kirchhoff} that
\begin{align}
    \psi = \bar{\psi} + \frac{1}{2\pi ( u-u') }\int \rmd^2 x' v^{(1)}_\mathrm{ret} \partial_{v'}^2 \psi (u', \bar{v}_\mathrm{ret}, x'^j) 
    \nonumber
    \\
    ~ + \mathcal{O}(H^2),
    \label{KirchhoffPert}
\end{align}
where
\begin{equation}
    \bar{\psi} \equiv \frac{ 1 }{ 2 \pi (u-u') } \int \rmd^2 x' \partial_{v'} \psi (u', \bar{v}_\mathrm{ret}, x'^j)
    \label{psiBar}
\end{equation}
denotes a flat-spacetime field which agrees with $\psi$ on the $u' = \text{constant}$ hypersurface $\Sigma$. The perturbation to the field due to a gravitational wave can therefore be written as an integral involving $v^{(1)}_\mathrm{ret}$. However, since \cref{vRet1} shows that $v^{(1)}_\mathrm{ret}$ is quadratic in $x'^i$, the perturbed field is really a linear combination of the zeroth, the first, and the second ``transverse moments'' of $\partial_{v'}^2 \psi$ on $\Sigma$. 

Interestingly, these moments can largely be computed simply by differentiating (rather than integrating) the background field $\bar{\psi}$. Using \cref{vBar,psiBar}, the zeroth and the first transverse moments of $\partial_{v'}^2 \psi$ are
\begin{align}
    \int\!\dd^2 x'\, \partial_{v'}^2 \psi &= 2 \pi (u-u') \partial_v \bar{\psi} ,
    \label{moment0}
    \\
    \int\!\dd^2 x'\, x'_i \partial_{v'}^2 \psi &=  2 \pi (u-u') \big[ x_i \partial_v \bar{\psi}  + (u-u') \partial_i \bar{\psi} \big].
    \label{moment1}
\end{align}
Although the second moment of the second derivative $\partial_{v'}^2 \psi$ cannot be written quite so simply, the second moment of the \emph{third} derivative $\partial_{v'}^3 \psi$ can be shown to be given by
\begin{align}
     \int\!\dd^2 x'\, x'_i x'_j \partial_{v'}^3 \psi =  2 \pi (u-u') \big[x_i x_j \partial_v^2 \bar{\psi} + (u-u') 
     \nonumber
     \\
      ~ \times \big( \delta_{ij} \partial_v \bar{\psi} + 2 x_{(i} \partial_{j)} \partial_v \bar{\psi} \big) 
      + (u-u')^2 \partial_i \partial_j \bar{\psi} \big].
     \label{moment2}
 \end{align}

Now, since $\ell^\alpha \partial_\alpha = \partial_v$ is Killing and $\psi$ is a solution to the scalar wave equation in a plane wave spacetime, $\Psi \equiv \mathcal{L}_\ell \psi = \partial_v \psi$ is also a solution in that spacetime. Similarly, $\bar{\Psi} \equiv \partial_v \bar{\psi}$ is a solution to the scalar wave equation in flat spacetime whenever $\bar{\psi}$ is a solution in flat spacetime. It then follows from \cref{KirchhoffPert} that
\begin{equation}
    \Psi = \bar{\Psi} + \frac{1}{2\pi ( u-u') }\int\!\dd^2 x'\, v^{(1)}_\mathrm{ret} \partial_{v'}^3 \psi (u', \bar{v}_\mathrm{ret}, x'^j) 
\end{equation}
is an approximate solution on a plane wave background. The integral here involves the zeroth, the first, and the second moments of $\partial_{v'}^3 \psi$ on $\Sigma$, all of which can be extracted from \cref{moment0,moment1,moment2}. If $\bar{\psi}$ is any solution to the flat-spacetime wave equation, then these results show that
\begin{align} \label{eq:wave_pert}
    \Psi = \partial_v \bar{\psi} - \frac{1}{2} \bigg( \int_{u'}^u \rmd w\, H^{ij}(w) [ x_i \partial_v + (u-w) \partial_i] 
    \nonumber
    \\
    ~ \times [ x_j \partial_v + (u-w) \partial_j] \bigg) \bar{\psi} 
\end{align}
is an approximate solution to the wave equation on a plane wave background. \emph{Scattered solutions can thus be generated simply by differentiating unscattered solutions}. However, we emphasize that the above equation describes a process in which the field before the arrival of the gravitational wave is $\partial_v \bar{\psi}$, not $\bar{\psi}$. Despite the roundabout nature in which we have obtained this result, it may be applied to any $\bar{\psi}$ that satisfies the massless scalar wave equation in flat spacetime. 

Our result can be simplified by evaluating $\Psi$ in the flat region $\mathcal{I}_+$ after the gravitational wave has left, in which case all integrals in \cref{eq:wave_pert} reduce to memory tensors. Applying \cref{memTensLin},
\begin{align}
    \Psi = \partial_v \bar{\psi} + \frac{1}{2} \big[ \Mvd_{ij} \partial^i \partial^j - 2 \Mdd_{ij} ( x^i \partial_v + u \partial^i) \partial^j 
    \nonumber
    \\
    ~ - \Mdv_{ij}  ( x^i \partial_v + u \partial^i)  ( x^j \partial_v + u \partial^j)  \big] \bar{\psi}.
    \label{scatteredField}
\end{align}
All derivatives which appear here may be seen to be Lie derivatives with respect to flat-spacetime Killing fields. The velocity-displacement memory thus contributes pairs of translations, the displacement-displacement memory contributes a combination of translations mixed with null rotations, and the displacement-velocity memory contributes pairs of null rotations. The bracketed differential operator in \cref{scatteredField} may be viewed as a kind of ``continuum'' memory effect. Various ``before and after'' comparisons could be performed with scattered fields, and these would be determined by that operator and by the initial field configuration.

A related result---with different assumptions, a different derivation, and involving Fourier components of the field---may be found in Ref.~\cite{Exirifard2021}. See also the construction in Ref.~\cite{Audagnotto2024}, where it has been shown how solutions of certain wave equations in plane wave spacetimes can be related to solutions of the corresponding charged fields in flat spacetime but in the presence of an electromagnetic plane wave.

\subsection{Scattering of plane waves}
\label{Sect:PWscatter}

\Cref{Sect:Geodesics,Sect:SH} above both consider the effects of a gravitational plane wave on highly localized objects: geodesics and massless spinning particles. However, we may also consider the effects of a gravitational sandwich wave on scalar plane waves, which are completely \emph{delocalized}. Suppose that for $u \in \mathcal{I}_-$, we have the ``incoming'' flat spacetime plane wave
\begin{align}
    \psi_- &= \exp \Big\{ \ii \omega_- ( t - z \cos \theta_- - n^-_i x^i \sin\theta_- ) \Big\}
    \nonumber \\
    &= \exp \bigg\{ \frac{\ii \omega_-}{\sqrt{2}} \Big[(1- \cos \theta_-) v + ( 1 + \cos \theta_-) u \nonumber\\
    & \qquad \qquad \qquad- \sqrt{2} n^-_i x^i \sin \theta_- \Big] \bigg\},
    \label{psiIn}
\end{align}
where $\omega_-$ denotes the initial frequency (as seen by the canonical observer at the origin), $\theta_-$ the initial angle of propagation relative to the gravitational wave, and $n_i^-$ a unit 2-vector which describes the initial direction of propagation in the transverse plane. At least in a weak field limit, we could apply \cref{scatteredField} in order to determine how such a scalar wave is scattered by a gravitational wave. However, we instead solve this problem \emph{exactly}\footnote{If \cref{scatteredField} is used to find a scattered scalar field $\Psi$, then it is natural to employ the seed field $\bar{\psi} = - \ii \sqrt{2} \psi_- /[ \omega_- (1 - \cos \theta_-)]$ so $\partial_v \bar{\psi} = \psi_-$. The result matches \cref{psiOut} below through first order in the memory tensors.} by employing Ward's progressing-wave solutions \cite{Ward1987}. This expands upon results which initially appeared in, e.g., Ref.~\cite{Adamo2017}.

As described \cref{App:Ward}, Ward found that the wave equation in an arbitrary plane wave spacetime can be solved by
\begin{equation}
    \psi = (\det \gamma_{ij})^{-1/4} f(S_P),
    \label{psiprogWave}
\end{equation}
where $f$ is an arbitrary function, the eikonal $S_P (\mathscr{U}, \mathscr{V}, \mathscr{X}^i)$ is given by \cref{SPDef} in terms of the Rosen coordinates $(\mathscr{U}, \mathscr{V}, \mathscr{X}^i)$, and $\gamma_{ij}(\mathscr{U})$ denotes the 2-metric in the Rosen line element \eqref{Rosen}. We would like to identify a solution in this class which is equal to $\psi_-$ before the gravitational wave arrives. However, $\psi$ is written in terms of Rosen coordinates while $\psi_-$ is written in terms of Brinkmann coordinates. 

In order to compare, first recall that $\gamma_{ij}$ is given, via \cref{transverseMetric}, by the square of some $E_{ij}$ which satisfies the differential equation \eqref{JacobiMatrix} and the constraint \eqref{Esym}. It is always possible to choose $E_{ij} = A_{ij}(u,u')$ for some fixed $u' \in \mathcal{I}_-$, which guarantees that the Rosen metric is trivial before the gravitational wave arrives: $\gamma_{ij} = \delta_{ij}$ in $\mathcal{I}_-$. Using that same $E_{ij}$ to transform from Rosen to Brinkmann coordinates using \cref{coordXformRosen}, both coordinate systems can be seen to coincide before the gravitational wave arrives. The progressing wave $\psi$ then matches the incoming plane wave $\psi_-$ in $\mathcal{I}_-$ when the $f$ in \cref{psiprogWave} is chosen to be
\begin{equation}
    f( S_P ) = \exp\left[  \frac{ \ii \omega_- }{ \sqrt{2} }   (1-\cos \theta_-) S_P  \right],
\end{equation}
while the constants $\mathscr{U}_0$, $h^{ij}_0$, and $P_i$ which appear in $S_P$ are given by
\begin{subequations}
\begin{align}
\mathscr{U}_0 &= u', \qquad h^{ij}_0 = u' \delta^{ij}, 
\\
P_i &= - \sqrt{2} \cot(\tfrac{1}{2} \theta_-) n_i^-.
\end{align}
\end{subequations}

Although the resulting $\psi$ coincides with $\psi_-$ before the gravitational wave arrives, it can differ at later times. In order to evaluate the scattered field more generally, note that it follows from \cref{SPDef} that $\psi$ depends on an integral of $(\gamma^{-1})^{ij} = (A^{-1})^{ik} (A^{-1})^{j}{}_{k}$. However, this integral can be evaluated by using \cref{ABidentity,SymMatrices} to see that 
\begin{equation}
    \partial_u [(A^{-1})^{ik} B_{k}{}^{j}] = (A^{-1})^{i}{}_{k} (A^{-1})^{jk}.
\end{equation}
The Brinkmann-to-Rosen coordinate transformation \eqref{coordXformRosen} then implies that as long as $\det A_{ij}(u'', u') \neq 0$ for all $u'' \in [u',u]$, which avoids so-called focal points \cite{Bilocal4}, the scalar field \eqref{psiprogWave} reduces to
\begin{widetext}
\begin{align}
    \psi = \frac{ 1 }{(\det A_{kl})^{1/2} } \exp\bigg\{ \frac{ \ii \omega_- }{ \sqrt{2} }  \bigg[ (1-\cos \theta_-) \Big( v
     - \frac{1}{2} (\partial_u A A^{-1} )_{ij} x^i x^j \Big) 
          - \sqrt{2}  (A^{-1})_{ij} n^i_- x^j \sin \theta_-   
    \nonumber
    \\
    ~ + (1+ \cos \theta_-)  \Big( ( A^{-1} B )_{ij} n^i_- n^j_- +u' \Big) \bigg] \bigg\}
    \label{psiOutExact}
\end{align}
in Brinkmann coordinates. This is an exact solution to the massless wave equation which agrees with the incoming plane wave $\psi_-$ when $u \in \mathcal{I}_-$. 

In a scattering context where $u$ lies in $\mathcal{I}_+$ while $u'$ lies in $\mathcal{I}_-$, the Jacobi propagators which appear here can be written in terms of the memory tensors using \cref{Bexpand,Aexpand}. It follows from those expressions that the outgoing wave does not depend on the precise value of $u'$. Working only to first order in the curvature for simplicity, we find that the outgoing wave $\psi_+ \equiv \psi|_{u \in \mathcal{I}_+}$ is given by
\begin{align}
    \psi_+ = \exp \bigg\{ \frac{\ii \omega_- }{ \sqrt{2} } \Big[ (1- \cos \theta_-) v + ( 1 + \cos \theta_- ) \Big( \Mvd_{ij} + ( \delta_{ij} - 2 \Mdd_{ij} ) u \Big) n^i_- n^j_- - \sqrt{2} (\delta_{ij} - \Mdd_{ij} ) n^i_- x^j \sin \theta_- 
    \nonumber
    \\
    ~  - \frac{1}{2} \Mdv_{ij} \Big(  (1-\cos \theta_-) x^i x^j  - 2 \sqrt{2} u n^i_- x^j \sin \theta_-  +  2 u^2 (1+\cos \theta_- ) n^i_- n^j_- \Big) \Big]\bigg\} + \mathcal{O}(H^2).
    \label{psiOut}
\end{align}
\end{widetext}
In general, the phase of the exponential here depends quadratically on both $u$ and $x^i$, implying that the outgoing wave is not necessarily a plane wave. However, quadratic terms arise only when $\Mdv_{ij} \neq 0$. At first order in $H_{ij}$, plane waves are therefore scattered into plane waves only when the displacement-velocity memory vanishes. This can be understood heuristically by noting that initially parallel null geodesics remain parallel only when $\Mdv_{ij}$ vanishes.

When the displacement-velocity memory \emph{does} vanish at leading order, which is what typically occurs in astrophysical applications, the outgoing wave \eqref{psiOut} can be written as
\begin{align}
    \psi_+ = \exp \bigg\{ \frac{\ii \omega_+}{\sqrt{2}} \Big[(1- \cos \theta_+) v + ( 1 + \cos \theta_+) u \nonumber\\
    ~ - \sqrt{2} n^+_i x^i \sin \theta_+ + \phi_+ \Big] \bigg\}  + \mathcal{O}(H^2).
    \label{psiOut2}
\end{align}
Comparing with the ingoing wave \eqref{psiIn}, this may be interpreted as a plane wave with the outgoing frequency
\begin{align}
    \omega_+ = \left[1 - (1 + \cos \theta_- ) \Mdd_{ij} n^i_- n^j_-  \right]  \omega_-,
    \label{domega}
\end{align}
and an outgoing propagation direction which is determined by the angles 
\begin{subequations}
\label{dAngsPW}
\begin{align}
    \theta_+ &= \theta_- + \Mdd_{ij} n^i_- n^j_-  \sin \theta_- ,
    \\
     n^i_+ &= n^i_- + ( n^i_- n^j_- - \delta^{ij} ) \Mdd_{jk} n^k_- .
\end{align}
\end{subequations}
There is also the phase shift 
\begin{equation}
    \phi_+ = ( 1 + \cos \theta_-) \Mvd_{ij} n^i_- n^j_- .
    \label{dphi}
\end{equation}
While the displacement-displacement memory perturbs both the frequency and the propagation direction of a scattered plane wave, the velocity-displacement memory imparts only a phase shift. More specifically, the plane wave deflections given by \cref{dAngsPW} agree with our results \eqref{dtheta} and \eqref{dn} for null geodesics. The frequency shift \cref{domega} can also be seen to agree with the frequency shifts that have been computed in geometric optics \cite{HarteOptics2, Hobbs2010, Detweiler1979}, often in connection with pulsar timing measurements. In these senses, (delocalized) plane waves behave like (localized) null geodesics, at least when $H_{ij}$ is small and $\Mdv_{ij}$ can be neglected.

\begin{figure*}
    \begin{minipage}[t]{0.4\columnwidth}
        \includegraphics[width=\textwidth]{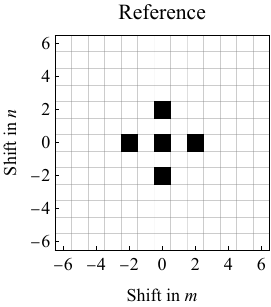}
    \end{minipage}
    \begin{minipage}[t]{0.4\columnwidth}
        \includegraphics[width=\textwidth]{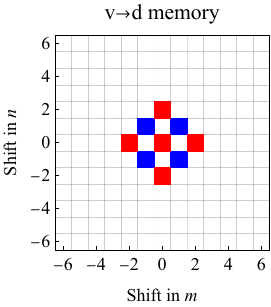}
    \end{minipage}
    \begin{minipage}[t]{0.4\columnwidth}
        \includegraphics[width=\textwidth]{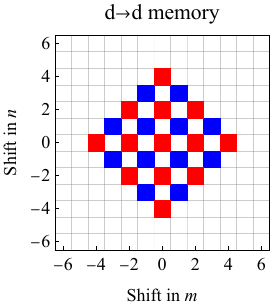}
    \end{minipage}
    \begin{minipage}[t]{0.4\columnwidth}
        \includegraphics[width=\textwidth]{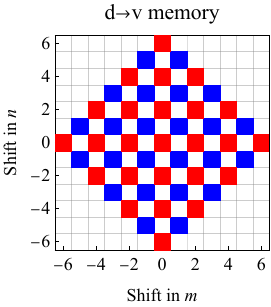}
    \end{minipage}
    \caption{
        Graphical representation of Hermite–Gauss modes that are generically excited by gravitational wave memory. The leftmost pane illustrates the HG modes which compose the ingoing field $\partial_v \psi^\text{HG}_{m,n}$. The three right panes indicate which modes are affected by different memory tensors in the outgoing field. Red squares are used to indicate that the corresponding mode is excited only by diagonal components of the corresponding memory tensor; blue squares are instead excited by off-diagonal components.
    }
    \label{fig:memory modes HG}
\end{figure*}

One other point to note is that our expressions are not meaningful when $\theta_- = 0$. Physically, this corresponds to the case in which a scalar wave is copropagating with a gravitational wave. However, there is no sense in which two copropagating waves can pass through each other; one cannot be scattered by the other. Mathematically, note that both $\psi_-$ and the ``outgoing wave'' given by \cref{psiOutExact} depend only on $u$ in this context. However, \emph{any} function of $u$ is a globally valid solution to the scalar wave equation. This ambiguity is related to the aforementioned fact that we are effectively treating the null $u=u'$ hypersurface as if it were a Cauchy surface even though it is not. Solutions are not unique because we have effectively chosen not to control radiation which is copropagating with the gravitational wave. 

\subsection{Scattering of higher-order Gaussian beams}
\label{sec:HG_LG_scattering}

Having now considered the scattering of both null geodesics, which are fully localized, and scalar plane waves, which are completely delocalized, we now discuss an intermediate between these two extremes: scalar beams that are strongly (but not ``fully'') localized in two spatial directions while being weakly localized in the third. In order to accommodate this generality without undue complication, we work only to first order in the memory tensors, applying the perturbative scattering formula \eqref{scatteredField}. We also assume for simplicity that the scalar beams are initially counterpropagating with respect to the background gravitational wave.

It follows from \cref{psiOut} that, at first order, a counterpropagating scalar wave which is initially planar generically has its wavefronts deformed into hyperbolic paraboloids. However, this effect is purely due to the displacement-velocity memory. When $\Mdv_{ij}$ vanishes at leading order, an incoming plane wave is actually unaffected by a counterpropagating gravitational sandwich wave. This contrasts with what happens for the localized beams we now consider, where nontrivial memory effects arise even in the absence of first order displacement-velocity memory.

These beams are constructed from the Hermite–Gauss (HG) and Laguerre–Gauss (LG) families   \cite{Kiselev2007,siegman1986lasers,ducharme2014}, which are exact solutions of the massless scalar wave equation in Minkowski spacetime. Choosing a direction of propagation opposite to that of the gravitational wave (whose effects are incorporated below), the HG solutions $\psi^\text{HG}_{m,n}$ and the LG solutions $\psi^\text{LG}_{l,p}$ are explicitly given by
\begin{subequations} 
\begin{align}
    \psi_{m,n}^\text{HG} 
    &\equiv \frac{\mathcal{N}_{m,n}}{w(v)}
        H_m\!\left(\frac{\sqrt{2} x}{w(v)}\right)
        H_n\!\left(\frac{\sqrt{2} y}{w(v)}\right)
        e^{\chi(u,v,r)},
    \label{6XII22.2}
\\
    \psi_{l,p}^\text{LG}
    &\equiv \frac{\mathcal{N}_{l,p}}{w(v)}
        \left(\frac{\sqrt{2} r}{w(v)}\right)^{|l|}
        L_p^{|l|}\!\left(\frac{2 r^2}{w(v)^2}\right)
         e^{\chi(u,v,r) - \ii l \phi} ,
    \label{6XII22.3}
\end{align}    
\end{subequations}
where $r^2 \equiv x^2 + y^2$ is the squared transverse radius, $\phi \equiv \arg(x + \ii y)$ is a polar coordinate along the beam axis, and 
\begin{align}
    \chi(u,v,r) \equiv  -\frac{r^2}{[w(v)]^2} +
            \ii \bigg[ \sqrt{2} k
        \left(u-
            \frac{r^2}{2 R(v)}  
        \right)
            \nonumber
            \\
            ~ + (m + n + 1) \arctan
            \left(
            \frac{\sqrt{2} v}{w_0^2 k}
            \right) \bigg].
\end{align}
The transverse beam width is characterized by the length scale
\begin{align}
    w(v) \equiv w_0 \sqrt{1+\frac{2 v^2}{w_0^4 k^2}},
\end{align}
while
\begin{align}
    R(v) \equiv v \left(1 + \frac{w_0^4 k^2}{2 v^2}\right)
\end{align}
may be interpreted as the curvature radius of the wave fronts. It is also convenient to introduce the normalization constants 
\begin{align}
\label{norms}
    \mathcal{N}_{m,n} \equiv \sqrt{\frac{2}{\pi 2^{m+n} m! n!}}, \qquad    \mathcal{N}_{l,p} \equiv \sqrt{\frac{2 p!}{\pi (p + |l|)!}},
\end{align}
which ensure that if an asterisk is used to denote complex conjugation, then
\begin{subequations}
    \begin{align}
    \int (\psi^\text{HG}_{m,n})^* \psi^{\text{HG}}_{m',n'} \rmd x \, \rmd y &= \delta_{mm'} \delta_{nn'},
    \\
     \int (\psi^\text{LG}_{l,p})^* \psi^{\text{HG}}_{l',p'} \rmd x \, \rmd y &= \delta_{ll'} \delta_{pp'}.
\end{align}
\end{subequations}
Both integrals here are to be evaluated on a screen with constant $u$ and $v$. On such a screen, the HG beams are naturally expressed in Cartesian coordinates $(x , y)$ in terms of Hermite polynomials $H_m$, whereas the LG beams are written in terms of polar coordinates $(r, \phi)$ and generalized Laguerre polynomials $L_p^l$. Each beam is uniquely characterized by its waist radius $w_0$, its wave number $k$, and the mode numbers $(m,n)$ or $(l,p)$, where $m, n, p \in \mathbb{N}$ and $l \in \mathbb{Z}$. The HG mode numbers $m$ and $n$ describe how many nodes there are in the $x$ and the $y$ directions, respectively. The LG mode number $l$ provides a measure of the orbital angular momentum carried by $\psi_{l,p}^\text{LG}$ \cite{AM_Light}. The cases $m=n=0$ and $l=p=0$ coincide, and are referred to as Gaussian beams. More general cases may be described as ``higher order'' Gaussian beams.

\begin{figure*}
    \begin{minipage}[t]{0.4\columnwidth}
        \includegraphics[width=\textwidth]{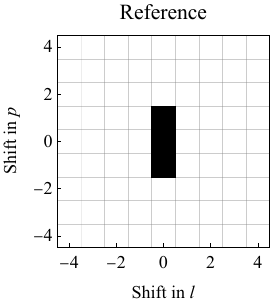}
    \end{minipage}
    \begin{minipage}[t]{0.4\columnwidth}
        \includegraphics[width=\textwidth]{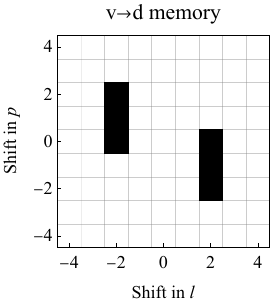}
    \end{minipage}
    \begin{minipage}[t]{0.4\columnwidth}
        \includegraphics[width=\textwidth]{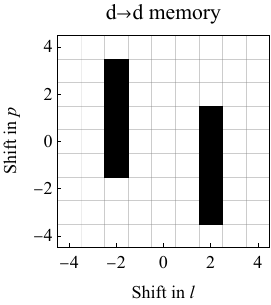}
    \end{minipage}
    \begin{minipage}[t]{0.4\columnwidth}
        \includegraphics[width=\textwidth]{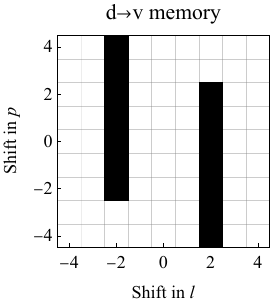}
    \end{minipage}
    \\
    \begin{minipage}[t]{0.4\columnwidth}
        \includegraphics[width=\textwidth]{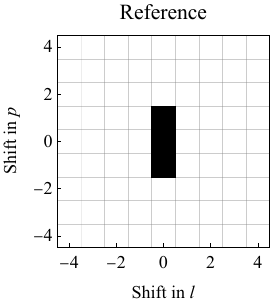}
    \end{minipage}
    \begin{minipage}[t]{0.4\columnwidth}
        \includegraphics[width=\textwidth]{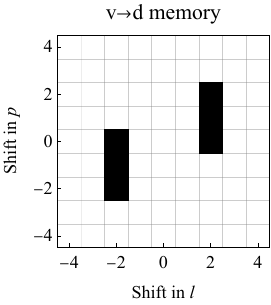}
    \end{minipage}
    \begin{minipage}[t]{0.4\columnwidth}
        \includegraphics[width=\textwidth]{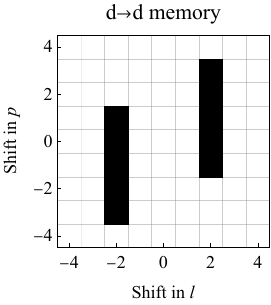}
    \end{minipage}
    \begin{minipage}[t]{0.4\columnwidth}
        \includegraphics[width=\textwidth]{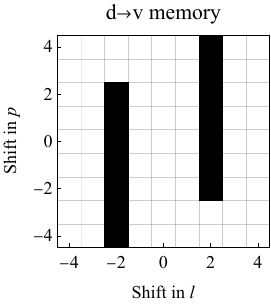}
    \end{minipage}

    \caption{
        Graphical representation of Laguerre–Gauss modes that are excited by gravitational wave memory. The leftmost panes illustrate the LG modes which compose the ingoing field $\partial_v \psi^\text{LG}_{l,p}$.
        The figures in the upper row correspond to $l \geq 2$, whereas figures in the lower row describe $l \leq -2$.
        In contrast to the case of Hermite–Gauss modes (as illustrated in \cref{fig:memory modes HG}), where different components of the memory tensors excite different modes, all modes marked here are excited by both diagonal and off-diagonal terms of the various memory tensors.
    }
    \label{fig:memory modes LG}
\end{figure*}

Both the HG and the LG solutions decay exponentially in the transverse $x$ and $y$ directions. The characteristic length scale for this decay is given by $w(v)$, which has a minimum value of $w_0$ at the $v=0$ ``focus,'' but it grows linearly with $v$ when $|v| \gg w_0^2 k$. The beam therefore focuses inward at an initial angle and then diverges outward at that same angle. A beam with a smaller $w_0 k$ spreads out more rapidly away from its focus. This transverse spread causes both the HG and the LG beams to decay longitudinally as well as transversely. However, the longitudinal decay is slower, being only polynomial in $v$. 

We now consider a gravitational sandwich-wave background and let the seed field $\bar{\psi}$ in the scattering formula \eqref{scatteredField} be equal either to $\psi^\text{HG}_{m,n}$ or to $\psi^\text{LG}_{l,p}$. Before the gravitational wave arrives, the ``incoming'' fields here reduce either to $\partial_v \psi^\text{HG}_{m,n}$ or to $\partial_v \psi^\text{LG}_{l,p}$, which are \emph{not} pure HG or LG modes. \cref{eq:memory HG dv,eq:memory LG dv} in \cref{app:beam_derivatives} nevertheless imply that these derivatives can be written as linear combinations of at most five HG or LG modes. The ``outgoing'' fields that appear after the gravitational wave has passed are more complicated, with the perturbed field involving both the memory tensors and the second partial derivatives of $\bar{\psi}$. However, it is shown in \cref{app:beam_derivatives} that all such derivatives can be explicitly written in terms of a finite number of undifferentiated HG or LG modes. Although it is possible to write down these combinations explicitly, the resulting expressions are unwieldy and we omit them.

It is nevertheless straightforward to explain at least which side modes are excited by different types of gravitational wave memory. In the HG case, the complete list is illustrated in \cref{fig:memory modes HG}. The incoming beam $\partial_v \psi^\text{HG}_{m,n}$ is, generically, a linear combination of HG fields with the five modes $(m,n)$, $(m \pm 2,n)$, and $(m,n \pm 2)$. If the velocity-displacement memory $\Mvd_{ij}$ is purely diagonal, then it affects only these five modes. However, an off-diagonal component of $\Mvd_{ij}$ can instead excite only the four new modes $(m \pm 1, n \pm 1)$. The effect of the displacement-displacement memory is more complicated, generally coupling to modes up to $(m \pm 4, n)$ and $(m, n \pm 4)$. The effect of the displacement-velocity memory is more complicated still, exciting side modes as high as $(m \pm 6, n)$ and $(m, n \pm 6)$. In all cases here, the diagonal and the off-diagonal components of each memory tensor affect disjoint sets of modes. However, this separation would not occur if the HG modes that we employed were not aligned with the $x$ and $y$ axes. 

A similar analysis can also be performed for LG beams, where the side modes excited by the various memory tensors are illustrated in \cref{fig:memory modes LG}.
An incoming beam $\partial_v \psi^\text{LG}_{l,p}$ is generally made up of a linear combination of the three modes $(l,p)$ and $(l,p \pm 1)$. Unlike in the HG case, the diagonal and the off-diagonal components of the memory tensors affect such beams similarly, essentially because $\psi^\text{LG}_{l,p}$ is rotationally symmetric except for the phase $l \phi$. In any case, the excitations generated by the different memory tensors have the form $(l \pm 2, p \pm \Delta p)$. For the velocity-displacement memory, $-2 \leq \Delta p \leq 2$, for the displacement-displacement memory, $-3 \leq \Delta p \leq 3$, and for the displacement-velocity memory, $-4 \leq \Delta p \leq 4$. 

\begin{figure*}
    \centering
    \begin{minipage}[t]{0.988\columnwidth}
    \includegraphics[width=\textwidth]{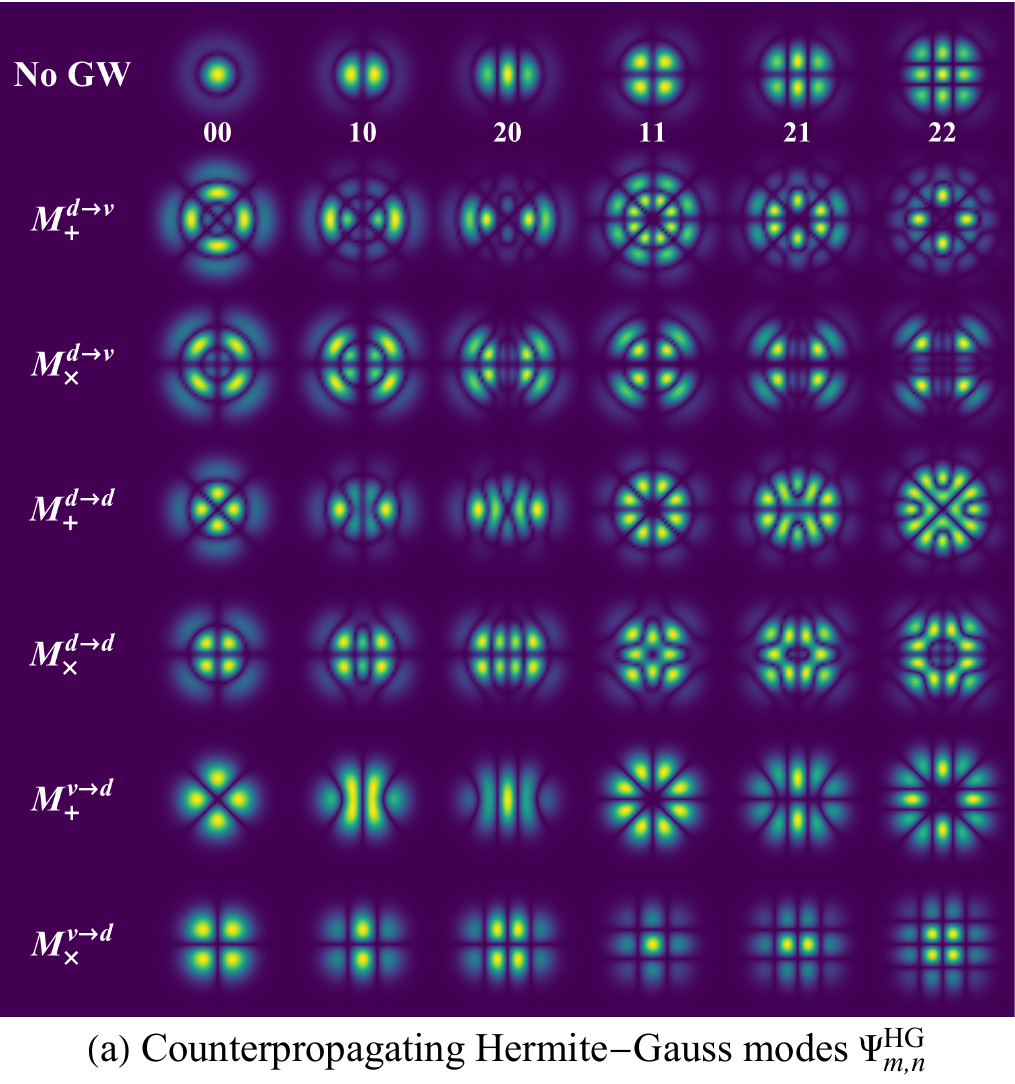}
    % \\(a) Counterpropagating Hermite–Gauss modes $\Psi^\text{HG}_{m,n}$
    \label{fig:modes HG counter}
    \end{minipage}
    \hspace{\columnsep}
    \begin{minipage}[t]{0.988\columnwidth}
    \includegraphics[width=\textwidth]{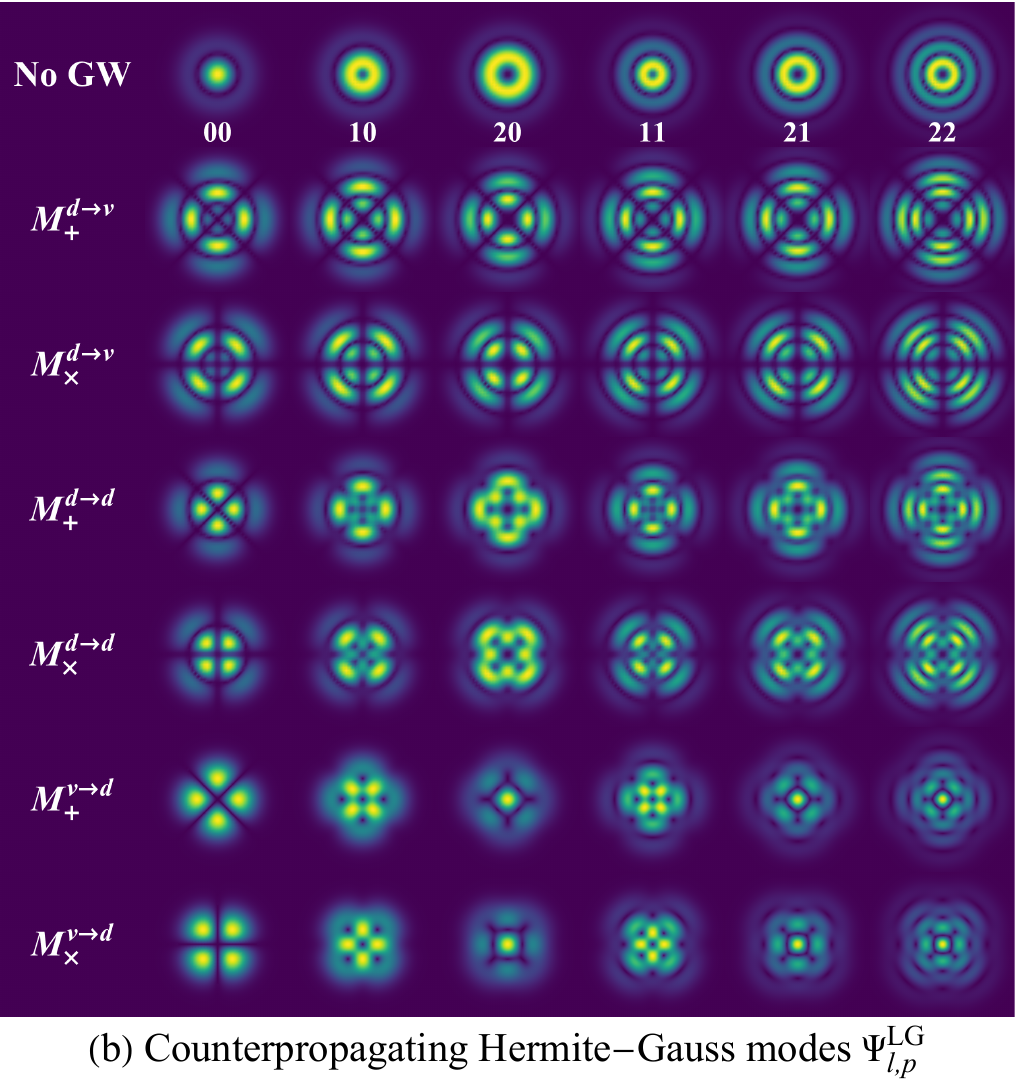}
    % \\(b) Counterpropagating  Laguerre–Gauss modes $\Psi^\text{LG}_{l,p}$
    \label{fig:modes LG counter}
    \end{minipage}
    \caption[Slices of fields seeded by HG modes (a) and LG modes (b) counterpropagating with the gravitational wave.]{Slices of fields seeded by HG modes (a) and LG modes (b) which are counterpropagating with the gravitational wave, all evaluated $t =z = 0$. It is assumed that $w_0 k = 100$. The first row corresponds to the magnitude of the unperturbed reference field $\partial_v \bar{\psi}$, while the numbers refer to the HG or LG mode used for the seed field $\bar{\psi}$. Subsequent rows correspond to the magnitudes of the first-order field perturbations associated with different memory tensors. For example, the row labeled by $\Mdv_+$ plots the difference fields associated with a purely diagonal displacement-velocity memory. The row labeled by $\Mdv_\times$ instead plots the results of a purely off-diagonal displacement-velocity memory. In the LG (but not the HG) cases, magnitudes of perturbations due to the diagonal and the off-diagonal components of a given memory tensor differ only by a $45^{\circ}$ rotation.}
    \label{fig:mode profiles}
\end{figure*}

Another way to illustrate memory effects in this context is to plot example cross sections of different scattered beams. This is done in \cref{fig:mode profiles}, which plots the magnitudes of the difference fields which are associated with the various memory tensors on a screen at $t=z=0$ (or equivalently, at $u=v=0$). Such a screen corresponds to the focus of the unperturbed beam at $t=0$. If one were to instead evaluate on a screen in which $|u|$ is sufficiently large, the $u$-dependent factors on the right-hand side of \cref{scatteredField} would cause the effects of $\Mdv_{ij}$ to dominate over those of $\Mdd_{ij}$ and the effects of $\Mdd_{ij}$ to dominate over those of $\Mvd_{ij}$. Plots analogous to \cref{fig:mode profiles} would then show no significant differences between the effects of the different memory tensors (at least when comparing only diagonal memory tensors or only off-diagonal ones). The strong difference between the effects of diagonal and off-diagonal terms in the various memory tensors suggests that the induced transverse deformations of the beams are highly dependent on the polarization of an intervening gravitational wave. A different discussion of deformed HG beams propagating in curved spacetimes has been reported in Ref.~\cite{Exirifard2021}.

\subsection{Scattering of wave packets and comparison with the spin Hall equations}
\label{Sect:SHcompare}

The analysis of scalar wave scattering in \cref{Sect:scatteredField,sec:HG_LG_scattering} is restricted to weak gravitational waves and perturbative methods. In the nonperturbative regime, wave propagation can be described by the Kirchhoff-like integral \eqref{Kirchhoff}. However, that integral cannot generally be evaluated in closed form. Nonperturbative scattering for generic scalar fields could instead be understood by summing the initially planar waves described in \cref{Sect:PWscatter}. However, using analytic methods to extract interesting features from such sums is again nontrivial. As a consequence, we now turn to a numerical analysis of scalar waves scattered by nonperturbative gravitational sandwich waves. This section presents results obtained from the numerical evaluation of a Kirchhoff integral which yields exact solutions to the complex scalar wave equation in specific plane wave spacetimes (up to numerical error). We consider the scattering of small, high-frequency wave packets and show that their dynamics are in good agreement with the ray description provided by the spin Hall equations \eqref{eq:Spin Hall}. This provides evidence that the spin Hall effect of wave packets carrying angular momentum is indeed described by those equations.

We choose here to use a “standard” Kirchhoff integral where the initial hypersurface $\Sigma$ that appears in \cref{Kirchhoff0} is spacelike rather than null. This is possible because (i) we work in a sufficiently small region that conjugate hyperplanes are not encountered, and (ii) the initial data considered here (very nearly) vanish outside of a sufficiently compact region that all relevant portions of $\Sigma$ can be placed in $\mathcal{I}_-$, before any significant portion of the gravitational wave arrives. In particular, we identify $\Sigma$ with the hypersurface $t' = \text{constant}$, where the time coordinate is again defined by \cref{tzDef}. It is then convenient to introduce spherical coordinates $(r', \theta',\varphi')$ for the source point $x'$ which are centered around the three spatial coordinates $x^I$ of the field point $x$. Combining \cref{JDef,Kirchhoff0,eq:Green function scalar} then yields the Kirchhoff integral
\begin{align}
\label{eq:numerics:Kirchhoff}
\begin{split}
    \psi(x) ={}&
    \frac{1}{4\pi}
        \int_{S_2} \frac{\dd \Omega'}{|\partial_{r'} \sigma| } \Big[ r'^2  \big( \sqrt{\Delta} \partial_{t'} \psi' - \psi' \partial_{t'} \sqrt{\Delta} \big) 
        \\&\hspace{3em}
        - \partial_{r'} \big( r'^2 \sqrt{\Delta} \psi' \partial_{t'} \sigma / \partial_{r'} \sigma \big) \Big]_{r' = r_\text{ret}} 
\end{split}
\end{align}
when $t > t'$, where $\dd \Omega' = \sin \theta' \dd \theta' \dd \varphi'$ is the standard measure on the round unit two-sphere $S_2$ and $r_\text{ret} = r_\text{ret} (x; t', \theta', \varphi')$ describes the radius of the past light cone with vertex $x$ on the initial $t' = \text{constant}$ hypersurface and in the direction determined by the angles $(\theta', \varphi')$. This last function can be obtained by transforming \cref{sigma} into the given coordinates and then solving $\sigma =0$. The Jacobi propagators $A_{ij}$ and $B_{ij}$ are computed numerically, and then $\sigma$ follows from \cref{sigma} and $\Delta$ from \cref{eq:van Vleck determinant}.

\begin{figure*}[ht]
    \centering
    \begin{minipage}[t]{0.988\columnwidth}
    \includegraphics[width=\textwidth]{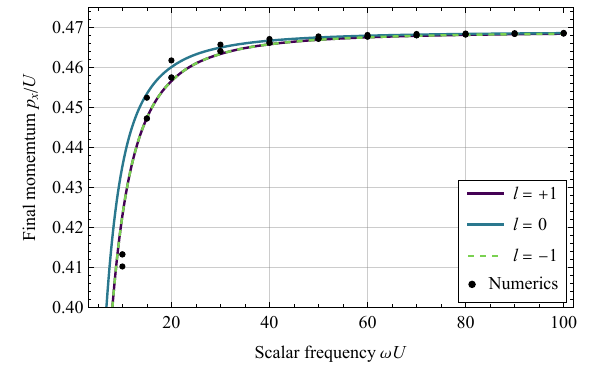}
    \end{minipage}
    \hspace{\columnsep}
    \begin{minipage}[t]{0.988\columnwidth}
    \includegraphics[width=\textwidth]{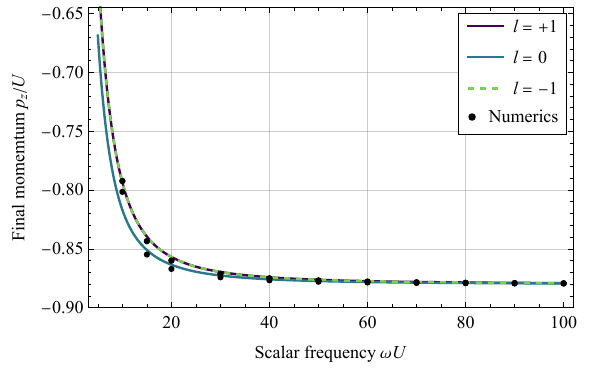}
    \end{minipage}
    \caption{
         Final momentum components $\t p{_x}$ and $\t p{_z}$ (both initially nonzero) due to scattering off a circularly polarized gravitational sandwich wave, as a function of the initial scalar wave frequency $\omega$. The dots indicate data obtained by numerical quadrature of \cref{eq:numerics:Kirchhoff}, while the solid curves were obtained by solving the spin Hall equations \eqref{eq:Spin Hall} using initial data derived from the initial scalar wave packet. The spin Hall equations are seen to agree with the full numerical evolution, especially when $\omega$ is large. It can also be seen that there is little dependence on $l$ at high frequencies. The gravitational wave here has amplitude $A = 1/2U^2$ and frequency $\nu = 5/U$. The scales in these plots use units in which $U=1$. The remaining momentum component $p_y$ is plotted in \cref{fig:numerics:frequency momentum y}.
    }
    \label{fig:numerics:frequency momentum xz}
\end{figure*}

Scalar wave propagation can be related to the predictions of the gravitational spin Hall equations by (i) prescribing initial data for a scalar field at $t' = \text{constant}$, before the gravitational wave arrives, (ii) computing the 4-momentum, the energy centroid, and the angular momentum of this data, (iii) evolving the spin Hall equations with these parameters as initial data, (iv) numerically evaluating the Kirchhoff integral \cref{eq:numerics:Kirchhoff} to find the field after the gravitational wave has left, and (v) computing the 4-momentum and the centroid of the scattered field. Finally, the 4-momenta and the centroids of steps (iii) and (v) can be directly compared.

In order to describe this in more detail, we first explain how to compute 4-momenta and centroids. These are derived from the generalized momentum given by \cref{consLaw,eq:momentum generalized}. Integrating over a constant-$t$ hypersurface while assuming that the stress-energy tensor of the scalar field $T^{\alpha\beta}$ is nonzero only where the spacetime is (very nearly) flat, the 4-momentum in $\mathcal{I}_-$ and in $\mathcal{I}_+$ can be computed using
\begin{align}
	p^\alpha &= \int T^{\alpha t} \dd^3 x.
	\label{pFlat}
\end{align}
We take $\psi$ to be a genuinely complex field, as opposed to a complex field interpreted so that only its real component is physical. Denoting by $\psi^*$ its complex conjugate, the stress-energy tensor is
\begin{equation}
    T_{\alpha\beta} = 
	\big( \delta^\gamma_{(\alpha} \delta^\lambda_{\beta)} - \tfrac{1}{2} g_{\alpha\beta} g^{\gamma \lambda} \big) \nabla_\gamma \psi^* \nabla_{\lambda} \psi .
 \label{Tscalar}
\end{equation}
Stress-energy conservation implies that although the 4-momentum can be changed by a passing gravitational wave, it is conserved within each of the two flat regions. Its components are the generalized momentum $\mathcal{P}_\xi$ evaluated for vector fields $\xi^\alpha$ which locally generate translations.

The centroid of a scalar field can be extracted by first noting from \cref{consLaw,eq:momentum generalized} that the angular momentum with respect to an origin with coordinates $\bar{x}^\alpha$ is given by
\begin{equation}
	S^{\alpha\beta} = 2\int (x - \bar{x})^{[\alpha} T^{\beta]t} \dd^3 x
\end{equation}
in any flat region. Applying the spin supplementary condition \eqref{SSC} with the frame vector given by \cref{tDef}, the spatial coordinates of the centroid in any flat region are therefore
\begin{equation}
	\bar{x}^I = \frac{1}{p^t} \int T^{tt} x^I \dd^3 x .
	\label{xFlat}
\end{equation}
Unlike in \cref{Sect:SH}, we now use $\bar{x}^I$ to denote the centroid position instead of $x^I$, reserving the latter for the spatial coordinates of a general field point. Regardless, comparison with the spin Hall equations requires that we also evaluate $\epsilon s$. This is done by computing $S^{\alpha\beta}$ and then applying \cref{Spin2}, which determines the absolute value of $\epsilon s$. Its sign is chosen to coincide with the sign of the angular momentum parameter $l$ introduced in \cref{eq:numerics:WKB amplitude} below. The $u$ derivatives needed to compute the stress-energy tensor were obtained using finite differences. All other derivatives were determined using the fact that derivatives along the four Killing vector fields described by \cref{Killing} also satisfy the scalar wave equation and can thus also be propagated forward in time using the Kirchhoff integral \eqref{eq:numerics:Kirchhoff}.

We work with gravitational waves that can be interpreted as circularly polarized with Gaussian profiles. More precisely, we consider the waveforms given by \cref{gw_profile}, with center point $U_0 = 10 U$, phase offsets $\phi_+ = \pi/2$ and $\phi_\times = 0$, and equal curvature length scales $l_+ = l_\times$. We also define the curvature amplitude 
\begin{equation}
	A \equiv \frac{ 2 }{ \sqrt{\pi} l_+^2 },
	\label{ADef}
\end{equation}
where this and the gravitational wave frequency $\nu$ are both kept as adjustable parameters. The given waveform never decays entirely to zero, so these are not technically sandwich waves. However, the “flat regions” $\mathcal{I}_\pm$ can be roughly identified as those in which $|u-U_0| \gg U$. 

Next, we impose initial data for $\psi(x')$ on the $t' = 0$ hypersurface, which amounts to specifying the field and its first time derivative there. Introducing an initial scalar amplitude profile $a(x^J)$, a scalar frequency $\omega$, and a unit 3-vector $n^I$ (unlike the unit 2-vector $n^i$ used above) which describes the initial direction of propagation, we let
\begin{subequations}
\label{eq:numerics:WKB data}
\begin{align}
    \left. \psi \right|_{t=0}
        &= a(x^J) e^{\ii \omega \t n{_I} \t x{^I}},
    \\
    \left. \partial_t \psi \right|_{t=0}
        &= -\big[ \ii \omega a(x^J) + \t n{^K} \t{\nabla\!}{_K} a(x^J) \big] e^{\ii \omega \t n{_I} \t x{^I}}.
\end{align}
\end{subequations}
This choice is motivated by the high-frequency approximation, in which a family of fields on spacetime has the form $[\mathfrak{a}(x) + \mathcal{O}(\omega^{-1})] e^{\ii \omega \mathfrak{u}(x)}$ as $\omega \to \infty$. Working in flat spacetime, planar wavefronts result from $\mathfrak{u} = n_I x^I - t$, and the scalar wave equation implies that $\partial_\alpha \mathfrak{u} \, \partial^\alpha \mathfrak{a} = 0$. Evaluating these results on the initial hypersurface while dropping the $\mathcal{O}(\omega^{-1})$ correction recovers the initial data above. Despite the approximation used to produce these data, their evolution into the future takes into account all finite-frequency effects. Up to numerical error, its use in the Kirchhoff integral \eqref{eq:numerics:Kirchhoff} produces an exact solution to the scalar wave equation in a sandwich wave background. 

\begin{figure}
    \centering
    \includegraphics[width=\columnwidth]{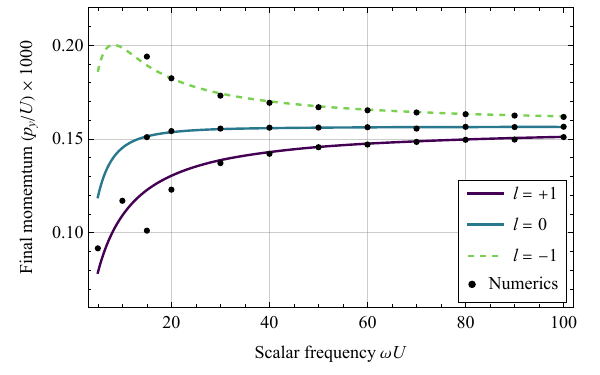}
    \caption{
        Final momentum component $\t p{_y}$ (which initially vanishes) due to scattering off a circularly polarized gravitational sandwich wave, as a function of the initial scalar wave frequency $\omega$. All conventions here are the same as in \cref{fig:numerics:frequency momentum xz}, which plots $p_x$ and $p_z$. Here, $p_y$ is seen to converge to the $l=0$ case of a null geodesic at large $\omega$, and the first corrections are well described by the spin Hall equations. Unlike for the $p_x$ and $p_z$ components, those corrections are nontrivial: There is a clear dependence on the angular momentum of the wave packet.
    }
    \label{fig:numerics:frequency momentum y}
\end{figure}

Comparison with the spin Hall equations requires a small wave packet, so we choose the initial amplitude profile of the scalar field to decay exponentially---though anisotropically---in all spatial directions. Introducing a constant, positive-definite matrix $W_{IJ}$ that controls this decay, 
\begin{align}
\label{eq:numerics:WKB amplitude}
    a(x^I) = \mathscr N |x|_\perp^2 \exp \big(
            - \t{W}{_J_K} \t x{^J} \t x{^K}
            + \ii l \phi \big),
\end{align}
where $\mathscr N$ is a normalization constant, $|x|_\perp^2 \equiv (\delta_{IJ} - n_I n_J) x^I x^J$ is the squared distance away from the propagation axis, $l$ is an integer controlling the orbital angular momentum (OAM) of the field\footnote{Using the given initial data to compute the angular momentum, it can be shown that $\epsilon s = (p^t/\omega) l + \mathcal{O}(\omega^{-3})$. That the leading term here is of order $1/\omega$ implies that the angular momentum of the wave packet first appears at one order beyond geometric optics.}, and $\phi$ is an angle around the propagation direction. Using rotated coordinates $\hat{x}^I$ that are aligned with the propagation direction, so $\hat{n}^3=1$, this last angle is more precisely defined to be $\operatorname{arg}(\hat{x}^1 + \ii \hat{x}^2)$. Although $\phi$ is uniquely defined only up to an overall additive constant, that constant is irrelevant to our discussion. In our implementation, we choose $\hat{W}_{IJ} = (2\pi/U^2) \operatorname{diag}[ 1,1,1/9]_{IJ}$ in these same rotated coordinates, so that the wave packet decays more rapidly away from its propagation direction than along it. All calculations are carried out with the propagation direction $(n_I) = (\sin \beta, 0, - \cos \beta)$, where $\beta = 1/2$, and, for convenience, $\mathscr N$ is chosen such that the initial energy $p^t |_{t=0}$ is equal to the gravitation wave width $U$.

\begin{figure}
    \centering
    \includegraphics[width=\columnwidth]{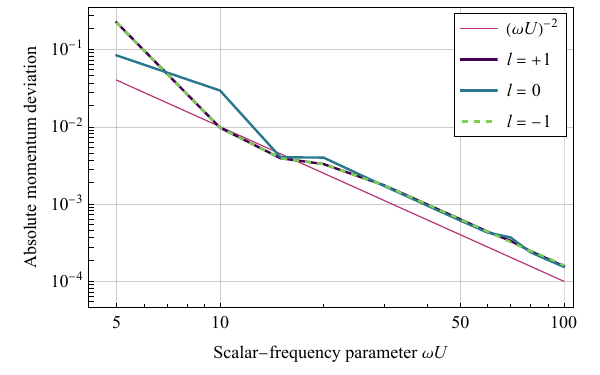}
    \caption{
        Frequency dependence of the differences between the final 3-momenta obtained using the Kirchhoff integral and the spin Hall equations. The vertical scale plots the norm of the difference between these two momenta. As the scalar frequency $\omega$ increases, this difference scales like $1/\omega^2$, which is plotted in red. All conventions here are the same as in \cref{fig:numerics:frequency momentum xz}.
    }
    \label{fig:numerics:frequency convergence}
\end{figure}

We implemented the above scheme using \texttt{Wolfram Mathematica} \cite{Mathematica}. As a first test, the scattering of scalar wave packets in a fixed gravitational wave background was considered for various values of the initial scalar wave frequency $\omega$. Since wave packets with different values of $l$ did not acquire significant spatial separation during the integration time $\Delta t = 20U$, after which they had propagated to the region $\mathcal I_+$, our analysis focuses mainly on the final 3-momentum $\t p{_I}$. \Cref{fig:numerics:frequency momentum xz,fig:numerics:frequency momentum y} show the final momentum of the wave packet as a function of its initial angular frequency $\omega$. In these plots, the dots represent the numerical results obtained from the Kirchhoff formula \eqref{Kirchhoff0}, applied to the initial data mentioned above, while the solid lines represent the numerical results obtained from numerical solutions to the spin Hall equations \eqref{eq:Spin Hall}. As demonstrated in \cref{fig:numerics:frequency momentum xz}, the momentum components that are initially nonzero show only little dependence on the OAM parameter $l$: Defining the relative momentum splitting as
\begin{align}
    \label{eq:momentum splitting}
    \delta p_I = \frac{\t p{_I}(l = +1) - \t p{_I}(l = -1)}{\operatorname{max}\{|\t p{_I}(l = +1)|, |\t p{_I}(l = -1)|\}},
\end{align}
we find that $\delta\t p{_x}$ and $\delta\t p{_y}$ do not exceed $1.6 \times 10^{-5}$ in the considered parameter regime. However, \cref{fig:numerics:frequency momentum y} shows a clear $l$ dependence of the final value of $\t p{_y}$ (which is initially zero): The relative splitting ranges from $\delta \t p{_y} \approx -2$ at $\omega = 5/U$ to $\delta \t p{_y} \approx -0.06$ at $\omega = 100/U$.

\begin{figure}
    \centering
    \includegraphics[width=\columnwidth]{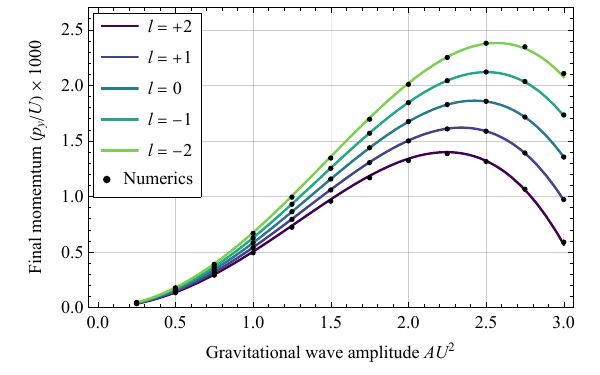}
    \caption{Variation of the final momentum $\t p{_y}$ (which initially vanishes) as a function of the gravitational wave amplitude $A$. Conventions are the same as in \cref{fig:numerics:frequency momentum xz}, except that $A$ is free in this plot. The initial frequency of the scalar wave is also fixed at $\omega = 50/U$. The spin Hall predictions are seen to remain valid even for very large gravitational wave amplitudes. 
    }
    \label{fig:numerics:amplitude momentum xz}
\end{figure}

\Cref{fig:numerics:frequency convergence} shows that the final momenta of the considered scalar wave packets converge to the spin Hall predictions at a rate that is proportional to $\omega^{-2}$. This result is to be expected from the construction of the spin Hall equations using high-frequency asymptotics, though the present comparison provides the first quantitative measures of the quality of the gravitational spin Hall model. To further analyze this quality in the present setup, the dependence of $\t p{_y}$ on the gravitational wave amplitude $A$ was investigated for a fixed value of the scalar wave frequency, $\omega = 50/U$, but different values of the OAM parameter $l$, namely $0$, $\pm 1$, and $\pm 2$. \Cref{fig:numerics:amplitude momentum xz} shows a clear $l$-dependent change in $\t p{_y}$ for gravitational wave amplitudes up to $A = 3/U^2$ that is consistent with the spin Hall predictions. All of these results were obtained without weak-field approximations. 

\section{Conclusions}
\label{sec:conclusions}

This paper discusses a wide range of memory effects in gravitational plane wave spacetimes, including effects on geodesics, spinning massless particles, and even full test fields. These effects are summarized in Table~\ref{tab:summary}, and all can be written entirely in terms of the Jacobi propagators $\t A{_i_j}(u, u')$ and $\t B{_i_j}(u, u')$ which characterize solutions to the geodesic deviation equation. Although there is an infinite-dimensional space of possible waveforms, and there are different Jacobi propagators for each waveform, there is only a finite-dimensional space of possibilities for the ``initial'' and the ``final'' Jacobi propagators associated with a gravitational wave burst. All memory effects we consider can thus be encoded in the four memory tensors introduced in \cref{Sect:PWs}; cf. \cref{Bexpand,Aexpand}. Since the memory tensors encode only a relatively small amount of information about the gravitational waveform, memory effects of the types considered here can probe only certain characteristics of a gravitational wave: in a weak-field limit, only the zeroth, the first, and the second moments of the curvature. 

Applied to geodesics, our memory tensors $\Mdd_{ij}$, $\Mdv_{ij}$, $\Mvd_{ij}$, and $\Mvv_{ij}$ determine the transverse displacement-displacement, displacement-velocity, velocity-displacement, and velocity-velocity memory effects, respectively; cf. \cref{Smatrix,SDef}. Although longitudinal memory effects occur as well---as discussed in \cref{Sec:NullMem} particularly for null geodesics---these effects are determined by the same four memory tensors.

The memory tensors also determine a number of memory effects which are not necessarily associated with geodesics. For example, we find in \cref{Sect:SH} that they completely characterize the scattering of massless particles (or compact wave packets) with longitudinal angular momentum, a process which is described by the spin Hall equations. Although spinning wave packets can be scattered differently than null geodesics, nothing more can be learned about a gravitational wave by using them to perform memory experiments.  

This is also true when considering the scattering of arbitrary test fields: The memory tensors determine Synge's world function $\sigma$, as well as the van~Vleck determinant $\Delta$, and these are all that enter the relevant Green functions for (massless or massive \cite{HarteTails}) scalar fields in plane wave spacetimes. Memory effects in the propagation of massless scalar fields are discussed from several perspectives in \cref{Sect:FieldProp}. One result is that we use the aforementioned Green function to derive a Kirchhoff-like integral formula for the final field in terms of an initial one. We also compute the exact scattering of an initially planar scalar wave by a gravitational wave. If the outgoing wave is also planar (which is not necessarily the case), the initial and the final waves are shown to be scattered similarly to a null geodesic.

We also show that when a gravitational wave is weak, its effect on an arbitrary scalar field can be obtained simply by applying a differential operator to an appropriate ``seed'' solution which satisfies the wave equation in flat spacetime. The perturbative portion of this operator involves pairs of Lie derivatives with respect to four of the relevant Killing fields, where these derivatives are weighted by the various memory tensors; cf. \cref{scatteredField}. We apply this result for seed solutions involving counterpropagating Hermite–Gauss and Laguerre–Gauss beams in flat spacetime, where different memory tensors are found to excite (a finite number of) different side modes. 

Lastly, we numerically analyze the behavior of massless scalar wave packets in fully nonlinear gravitational waves. Our main result is that the momenta of these wave packets are well approximated by the spin Hall equations, which are described and solved in \cref{Sect:SH}. This provides the first numerical evidence for the convergence of scalar-wave-packet dynamics to the behavior predicted by the spin Hall equations.

This paper has focused on memory effects in plane wave spacetimes. Physically, plane wave geometries might be expected to approximate the geometry in a spatially small region far from a gravitating source. Given the Penrose limit, plane waves might also be expected to be relevant for memory effects which involve systems confined to remain near a null geodesic in an arbitrary spacetime. However, precise translations to these contexts are left for later work. 

\begin{table*}
  \centering
  \setlength{\tabcolsep}{9pt}

  \begin{tabular}{@{}p{0.6 \textwidth} p{0.2\textwidth}@{}}
    \toprule
    Physical description & Results \\
    \midrule
    Transverse properties of arbitrary geodesics 
    & Eqs.~\eqref{Smatrix}, \eqref{SDef}
    \\
    Transverse \& longitudinal properties of slow timelike geodesics (IC)    & Eqs.~\eqref{dZSlow}--\eqref{dVSlow}
    \\
    Deflections of null geodesics (IC)  & Eqs.~\eqref{dtheta}, \eqref{dn}
    \\
    Displacements of null geodesics (IC)   & Eqs.~\eqref{dZNull},  \eqref{dXNull}    
    \\
    World function		&	Eq.~\eqref{sigmaPert2}
    \\
    Perturbed light cones (IC)	 & Eqs.~\eqref{sigma}, \eqref{lightCone}	
    \\
    Transverse properties of spin Hall solutions & Eqs.~\eqref{Smatrix2}, \eqref{SDef2}
    \\
    \midrule
    Generic scattered scalar fields 	& Eqs.~\eqref{Kirchhoff}, \eqref{scatteredField}
    \\
    Scattered scalar plane waves: Frequencies, deflections, \& phases 		& Eqs.~\eqref{psiOut}--\eqref{dphi}
    \\ 
    Scattered Hermite-Gauss modes	&	 Fig.~\ref{fig:memory modes HG}
    \\
    Scattered Laguerre-Gauss modes	& Fig.~\ref{fig:memory modes LG}
    \\
    \bottomrule
  \end{tabular}
  
  \caption[Summary of the effects of gravitational wave memory on different physical systems.]{Summary of the effects of gravitational wave memory on different physical systems.  Results are stated in Brinkmann coordinates unless indicated with ``IC'' (inertial coordinates). All effects here are determined by the same four memory tensors $\Mvd_{ij}$, $\Mdd_{ij}$, $\Mvv_{ij}$, and $\Mdv_{ij}$.}
  
  \label{tab:summary}
\end{table*}

\section*{Acknowledgments}

We are grateful to Tim Adamo, Andrea Cristofoli, Peter Horvathy, Anton Ilderton, and Sonja Klisch for helpful discussions. E.S. was funded in part by the Austrian Science Fund (FWF), Project No. P34274.
Research by T.B.M. is funded by the European Union (ERC, GRAVITES, Project No. 101071779). Views and opinions expressed are however those of the authors only and do not necessarily reflect those of the European Union or the European Research Council Executive Agency. Neither the European Union nor the granting authority can be held responsible for them.

\appendix
\renewcommand{\theequation}{\thesection.\arabic{equation}}

\section{Notation}
\label{app:notation}

We work on Lorentzian spacetimes $(M, g_{\alpha \beta})$, where the metric tensor $g_{\alpha \beta}$ has signature $(-\,+\,+\,+)$. Greek letters are used for spacetime indices and range from $0$ to $3$. We mostly describe plane wave spacetimes in Brinkmann coordinates $(u, v, x^i)$, although Rosen coordinates $(\mathscr{U}, \mathscr{V}, \mathscr{X}^i)$ are also used. In this context, overset dots are used to denote derivatives with respect to $u$ or $\mathscr{U}$. Also, Latin letters from the middle of the alphabet, $(i, j, k, \ldots)$, which run from $1$ to $2$, are used to label the transverse and spacelike coordinates in these spacetimes. The components of $3$-vectors are labeled with capitalized Latin letters from the middle of the alphabet, $(I, J, K, \ldots)$, and range from $1$ to $3$. We work in geometrized units in which $G=c=1$, the Einstein summation convention is assumed, and we use the notation $a_\alpha b^\alpha = a \cdot b$. The Riemann tensor is defined so that $2 \nabla_{[\alpha} \nabla_{\beta]} \omega_\gamma = R_{\alpha\beta\gamma}{}^{\lambda} \omega_\lambda$ for any smooth $\omega_\gamma$. A summary of the important symbols used throughout the paper is given in \cref{Table:Notation}.

\begin{table*}
  \centering
  \setlength{\tabcolsep}{9pt}

  \begin{tabular}{@{}p{0.22 \textwidth} p{0.45\textwidth} p{0.2\textwidth}@{}}
    \toprule
    Symbol & Description & Reference \\
    \midrule
    $u$, $v$, $x^1 = x$, $x^2 =y$  & Brinkmann coordinates & \cref{Brinkmann}
    \\
    $t$, $z$    &   Time and longitudinal coordinates & \cref{tzDef}  
    \\
    $\mathscr{U}$, $\mathscr{V}$, $\mathscr{X}^i$  &    Rosen coordinates  &  \cref{Rosen}
    \\
    $H_{ij}$	&	Plane wave waveform (Brinkmann) 	&	Eqs.~\eqref{Brinkmann}, \eqref{pwRiemann}
    \\
    $\gamma_{ij}$    & Plane wave waveform (Rosen)   &   Eqs.~\eqref{Rosen}, \eqref{transverseMetric} 
    \\
    $h_{ij}$    & Transverse-traceless metric perturbation   &  Eq.~\eqref{hToH}
    \\
    $\ell^\alpha$   &  Plane wave propagation direction & \cref{lDef}
    \\
    $\lambda^\alpha$    & Homothety & \cref{homothety}
    \\
    $\Xi^i$    & Solution to transverse vector Jacobi equation   &  Eqs.~\eqref{Killing}, \eqref{JacobiBasic}
    \\
    $E_{ij}$    & Solution to transverse matrix Jacobi equation &  Eqs.~\eqref{JacobiMatrix}, \eqref{transverseMetric}
    \\
    $A_{ij}$, $B_{ij}$  & Transverse Jacobi propagators & \cref{JacobiInit}
    \\
    $\Mdv_{ij}$, $\Mdd_{ij}$, $\Mvv_{ij}$, $\Mvd_{ij}$  & Memory tensors  & Eqs.~\eqref{Bexpand}, \eqref{memTensLin}, \eqref{SDef}
    \\
    $\mathcal{I}_\pm$, $\mathcal{I}_C$  & Flat and curved regions of a sandwich wave & \cref{fig:SandWave}
    \\
    $\gamma$, $u_0$, $v_0$, $r^{i}{}_{j}$   & Transformations between Brinkmann coordinates & \cref{coordXform}
    \\
    \midrule
    $\kappa$   & Geodesic constant  & Eqs.~\eqref{kappaDef}, \eqref{kappaV}
    \\
    $x^i_\pm$  & Positions projected on $u=0$ hyperplane &  Eqs.~\eqref{xPM}, \eqref{Smatrix}, \cref{fig:coord}
    \\
    $X^I_\pm$ &  Positions projected on $t=0$ hyperplane & \cref{geoInertial}, \cref{fig:coord}
    \\
    $\dot{x}^i_\pm$ & Initial and final transverse velocities & Eq.~\eqref{Smatrix}
    \\
    $V_\pm^I$  &  Initial and final 3-velocities & \cref{geoInertial}
    \\
    $n^i_\pm$ &  Initial and final unit propagation directions & Eqs.~\eqref{nV}, \eqref{psiIn}, \eqref{psiOut}
    \\
    $\theta_\pm$ & Initial and final angles away from the gravitational wave & Eqs.~\eqref{nV}, \eqref{psiIn}
    \\
    $p_\alpha$ & Linear momentum & Eqs.~\eqref{eq:Spin Hall}, \eqref{consLaw}
    \\
    $S^{\alpha\beta}$   & Angular momentum  & Eqs.~\eqref{SSC}, \eqref{eq:Spin Hall}, \eqref{consLaw}
    \\
    $\epsilon s$, $l$
        & Angular momentum parameters
        & Eqs.~\eqref{Spin2}, \eqref{6XII22.3}, \eqref{eq:numerics:WKB data}
    \\    
    $\mathcal{P}_\xi$ 
        & $\xi^\alpha$ component of generalized momentum
        & Eqs.~\eqref{consLaw}, \eqref{eq:momentum generalized}
    \\
    $\mathcal{S}$, $\mathcal{S}_1$
        & Scattering matrices
        & Eqs.~\eqref{Smatrix}, \eqref{SDef}, \eqref{Smatrix2}
    \\
    \midrule
    $\sigma$
        & Synge's world function
        & \cref{sigma}
        \\
    $\Delta$
        & van~Vleck determinant
        & \cref{eq:van Vleck determinant}
        \\
    $G$
        & Scalar Green function
        & \cref{eq:Green function scalar}
        \\
    $\Sigma$, $\Sigma_\tau$
        & Hypersurfaces
        & Eqs.~\eqref{eq:momentum generalized}, \eqref{Kirchhoff0}
        \\
    \midrule
    $\psi_{m,n}^\text{HG}$
        & Hermite–Gauss modes
        & \cref{6XII22.2}
        \\
    $\psi_{l,p}^\text{LG}$
        & Laguerre–Gauss modes
        & \cref{6XII22.3}
        \\
    \midrule
    $T^{\alpha\beta}$
        & Stress-energy tensor
        & Eqs.~\eqref{eq:momentum generalized}, \eqref{Tscalar}
    \\
    $a$, $\omega$
        & Initial scalar wave packet amplitude and frequency
        & Eqs.~\eqref{eq:numerics:WKB data}, \eqref{eq:numerics:WKB amplitude}
        \\
    \midrule
    $U$, $U_0$
        & Gravitational wave width and center
        &  \cref{gw_profile}
    \\
    $\nu$, $\phi_+$, $\phi_\times$
        & Gravitational wave frequency and phase offsets
        & \cref{gw_profile}
    \\
    $l_+$, $l_\times$
        &  Curvature length scales
        & Eqs.~\eqref{constPulse}, \eqref{gw_profile}
    \\
    $A \equiv 2/\sqrt{\pi} l_+^2$
        &   Curvature amplitude
        & Eq.~\eqref{ADef}, \cref{fig:MvvNonlinear}
    \\
    \bottomrule
  \end{tabular}
  
  \caption{\label{tab:symbols} Table of symbols. }
  \label{Table:Notation}
\end{table*}

\section{Example plane waves and their memory tensors}
\label{App:Examples}

This paper mostly takes the perspective that the precise waveform of a gravitational plane wave is less relevant than its associated memory tensors. However, it is always possible to compute the latter from the former, at least numerically. This Appendix provides three examples of specific plane wave spacetimes in which the memory tensors can be computed analytically: one exactly and two approximately. Our first example is a constant, finite-width pulse. Although this does not model any astrophysically relevant radiation, all memory tensors can easily be computed in the fully nonlinear regime. Our latter two examples are treated only in the linear approximation. One of these describes an oscillating burst while the other models the radiation which is emitted by the head-on collision of two masses.

\subsection{Exact constant pulse} \label{app:exact_pulse}

Perhaps the simplest nontrivial examples of vacuum sandwich waves are constant pulses with duration $U > 0$ and curvature length scale $l > 0$ (which is not to be confused with the angular momentum parameter $l$ employed in \cref{sec:HG_LG_scattering,Sect:SHcompare}). Any such pulse must be linearly polarized, so coordinates can be chosen in which its waveform is diagonal: Letting $\Theta$ denote the Heaviside step function, 
\begin{equation}
    H_{ij}(u) = 
    l^{-2} \Theta(u) \Theta(U-u) \, \mathrm{diag}[1,-1]_{ij} .
    \label{constPulse}
\end{equation}
Gravitational waves of this kind are not reasonable models for the far-field radiation emitted by any compact system in an asymptotically flat context. However, if the Heaviside functions in \cref{constPulse} were omitted, the resulting constant-curvature wave could be obtained by applying a Penrose limit to the Schwarzschild light ring; cf. Ref.~\cite{BlauNotes} and also \cref{HSchw}. Our waveform might, therefore, be viewed as a crude model for the Penrose limit of a null geodesic in the Schwarzschild spacetime which comes in from infinity, circles multiple times near the light ring, and then leaves. Regardless of interpretation, the waveform \eqref{constPulse}  provides a simple example in which all memory tensors can be computed exactly. Square pulses of this kind have also been considered, from a different perspective, in Ref. \cite{Chakraborty2022}.

In order to compute the memory tensors here, it is sufficient to obtain the Jacobi propagator $B_{ij}(u,u')$ as described in \cref{Sect:Jacobi}. Supposing that $u$ and $u'$ both lie in $\mathcal{I}_- = (-\infty,0)$, this is simply $(u-u') \delta_{ij}$. However, if $u' \in \mathcal{I}_-$ while $u \in \mathcal{I}_C = (0,U)$, solving \cref{JacobiMatrix} with $E_{ij} = B_{ij}$ results in 
\begin{widetext}
\begin{align}
    B_{ij} (u,u') = l \, \mathrm{diag} \big[ \sinh (u/l)  - (u'/l) \cosh (u/l) , \sin (u/l)  -(u'/l) \cos ( u/l)  \big]_{ij} .
\end{align}
Finally, if $u' \in \mathcal{I}_-$ while $u \in \mathcal{I}_+ = (U,\infty)$, 

\begin{align}
    B_{ij}(u,u') &= l \, \mathrm{diag} \big[ \sinh (U/l)  -(u'/l) \cosh (U/l) + \big( \cosh (U/l) - (u'/l) \sinh (U/l) \big) (u-U)/l , \nonumber
    \\ & \qquad\qquad\qquad\qquad \sin (U/l)  -(u'/l) \cos (U/l) + \big( \cos (U/l) + (u' /l) \sin (U/l) \big) (u-U)/l  \big]_{ij} .
    \label{Bpulse}
\end{align}
Applying \cref{AfromB} allows us to use this to obtain the other Jacobi propagator,
\begin{equation}
    A_{ij} (u,u') = \mathrm{diag} \big[ \cosh (U/l) + l^{-1} (u-U) \sinh (U/l) , \, \cos (U/l) - l^{-1} (u-U) \sin (U/l) \big]_{ij}.
    \label{Apulse}
\end{equation}
Comparison with \cref{Bexpand} then shows that the memory tensors are given by
\begin{subequations}
    \begin{gather}
    \Mvv_{ij} = \mathrm{diag} \big[ 
        \cosh (U/l) - 1  , \, \cos (U/l) - 1 \big]_{ij}, \qquad 
    \Mdv_{ij} = l^{-1} \, \mathrm{diag} \big[
        \sinh (U/l) , \, - \sin (U/l) \big]_{ij},
    \\
    \Mdd_{ij} = \mathrm{diag} \big[
        \cosh (U/l) - (U/l) \sinh (U/l) - 1     , \,  \cos (U/l) + (U/l) \sin (U/l) - 1 \big]_{ij} ,
    \\
    \Mvd_{ij} =l \, \mathrm{diag} \big[
         \sinh (U/l)  - (U/l) \cosh (U/l)   , \,  \sin ( U/l) - (
        U/l) \cos (U/l) \big]_{ij}.
\end{gather}
\end{subequations}
\end{widetext}
These expressions are exact. Unlike in the linear context, it is clear that none of these tensors are trace-free and also that $\Mdd_{ij} \neq - \Mvv_{ij}$. 

If we linearize in $H_{ij}$, which is equivalent to expanding through quadratic order in $1/l$, then the memory tensors here reduce to
\begin{subequations}
\begin{align}
    \Mdv_{ij} &= U^{-1} (U/l)^2 
    \, \mathrm{diag}[1,-1]_{ij} + \mathcal{O}(l^{-4}),
    \\
    \Mvv_{ij} &= - \Mdd_{ij} + \mathcal{O}(l^{-4})
    \nonumber
    \\
    &= \frac{1}{2} (U/l)^2 
   \, \mathrm{diag}[1,-1]_{ij} + \mathcal{O}(l^{-4}),
    \\
    \Mvd_{ij} &= - \frac{1}{3} U (U/l)^2
   \, \mathrm{diag}[1,-1]_{ij} + \mathcal{O}(l^{-4}).
\end{align}
\end{subequations}
These expressions \emph{are} trace-free, as expected. They can also be interpreted as the leading-order expansions in $U$. Of course, these tensors (and their nonperturbative counterparts) depend on a choice of origin for the $u$ coordinate. It follows from \cref{memXForm} that although no shift in the $u$ coordinate can be used to eliminate $\Mvd_{ij}$ or $\Mdv_{ij}$, both $\Mdd_{ij}$ and $\Mvv_{ij}$ can be simultaneously eliminated in the weak-field limit. Doing so has the consequence of leaving $\Mdv_{ij}$ unchanged while dividing $\Mvd_{ij}$ by 4.

If the origin of the $u$ coordinate is left as-is, then the first nonlinearities here result in, e.g.,
\begin{gather}
    \Mdd_{ij} + \Mvv_{ij} = - \frac{1}{12} ( U/l )^4 \delta_{ij} + \mathcal{O}(l^{-6}).
\end{gather}
More general results on how perturbative and nonperturbative nonlinearities affect the Jacobi propagators may be found in Ref.~\cite{HartePWOptics2}. 

Another point which can be noted here is that there may be cases in which $B_{ij}(u,u')$ fails to be invertible. Recalling the discussion surrounding \cref{detB}, the corresponding 
values of $u$ and $u'$ fix the conjugate hyperplanes \cite{HarteDrivas}. These are associated with geodesic focusing, as well as, e.g., the breakdown of both the Green function \eqref{eq:Green function scalar} and of the Kirchhoff-like formula \eqref{Kirchhoff}. For the constant pulse considered here, it follows from \cref{Bpulse} that if $u' \in \mathcal{I}_-$ is fixed, there exists some $u \in \mathcal{I}_+$ which is conjugate to $u'$ whenever
\begin{equation}
    \frac{ l \tan (U/l) - u' }{ 1 + (u'/l) \tan (U/l) } < 0 .
    \label{conjPoints}
\end{equation}
There are two possibilities here, depending on the sign of $\tan (U/l)$. If $\tan (U/l) > 0$, which occurs, e.g., for short or weak pulses, the relevant criterion reduces to $u'/l < - \cot (U/l)$. All geodesics which begin somewhere in the sufficiently distant past are thus focused in the future of the gravitational wave pulse. For pulses where $\tan (U/l)$ is instead negative, focusing from $\mathcal{I}_-$ to $\mathcal{I}_+$ occurs only from the finite interval in which $\tan (U/l) < u'/l < 0$. 

A similar analysis may be used to find when $A_{ij}(u,u')$ fails to be invertible, which signals the divergence of the initially planar scalar wave given by \cref{psiOutExact}. Using \cref{Apulse}, such waves diverge for some $u \in \mathcal{I}_+$ only when
\begin{equation}
    \tan (U/l) > 0.
\end{equation}
This criterion also describes the existence of a focal point \cite{Bilocal4}, a location where an initially comoving collection of geodesics would be focused after passing through the given gravitational wave (as in, e.g., \cref{fig:null_geos}).

\subsection{Oscillating waves} \label{app:weak_waves}

The list of waveforms whose effects can be understood nonperturbatively (and nonnumerically) is rather short. However, it is straightforward to perturbatively compute memory tensors in a wide range of scenarios. One class of possibilities are the waveforms
\begin{widetext}
\begin{equation}
   H_{ij}(u) = \frac{ e^{-[(u - U_0)/2 U]^2} }{2 \sqrt{\pi} }
   \begin{pmatrix*}[r]
      l_+^{-2} \sin (\nu u + \phi_+) &   l_\times^{-2} \sin(\nu u +\phi_\times)  \\
       l_\times^{-2} \sin(\nu u + \phi _\times) &   -l_+^{-2} \sin (\nu u +\phi_+)
   \end{pmatrix*}_{ij}.
   \label{gw_profile}
\end{equation}
A wave in this class oscillates with angular frequency $\nu$, and those oscillations are modulated by a Gaussian centered at $u=U_0$ and with width $U$. The curvature lengths of the + and the $\times$ polarized components of the wave are parametrized by $l_+$ and $l_\times$, and these components can also have the constant phase offsets $\phi_+$ and $\phi_\times$ [where the $\phi_+$ here should not be confused with the phase in \cref{dphi}]. Such a wave is + polarized when $l_\times \to \infty$ and $\times$ polarized when $l_+ \to \infty$. More generally, it is linearly polarized in both of these cases but also whenever $\phi_+ = \phi_\times$.  The wave may be described as circularly polarized whenever $l_\times = l_+$ and $\phi_\times = \phi_+ \pm \pi/2$.

Although these are technically not sandwich waves, they decay sufficiently fast that memory tensors remain a useful concept. If we assume that the amplitudes here are sufficiently small, then \cref{memTensLin} can be used to show that
\begin{subequations}
\label{memOscwaves}
\begin{gather}
    \Mdv_{ij} = U e^{- \nu^2 U^2}
    \begin{pmatrix*}[r]
        l_+^{-2} \sin (\nu U_0 + \phi_+) &       l_\times^{-2} \sin (\nu U_0 + \phi_\times) \\
        l_\times^{-2} \sin (\nu U_0 + \phi_\times) &
        -l_+^{-2} \sin (\nu U_0 + \phi_+)
    \end{pmatrix*}_{ij} + \mathcal{O}(H^2),
    \\
    \Mdd_{ij} = - 2 \nu U^3 e^{-\nu^2 U^2} 
     \begin{pmatrix*}[r]
        l_+^{-2} \cos (\nu U_0 + \phi_+) &       l_\times^{-2} \cos (\nu U_0 + \phi_\times) \\
        l_\times^{-2} \cos (\nu U_0 + \phi_\times) &
        -l_+^{-2} \cos (\nu U_0 + \phi_+)
    \end{pmatrix*}_{ij}
    - U_0 \Mdv_{ij} + \mathcal{O}(H^2),
    \label{Mddexample}
    \\
    \Mvd_{ij} = \big[ 2 U^2 ( 2 \nu^2 U^2 - 1 ) + U_0^2 \big] \Mdv_{ij} + 2 U_0 \Mdd_{ij} + \mathcal{O}(H^2),
\end{gather}
\end{subequations}
\end{widetext}
and $\Mvv_{ij} = - \Mdd_{ij} + \mathcal{O}(H^2)$. All memory tensors are generically nonzero at this order. However, some of these tensors can be made to vanish in special cases. For example, $\Mdv_{ij}$ vanishes at linear order when, e.g., $\phi_+ = \phi_\times = - \nu U_0$, which would correspond to a particular class of linearly polarized waves. If one would like to construct a more general class of oscillating bursts in which $\Mdv_{ij}$ vanishes at linear order, one way to do so would be to let $H_{ij}$ be proportional to a $u$-derivative of the right-hand side of \cref{gw_profile}; cf. \cref{Hsimp}.

\begin{figure}
    \centering
    % eigenvalues of Mvv
    \includegraphics[width=\columnwidth]{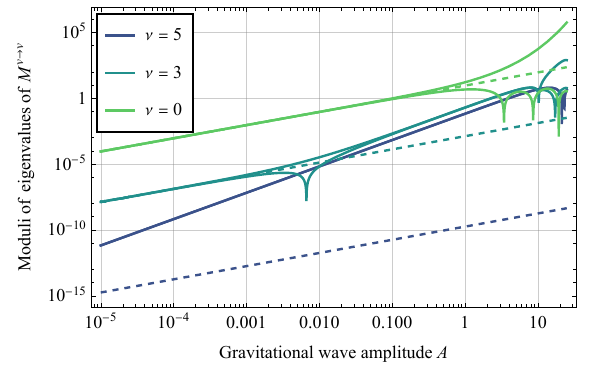}
    \caption[Absolute values of the eigenvalues of the velocity-velocity memory]{Absolute values of the eigenvalues of $\Mvv_{ij}$, as a function of amplitude, for circularly polarized gravitational plane waves with Gaussian profiles. Solid lines are obtained by numerical integration of the Jacobi equation, while dashed lines are linear approximations obtained from \cref{Mddexample} and $\Mvv_{ij} = - \Mdd_{ij} + \mathcal{O}(H^2)$. In units where $U=1$, all waveforms here are given by \cref{gw_profile} with $U_0 = 10$, $\phi_+ = 0$, $\phi_\times = \pi/2$, and $l_+ = l_\times$, which matches the parameters considered in \cref{Sect:SHcompare}. The amplitude plotted here is defined to be $A \equiv 2/\sqrt{\pi} l_+^2$. At larger gravitational wave frequencies $\nu$, the linear approximation is seen to fail even for relatively small  amplitudes. It may also be seen that at least in the $\nu =0$ and $\nu =3$ cases, the two eigenvalues differ in magnitude at large $A$, implying that $\Mvv_{ij}$ acquires a significant trace. 
    }
    \label{fig:MvvNonlinear}
\end{figure}

It may be noted that all of the linearized memory tensors given by \cref{memOscwaves} are exponentially suppressed when $\nu U \gg 1$. In these cases, the waveform nearly averages to zero over many oscillations. However, a similar suppression does not occur at higher orders, essentially because the squared waveform does not average to zero. This is illustrated in \cref{fig:MvvNonlinear}. Nonlinearity might therefore be considerably more important in these cases than simple estimates suggest. See also \cite{HartePWOptics2, HartePWMemory} for other cases in which nonlinear effects can be relatively large.

\subsection{Weak gravitational waves from colliding masses}

Our final example concerns the gravitational wave burst that is emitted by two massive objects which collide head-on, leaving a single static remnant. Idealizing such a process, the quadrupole formula suggests that it can be modeled by an impulsive plane wave with the waveform
\begin{equation}
	H_{ij}(u) = \alpha \delta'(u) \mathrm{diag}[1,-1]_{ij},
 \label{impulsiveWave}
\end{equation}
where $\alpha$ is a constant which can be related to the initial energies of the two objects and their final distance from the observer. A similar waveform arises if, instead, one mass splits into two. Regardless, applying \cref{memTensLin} to \cref{impulsiveWave}, one finds that $\Mdv_{ij}$  vanishes at linear order, as expected for an “astrophysically reasonable” gravitational wave. In fact, $\Mvd_{ij}$ vanishes as well. The only nontrivial memory tensors here are
\begin{equation}
	\Mdd_{ij} = \alpha \, \mathrm{diag}[1,-1]_{ij}  + \mathcal{O}(\alpha^2)
\end{equation}
and $\Mvv_{ij} = - \Mdd_{ij} + \mathcal{O}(\alpha^2)$.
In this weak-field approximation, there are no conjugate hyperplanes. 

\section{Ward’s formula for scalar fields on plane wave backgrounds}
\label{App:Ward}

In \cref{sec:Green_function}, we have used a Green function for the scalar wave equation in order to derive the Kirchhoff-like scattering formula \eqref{Kirchhoff} for scalar fields on gravitational plane wave backgrounds. Although the relation is not obvious, this may be shown to be a special case of an integral representation that has been previously derived by Ward \cite{Ward1987} using different methods (see also Ref.~\cite{Mason_Ward} for an extension to electromagnetic fields). This Appendix explains how the two representations are related and uses that relation to physically interpret the parameters in Ward's construction.  

Ward worked in Rosen coordinates $(\mathscr{U}, \mathscr{V} , \mathscr{X}^i)$, where the line element takes the form
\begin{equation}
    \rmd s^2 = - 2 \rmd \mathscr{U} \rmd \mathscr{V} + \gamma_{ij}( \mathscr{U} ) \rmd \mathscr{X}^i \rmd \mathscr{X}^j.
    \label{Rosen}
\end{equation}
In these coordinates, it is the 2-metric $\gamma_{ij}(\mathscr{U} )$ that encodes the waveform of the gravitational wave [rather than the $H_{ij}(u)$ that appears in the Brinkmann line element in \cref{Brinkmann}]. In terms of a transverse-traceless metric perturbation $h_{ij}$, the Rosen waveform might be taken to have the form $\gamma_{ij} = \delta_{ij} + h_{ij} + \mathcal{O}(h^2)$. In any case, Ward  introduced the family of scalar fields
\begin{align}
    S_P (\mathscr{U}, \mathscr{V} , \mathscr{X}^i) \equiv \mathscr{V} + P_i \mathscr{X}^i + \frac{1}{2} P_i P_j \bigg( h_0^{ij} + 
    \nonumber
    \\
    ~ + \int_{\mathscr{U}_0}^\mathscr{U} \!\! [\gamma^{-1}(\mathscr{W})]^{ij}  \,\rmd \mathscr{W} \bigg),
    \label{SPDef}
\end{align}
where $\mathscr{U}_0$, $h_0^{ij}$, and $P_i$ are arbitrary constants. These scalars were shown to satisfy the eikonal equation 
\begin{equation}
    \nabla^\alpha S_P \nabla_\alpha S_P = 0,
\end{equation}
so constant-$S_P$ hypersurfaces may be viewed as wavefronts in, e.g., geometric optics. However, much more can be said: If $f$ is any scalar function of one variable,
\begin{equation}
   (\det \gamma_{ij})^{-1/4} f( S_P )
    \label{progWave}
\end{equation}
is an \emph{exact} solution to the massless scalar wave equation. Since $f$ is arbitrary here, solutions with this form are examples of ``progressing waves'' \cite{Ward1987, Friedlander, FriedlanderProgWaves}. 

Summing up solutions with different $P_i$, it follows that given any $F = F(S_P, P_i)$,
\begin{equation}
    \psi = \frac{ 1 }{(\det \gamma_{ij})^{1/4}}  \int F ( S_P, P_k) \,\rmd^2 P
    \label{psiWard}
\end{equation}
is a solution to the massless scalar wave equation. In the flat-spacetime case where coordinates can be chosen such that $\gamma_{ij} = \delta_{ij}$, this reduces to a certain integral formula for solutions to the wave equation which was originally obtained by Whittaker \cite{Whittaker1903, Ward1987}.  

However, in order to relate Ward's integral \eqref{psiWard} to our representation \eqref{Kirchhoff}, we must first translate the former into the Brinkmann coordinates $(u,v,x^i)$ which are used in the latter. Letting $u= \mathscr{U}$, the appropriate transformation is given by \cite{HarteDrivas, HartePWOptics2, Adamo2017}
\begin{equation}
    x^i = E^{i}{}_{j} (\mathscr{U}) \mathscr{X}^j, \quad v = \mathscr{V} + \frac{1}{2} E^{k}{}_{i} (\mathscr{U}) \dot{E}_{kj} (\mathscr{U}) \mathscr{X}^i \mathscr{X}^j,
    \label{coordXformRosen}
\end{equation}
where $E_{ij}(\mathscr{U})$ is any solution to the matrix Jacobi equation \eqref{JacobiMatrix} in which 
\begin{equation}
    E_{k[i} \dot{E}^{k}{}_{j]} = 0.
    \label{Esym}
\end{equation}
If this latter constraint is satisfied at any one value of $\mathscr{U}$, then it may be shown to be satisfied for all $\mathscr{U}$. Applying the given coordinate transformation to the Brinkmann line element \eqref{Brinkmann} recovers the Rosen line element \eqref{Rosen} with the transverse metric
\begin{equation}
    \gamma_{ij} = E^{k}{}_{i} E_{kj}.
    \label{transverseMetric}
\end{equation}
Different choices for $E_{ij}$ result in different Rosen coordinate systems, with different (though physically equivalent) transverse metrics. It is clear from the discussion in \cref{Sect:Jacobi} that if $u'$ is viewed as a fixed parameter, $E_{ij}(u)$ must be a linear combination of the Jacobi propagators $A_{ij}(u,u')$ and $B_{ij}(u,u')$.

Elsewhere in the literature, it has been common to identify $E_{ij}$ with $A_{ij}$ \cite{Ward1987, Adamo2017, Adamo2017_2, Adamo2020}. We also do so in \cref{Sect:PWscatter} above in order to understand how planar test fields are scattered by plane gravitational waves. However, relating Ward's formula to our Kirchhoff integral \eqref{Kirchhoff} instead requires the choice 
\begin{equation}
    E_{ij} = B_{ij}.    
    \label{EB}
\end{equation}
The resulting Rosen coordinate system is then singular at least at $u = u'$, but this is not problematic in Brinkmann coordinates (which are valid everywhere). We also note that from \cref{ABidentity,transverseMetric}, as well as from the symmetry of $\partial_u B_{i}{}^{k} (B^{-1})_{kj}$ which follows from \cref{SymMatrices},
\begin{equation}
    \partial_u [ (B^{-1} A)^{ij} ] = - (B^{-1})^{ik} (B^{-1})_{k}{}^j = - (\gamma^{-1})^{ij}.
\end{equation}
Choosing $h_0^{ij} = - \left. (B^{-1})^{ik} A_{k}{}^{j} \right|_{(\mathscr{U}_0,u')}$ in  \cref{SPDef}, use of \cref{coordXformRosen} with $P_i = x'_i$ then shows that
\begin{equation}
    S_{x'} = v_\mathrm{ret},
\end{equation}
where $v_\mathrm{ret}$ is given by \cref{vRet}. This provides a clear physical interpretation for the eikonal here: Given an event with Brinkmann coordinates $(u,v,x^i)$, the event $(u', S_{x'}, x'^i)$ is null separated from it. Furthermore, it follows from \cref{eq:van Vleck determinant} that 
\begin{equation}
     ( \det \gamma_{ij} )^{-1/4} = ( \det B_{ij} )^{-1/2} = \frac{\Delta^{1/2}(u,u')}{u-u'}
\end{equation}
in this case. Our integral \eqref{Kirchhoff} is therefore equivalent to Ward's \eqref{psiWard} when $E_{ij} = B_{ij}$, when the constants $h_0^{ij}$ and $\mathscr{U}_0$ are chosen appropriately, and when $2\pi F(S_{x'}, x') = \left. \partial_{v'} \psi(u', v', x'^i) \right|_{v' = S_{x'}}$. The parameters $P_i$ which appear in Ward's original result are thus seen to be interpreted as transverse Brinkmann coordinates $x'^i$ on an ``initial'' null hypersurface. Furthermore, the function $F$ is seen to be a first derivative of the field on that hypersurface. Note however that these simple interpretations arise only when the Rosen coordinate system which is used in the original construction breaks down on the initial hypersurface. Interpretations  differ when, e.g., $E_{ij} = A_{ij}$.

\section{Charged particles in an electromagnetic sandwich wave}
\label{Sect:EMmemory}

The majority of this paper is concerned with gravitational wave memory. However, there are also electromagnetic memory effects, and it is instructive to compare with them. As in the gravitational context, electromagnetic memory can be considered either as an effect which arises far away from a compact source, or as an effect which is associated with a local plane wave (without reference to any particular source). In the former case, charged particles have been claimed to generically experience a change in velocity due to the passage of electromagnetic radiation \cite{Bieri2013}. However, this has recently been challenged \cite{noEMkick} with the statement that there can be a displacement memory but no velocity memory.  Following the gravitational discussion in this paper, which focuses on arbitrary plane waves without reference to their sources, we do not enter here into this disagreement, which depends on the asymptotic properties of electromagnetic fields.

Instead, we consider the motion of a charged particle in flat spacetime which is accelerated by an arbitrary electromagnetic ``sandwich wave.'' The position $x^\alpha(\tau)$ of a particle with charge $q$ and mass $m$ obeys the Lorentz force equation
\begin{equation}
    % \frac{\mathrm{D}^2 x^\alpha}{\rmd \tau^2} 
    \frac{\mathrm{D}}{\rmd \tau} 
    \frac{\rmd x^\alpha}{\rmd \tau} 
    = \frac{q}{m} F^{\alpha}{}_{\beta} \frac{ \rmd x^\beta }{ \rmd \tau } ,
\end{equation}
where $\tau$ denotes a proper time along the particle's worldline. Furthermore, the electromagnetic field of an arbitrary plane wave can be described by \cref{EMPW}. Combining these two expressions shows that
\begin{equation}
    \ell_\alpha \frac{ \rmd x^\alpha }{ \rmd \tau } = - \frac{ \rmd u }{ \rmd \tau }  = \mathrm{constant}  .
\end{equation}
If Brinkmann coordinates are adopted so $\rmd s^2=-2 \rmd u \rmd v + \rmd x^2 + \rmd y^2$, then the two transverse components of the Lorentz force equation reduce to 
\begin{equation}
    \frac{ \rmd^2 x^i }{ \rmd u^2 } = \frac{q}{2m} \frac{ \rmd \tau }{ \rmd u} \mathcal{A}^i(u),
\end{equation}
where $\mathcal{A}^i(u)/2 \sqrt{2}$ is the electric field which would be seen by an observer at constant $x,y,z$. This can be solved to yield
\begin{align}
    x^i(u) &= x^i(u') + (u-u') \dot{x}^i(u') 
    \nonumber
    \\
    & ~ + \frac{q}{2m} \frac{ \rmd \tau }{ \rmd u} \int_{u'}^u \rmd w\, (u-w) \mathcal{A}^i(w),
    \label{xEM}
\end{align}
which is exact for arbitrary electromagnetic waveforms.

If we consider a sandwich wave, so $F_{\alpha\beta}$ vanishes for all $u$ outside some finite interval, we can now ask how charged particles are scattered. It follows from \cref{xEM} that all relevant information is captured by the memory vectors
\begin{align}
    M_{\mathrm{v}}^i \equiv \int_{-\infty}^\infty \rmd u\, \mathcal{A}^i (u), \quad M_{\mathrm{d}}^i \equiv -\int_{-\infty}^\infty \rmd u\, u \mathcal{A}^i (u),
\end{align}
which are essentially the zeroth and the first moments of the electric field. If $u'$ lies before the wave has arrived and $u$ after it has left, then the transverse scattering variables $x^i_\pm$ and $\dot{x}^i_\pm$ which are defined by \cref{xPM} are related via
\begin{equation}
      \begin{pmatrix}
          x^i_+ \\ \dot{x}^i_+ 
      \end{pmatrix}  = 
      \begin{pmatrix}
          x^i_- \\ \dot{x}^i_- 
      \end{pmatrix} + 
      \frac{q}{2m} \frac{ \rmd \tau }{ \rmd u}
      \begin{pmatrix}
          M_\mathrm{d}^i \\ M_\mathrm{v}^i 
      \end{pmatrix}.
\end{equation}
This may be compared with \cref{Smatrix,SDef}, which describe the transverse scattering of geodesics by a gravitational plane wave. One similarity is that the memory effects in both cases depend on the lowest moments of the relevant waveform (at least when linearizing in the gravitational case). However, this similarity also leads to a difference: Electromagnetic memory is parametrized by memory \emph{vectors} rather than second-rank memory \emph{tensors}. This is essentially because the electromagnetic waveform is a vector while the gravitational waveform is a second-rank symmetric tensor. An even more striking difference is that although the gravitational memory is linear in the initial transverse state, the electromagnetic memory is independent of it. All charged particles with the same $(q/m)\rmd \tau / \rmd u$ experience the same transverse kicks and the same transverse displacements.

\begin{widetext}
\section{Derivatives of HG and LG beams}
\label{app:beam_derivatives}

It is possible to use \cref{scatteredField} to understand the scattering of the Hermite–Gauss and Laguerre–Gauss beams analytically, at least in the linear approximation. In order to do so, we now give analytical expressions for the various derivatives appearing in that equation when HG or LG solutions are used as seed fields. Our results follow easily by applying the standard properties of Hermite and Laguerre polynomials \cite{andrews_askey_roy_1999}. They are used in \cref{sec:HG_LG_scattering} to understand how gravitational wave memory affects localized scalar beams.

\subsection{Counterpropagating HG beams}

For counterpropagating HG beams $\psi_{m,n}^\text{HG}$ defined in \cref{6XII22.2}, the relevant partial derivatives for calculating the scattered beam are 
\begin{subequations}
\begin{align}
    \label{eq:memory HG dv}
    \partial_v  \psi_{m,n}^\text{HG} &= \frac{\ii}{2 \sqrt{2} k w_0^2} \bigg[ 2(m+n+1) \psi_{m,n}^\text{HG} -\sqrt{m(m-1)} \psi_{m-2,n}^\text{HG} -\sqrt{n(n-1)} \psi_{m,n - 2}^\text{HG} \nonumber \\
    &\qquad\qquad\qquad\qquad\qquad -\sqrt{(m+1)(m+2)} \psi_{m+2,n}^\text{HG} -\sqrt{(n+1)(n+2)} \psi_{m,n + 2}^\text{HG} \bigg], \\
    \partial_x  \psi_{m,n}^\text{HG} &= \frac{1}{w_0} \left( \sqrt{m} \psi_{m-1,n}^\text{HG} - \sqrt{m+1} \psi_{m+1,n}^\text{HG} \right), \\
    \partial_y  \psi_{m,n}^\text{HG} &= \frac{1}{w_0} \left( \sqrt{n} \psi_{m,n-1}^\text{HG} - \sqrt{n+1} \psi_{m,n+1}^\text{HG} \right).
\end{align}  
\end{subequations}
The remaining terms can in \cref{scatteredField} be calculated using the above expressions, together with
\begin{subequations}
\begin{align}
    x \psi_{m,n}^\text{HG} &= \frac{1}{2 k w_0} \bigg[ \sqrt{m} \left( w_0^2 k + \ii \sqrt{2} v \right) \psi_{m-1,n}^\text{HG} + \sqrt{m+1} \left( w_0^2 k - \ii \sqrt{2} v \right) \psi_{m+1,n}^\text{HG} \bigg], \\
    y \psi_{m,n}^\text{HG} &= \frac{1}{2 k w_0} \bigg[ \sqrt{n} \left( w_0^2 k + \ii \sqrt{2} v \right) \psi_{m,n-1}^\text{HG} + \sqrt{n+1} \left( w_0^2 k - \ii \sqrt{2} v \right) \psi_{m,n+1}^\text{HG} \bigg].
\end{align}
\end{subequations}
Using these equations, together with \cref{scatteredField}, we can easily determine the side modes excited by different components of the memory tensors when the ingoing beam before the gravitational wave is $\partial_v \psi_{m,n}^\text{HG}$. The excited side modes are summarized in \cref{fig:memory modes HG}.

\subsection{Counterpropagating LG beams}

For counterpropagating LG beams $\psi_{l,p}^\text{LG}$ defined in \cref{6XII22.3}, the relevant partial derivatives for calculating the scattered beam are 
\begin{subequations}
\begin{align}
    \label{eq:memory LG dv}
    \partial_v  \psi_{l,p}^\text{LG} &= \frac{\ii}{\sqrt{2} k w_0^2} \left[ \sqrt{p(p+|l|)} \psi_{l,p-1}^\text{LG} + (2p + |l| + 1) \psi_{l,p}^\text{LG} + \sqrt{(p+1)(p+|l|+1)}\psi_{l,p+1}^\text{LG} \right], \\
    \partial_x  \psi_{l,p}^\text{LG} &= \left[ \frac{e^{+\ii \theta}}{2} \left( \partial_r + \frac{\ii}{r} \partial_\theta \right) + \frac{e^{-\ii \theta}}{2} \left( \partial_r - \frac{\ii}{r} \partial_\theta \right) \right] \psi_{l,p}^\text{LG}, \\
    \partial_y  \psi_{l,p}^\text{LG} &= \left[ \frac{e^{+\ii \theta}}{2 \ii} \left( \partial_r + \frac{\ii}{r} \partial_\theta \right) - \frac{e^{-\ii \theta}}{2 \ii} \left( \partial_r - \frac{\ii}{r} \partial_\theta \right) \right] \psi_{l,p}^\text{LG},
\end{align}  
\end{subequations}
where
\begin{subequations} \label{eq:LG_der_1}
\begin{align}
    \frac{e^{+\ii \theta}}{2} \left( \partial_r + \frac{\ii}{r} \partial_\theta \right)  \psi_{l,p}^\text{LG} &= \frac{1}{\sqrt{2} w_0}
        \begin{cases}
		  - \sqrt{p + |l| + 1} \psi_{l-1,p}^\text{LG} - \sqrt{p} \psi_{l-1,p-1}^\text{LG}  
			& l \leq 0,
				\\
			+ \sqrt{p + |l|} \psi_{l-1,p}^\text{LG} + \sqrt{p+1} \psi_{l-1,p+1}^\text{LG}
            & l > 0. 
		\end{cases} \\
    \frac{e^{-\ii \theta}}{2} \left( \partial_r - \frac{\ii}{r} \partial_\theta \right)  \psi_{l,p}^\text{LG} &= \frac{1}{\sqrt{2} w_0}
        \begin{cases}
	       +  \sqrt{p + |l|} \psi_{l+1,p}^\text{LG} + \sqrt{p+1} \psi_{l+1,p+1}^\text{LG}
	       & l < 0, \\
	       - \sqrt{p + |l| + 1} \psi_{l+1,p}^\text{LG} -\sqrt{p} \psi_{l+1,p-1}^\text{LG}
	       & l \geq 0.
    \end{cases}
\end{align}    
\end{subequations}
The remaining types of terms that appear in \cref{scatteredField} are
\begin{subequations}
\begin{align}
    x \psi_{l,p}^\text{LG} &= r \cos \theta \psi_{l,p}^\text{LG} = r \frac{e^{+\ii \theta} + e^{-\ii \theta}}{2} \psi_{l,p}^\text{LG}, \\
    y \psi_{l,p}^\text{LG} &= r \sin \theta \psi_{l,p}^\text{LG} = r \frac{e^{+\ii \theta} - e^{-\ii \theta}}{2 \ii} \psi_{l,p}^\text{LG},
\end{align}
\end{subequations}
where 
\begin{subequations} \label{eq:LG_der_2}
\begin{align}
    r e^{+\ii \theta} \psi_{l,p}^\text{LG} &=
        \frac{1}{\sqrt{2} k w_0}
        \begin{cases}
                  \sqrt{p + |l|+1} \left( k w_0^2 - \ii \sqrt{2} v \right) \psi_{l-1,p}^\text{LG}
                - \sqrt{p} \left( k w_0^2 + \ii \sqrt{2} v \right) \psi_{l-1,p-1}^\text{LG}
				& l \leq 0,
				\\
		          \sqrt{p + |l|} \left( k w_0^2 + \ii \sqrt{2} v \right) \psi_{l-1,p}^\text{LG} 
                - \sqrt{p + 1} \left( k w_0^2 - \ii \sqrt{2} v \right) \psi_{l-1,p+1}^\text{LG}
				& l > 0. 
		\end{cases} \\
    r e^{-\ii \theta} \psi_{l,p}^\text{LG} &= \frac{1}{\sqrt{2} k w_0} \begin{cases}
		          \sqrt{p + |l|} \left( k w_0^2 + \ii \sqrt{2} v \right) \psi_{l+1,p}^\text{LG} 
                - \sqrt{p + 1} \left( k w_0^2 - \ii \sqrt{2} v \right) \psi_{l+1,p+1}^\text{LG}
				& l < 0,
				\\
                  \sqrt{p + |l| + 1} \left( k w_0^2 - \ii \sqrt{2} v \right) \psi_{l+1,p}^\text{LG}
                - \sqrt{p} \left( k w_0^2 + \ii \sqrt{2} v \right) \psi_{l+1,p-1}^\text{LG}
				& l \geq 0. 
		\end{cases} 
\end{align}
\end{subequations}
Using these equations, together with \cref{scatteredField}, we can easily determine the side modes excited by different components of the memory tensors when the ingoing beam before the gravitational wave is $\partial_v \psi_{l,p}^\text{LG}$. The excited side modes are summarized in \cref{fig:memory modes LG}.
\end{widetext}

\bibliography{references}

\end{document}